\documentclass[12pt]{article}
\pdfoutput=1
\usepackage{jheppub}
\usepackage[utf8]{inputenc}
\usepackage{verbatim}
\usepackage{amsmath}
\usepackage{amssymb}
\usepackage{amsthm}
\usepackage{slashed}
\usepackage{amsfonts}

\newcommand{\be}{\begin{equation}}
\newcommand{\ee}{\end{equation}}
\newcommand{\bfig}{\begin{figure}\begin{center}}
\newcommand{\efig}{\end{center}\end{figure}}
\newcommand{\bi}{\begin{itemize}}
\newcommand{\ei}{\end{itemize}}

\newcommand{\lan}{\langle}
\newcommand{\ran}{\rangle}

\newcommand{\wt}{\widetilde}

\theoremstyle{definition}

\begin{document}
\title{Algebra of diffeomorphism-invariant observables in Jackiw-Teitelboim gravity}
\author[a]{Daniel Harlow}
\author[a,b]{and Jie-qiang Wu\footnote{Corresponding author}}
\affiliation[a]{Center for Theoretical Physics\\ Massachusetts Institute of Technology, Cambridge, MA 02139, USA}
\affiliation[b]{Department of Physics\\ University of California, Santa Barbara, CA, 93106, USA}
\emailAdd{harlow@mit.edu, jieqiang@ucsb.edu}
\abstract{In this paper we use the covariant Peierls bracket to compute the algebra of a sizable number of diffeomorphism-invariant observables in classical Jackiw-Teitelboim gravity coupled to fairly arbitrary matter.  We then show that many recent results, including the construction of traversable wormholes, the existence of a family of $SL(2,\mathbb{R})$ algebras acting on the matter fields, and the calculation of the scrambling time, can be recast as simple consequences of this algebra.  We also use it to clarify the question of when the creation of an excitation deep in the bulk increases or decreases the boundary energy, which is of crucial importance for the ``typical state'' versions of the firewall paradox.  Unlike the ``Schwarzian'' or ``boundary particle'' formalism, our techniques involve no unphysical degrees of freedom and naturally generalize to higher dimensions.  We do a few higher-dimensional calculations to illustrate this, which indicate that the results we obtain in JT gravity are fairly robust.}
\maketitle

\section{Introduction}
Diffeomorphism symmetry, also called general covariance, is an essential feature of Einstein's theory of gravity \cite{einstein2014meaning}.  It is the mathematical expression of the equivalence principle, which says that there should be no preferred frames of reference in the laws of physics.  In more modern language, it says that the only background fields in a theory of gravity should be constant scalars (also known as coupling constants).\footnote{String theory and holography go even further, saying that such coupling constants are really the expectation values of dynamical fields.}

It has long been understood that diffeomorphism symmetry must be a gauge symmetry, and that physical observables must therefore be invariant under almost all diffeomorphisms \cite{diraclectures,DeWitt:1967yk,Weinberg:1972kfs,Henneaux:1992ig,Giddings:2011xs}.  The only exceptions are those diffeomorphisms which are non-vanishing at the boundary of spacetime: depending on the choice of boundary conditions these can be physically meaningful, in which case they act nontrivially on the observables.  There is an analogous situation in electromagnetism: fields which carry electric charge are unphysical unless they are ``dressed'' with Wilson lines attaching them either to other fields with the opposite charge or to the boundary of spacetime \cite{Dirac:1955uv}.  Charged fields which are dressed by Wilson lines reaching a spatial boundary transform nontrivially under gauge transformations which approach a constant at that boundary, and those gauge transformations are generated by the electric flux through it.

One way to think about dressed observables in electromagnetism is that they create both a charged particle and its associated Coulomb field, which ensures that the resulting configuration obeys the Gauss constraint.  Similarly in gravity any local observable by itself will not be diffeomorphism-invariant, so we must dress it with some kind of ``gravitational Wilson line'' which creates for it a gravitational field that obeys the constraint equations of gravity.  In practice such observables are usually constructed by a ``relational'' approach: rather than saying we study an observable at some fixed coordinate location, we instead define its location relative to some other features of the state \cite{DeWitt:1967yk,Page:1983uc,Banks:1984cw,Kuchar:1991qf,Marolf:1994nz,Giddings:2005id}.  For example in computing the spectrum of density perturbations in the early universe, it is convenient to use the inflaton field as a clock \cite{Maldacena:2002vr}.  The situation is better when there is some kind of asymptotic boundary, as one can then define observables relative to that boundary. This is easiest if the observables live directly at that boundary such as the S-matrix in asymptotically-Minkowski space or boundary correlators in asymptotically-AdS space \cite{Giddings:2011xs,Polchinski:1999ry,Susskind:1998vk}. More generally one can use the boundary as a reference point for defining diffeomorphism-invariant observables which extend deeper into the spacetime \cite{Heemskerk:2012np,Kabat:2013wga,Donnelly:2015hta,Giddings:2018umg,Harlow:2018tng,Giddings:2019wmj}.

In $3+1$ dimensions gravitationally-dressed observables are somewhat difficult to manipulate, and so far basically all calculations have been limited to the first nontrivial order of gravitational perturbation theory \cite{Heemskerk:2012np,Kabat:2013wga,Donnelly:2015hta,Giddings:2018umg,Giddings:2019wmj}.  The situation however is better in lower numbers of spacetime dimensions, where in some cases the gravitational dynamics are exactly solvable, and thus a more complete treatment of the problem should be achievable.  The main goal of this paper is to give a fairly thorough treatment of the problem for the case of the Jackiw-Teitelboim (JT) gravity theory \cite{Teitelboim:1983ux,Jackiw:1984je,Almheiri:2014cka} coupled to matter fields in $1+1$ dimensions.  This theory has been extensively studied in recent years, mostly using a ``Schwarzian'' or ``boundary particle'' formalism which suppresses the usual gravity variables and also introduces additional redundancies \cite{Engelsoy:2016xyb,Jensen:2016pah,Maldacena:2016upp,Kitaev:2017awl,Almheiri:2018xdw,Lin:2019qwu}.  Following \cite{Harlow:2018tqv} we will instead study it directly in the gravity variables, where we will see that the algebra of a wide variety of diffeomorphism-invariant observables, including both dressed matter observables and also intrinsically gravitational observables such as the Hamiltonian, is computable in closed form.  With this algebra in hand we are then able to straightforwardly reproduce many of the interesting results in this theory, and we will also use it to learn some new things.  Our approach has the intrinsic advantage of not involving any unphysical degrees of freedom. It also  eases the transition to higher dimensions, where in general there is no natural analogue of the Schwarzian/boundary particle formalism, and we do some initial calculations to highlight this.

\bfig
\includegraphics[height=4cm]{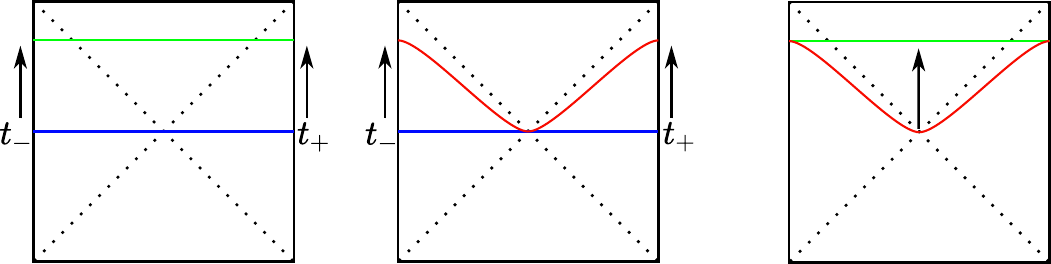}
\caption{Different kinds of bulk time evolution in the AdS-Schwarzschild geometry (figure adapted from \cite{Harlow:2018tqv}).  The ADM Hamiltonians $H_{\pm}$ move the boundary edges of a bulk Cauchy slice up and down, as in the left two diagrams, but the evolution of the interior of the slice, as in going from the red slice to the green slice in the right diagram, is generated by the Hamiltonian constraint and thus is trivial on gauge-invariant states.}\label{slicesfig}
\efig
The main question motivating our work is the nature of the interiors of black holes in quantum gravity.  A complete understanding of this problem surely requires a non-perturbative description of quantum gravity which goes beyond what is currently available, but some aspects of the problem are puzzling already at the classical level and it is those aspects which we aim to clarify in this paper.  In particular we have the following issues (see also \cite{Harlow:2018tqv}):
\bi
\item Naively one might guess that to evolve up into the black hole interior in AdS/CFT we should use the CFT time evolution, as suggested in the left diagram of figure \ref{slicesfig}.  But in fact we can just as easily understand this boundary evolution as implementing the bulk evolution in the middle diagram, where the Cauchy slice never enters the interior region.  The evolution from the red slice to the green slice is implemented by the Hamiltonian constraint, which acts trivially on the physical Hilbert space.  On the other hand there is clearly some physical distinction between observables which are behind the horizon and observables which are not, we just need to find a diffeomorphism-invariant way of discussing this (see \cite{Jafferis:2020ora,Giddings:2020usy} for other views on this).
\item It has been argued that black holes in sufficiently generic states possess singular ``firewalls'' at their horizons \cite{Almheiri:2012rt,Almheiri:2013hfa,Marolf:2013dba}.  These arguments are not rock-solid, but as of now the possibility can't be ruled out.  On the other hand if there really are firewalls, it seems implausible that they should really be found at the event horizon: the event horizon is defined acausally, and in particular can be moved by the presence of a large shell of matter very far away.    But where then should the firewall live?  A possibility which avoids this problem is the apparent horizon, but that depends on a choice of Cauchy slice and thus is not diffeomorphism-invariant \cite{Eardley:1997hk}.  Another possibility which is perhaps more appealing is that firewalls should instead live on the boundary of the future of the quantum extremal surface for the black hole exterior, sometimes called the \textit{entanglement horizon}. This in general is quite different from the event horizon, which could be a good thing, but it is also a non-linear function of the state of the system and thus doesn't correspond to any standard observable.\footnote{Some attempts to get rid of firewalls use non-linear observables, but here one would have to tolerate both non-linear observables \textit{and} firewalls, which seems like a bad deal.}   So far there has been no natural proposal for where firewalls should form which passes the tests of diffeomorphism-invariance, basic respect for locality, and linear dependence on the state.
\item A key element in any argument for firewalls is that a black hole can have a firewall in its interior without paying any significant cost in its energy \cite{Almheiri:2013hfa,Marolf:2013dba,Harlow:2014yka}.  Were this not the case, then firewalls would be Boltzmann-suppressed in the same way they are in daily life.  The standard way of explaining this is to note that the change in signature of the Killing symmetry of the Schwarzschild geometry as we cross the horizon causes the creation operators for outgoing modes behind the horizon to lower the energy defined with respect to Schwarzschild time, and thus right at the horizon we can excite these modes substantially without a big energy cost \cite{Almheiri:2013hfa}.  This argument however is rather specific to states which have this symmetry. Since firewalls are only supposed to be present in sufficiently generic states, and sufficiently generic states will not necessarily have this symmetry in their interiors, it is important to understand how the boundary energy changes as we create excitations in various locations in black hole geometries which are more complex than the pure Schwarzschild geometry.  If there is indeed a general diffeomorphism-invariant location where excitations can be created without substantially modifying the energy, and which does not rely on special features of the state, then this could be a natural location for a firewall.
\ei
We will not be able to give a complete resolution of any of these issues in this paper, but we are able to learn a fair bit about each of them in the special case of JT gravity coupled to matter.  The main results will be the following:
\bi
\item [(1)] Using the covariant Peierls bracket \cite{Peierls:1952cb}, we compute the algebra of a wide set of diffeomorphism-invariant observables in JT gravity plus matter.  Some of these are purely gravitational, while others are dressed matter observables.  We also show that the time-dependence of gravitational part of the algebra is explicitly solvable.  We compute the bracket at any point in phase space (quantum mechanically what we derive are operator equations, not just statements about expectation values in certain states), so they can be used to probe the black hole interior in a wide variety of states in a diffeomorphism-invariant way.
\item[(2)] We explain how recent ideas on traversable wormholes \cite{Gao:2016bin} and scrambling \cite{Shenker:2013pqa} can in this system be understood directly as consequences of our algebra.
\item[(3)] We explain how the $SL(2,\mathbb{R})$ algebras of \cite{Lin:2019qwu}, which roughly speaking implement $AdS$ isometries on the matter fields while doing nothing to the metric and minimally modifying the dilaton, fit into the gravitational sector of this algebra.
\item[(4)] We show that in JT gravity coupled to matter, there exist diffeomorphism-invariant operators which create quanta just behind the event horizon in any of a wide variety of null shockwave states of a given energy.   Moreover acting with these operators decreases the energy, just as argued in \cite{Almheiri:2013hfa,Marolf:2013dba}, but now in a situation where most of these states do not have any Killing symmetry.  Thus in JT gravity it seems that the true event horizon is indeed a viable location for firewalls under the above criteria.  On the other hand, these results rely on special features of JT gravity which we do not expect to hold for black holes in higher dimensions.  As a first step to understanding this, we perform some computations for higher-dimensional black holes in the presence of shockwaves.  Qualitatively our results continue to support the ideas of \cite{Almheiri:2013hfa,Marolf:2013dba}, suggesting that the existence of diffeomorphism-invariant operators which decreases the energy is a robust feature of quantum gravity in the presence of horizons.
\ei

The plan for the rest of our paper is as follows: in section \ref{peierlssec} we review the Peierls bracket, which is an old proposal for covariantly computing Poisson brackets directly in the Lagrangian formalism.  In section \ref{JTsec} we briefly review JT gravity, and then introduce the set of diffeomorphism-invariant observables we study.  In section \ref{algsec}, the main technical section of the paper, we use the Peierls bracket to compute the algebra of these observables.  In section \ref{appsec} we present the applications to traversable wormholes, scrambling, and $SL(2,\mathbb{R})$ charges.  In section \ref{energysec} we discuss the energy cost of exciting a quanta.  Finally in section \ref{dsec} we make some general remarks about the interpretation of our results.  Various technical points are discussed in appendices.  Readers who wish only to apply the algebra without worrying how it is computed can skip sections \ref{peierlssec} and \ref{algsec}, which should make the paper more digestible on a first pass.

\section{Peierls bracket review}\label{peierlssec}
A standard ingredient in Hamiltonian mechanics is the Poisson bracket, which for two functions $f$ and $g$ on phase space is given by
\be\label{poisson}
\{f,g\}=\frac{\partial f}{\partial q^a}\frac{\partial g}{\partial p_a}-\frac{\partial f}{\partial p_a}\frac{\partial g}{\partial q^a}.
\ee
Here $q^a$ and $p_a$ are coordinates on phase space such that the symplectic form is
\be
\Omega=dp_a\wedge dq^a.
\ee
This bracket is an essential feature of any dynamical theory, but unfortunately if we wish to compute it starting from a Lagrangian presentation of the theory some nontrivial work is necessary.  For two-derivative Lagrangian theories with actions of the form
\be
S=\int dt L(q,\dot{q}),
\ee
the steps are well-known: we define the canonical momenta
\be
p_a\equiv \frac{\partial L}{\partial \dot{q}^a},
\ee
change coordinates on phase space from $(q,\dot{q})$ to $(q,p)$, and then construct the Hamiltonian via
\be
H(q,p)=p_a\dot{q}^a(q,p)-L(q,\dot{q}(q,p)).
\ee
To compute the Poisson bracket of a pair of functions $f(q,\dot{q})$, $g(q,\dot{q})$ using this method, we rewrite them as functions of $(q,p)$ and then use \eqref{poisson}.

This algorithm suffers from a number of well-known problems:
\bi
\item It destroys manifest Lorentz covariance, as we need to choose a time coordinate in making the transformation from $(q,\dot{q})$ to $(q,p)$.
\item The extension to theories where the Lagrangian depends on higher time derivatives of $q^a$ is unpleasant.
\item In the presence of constraints and/or gauge symmetries the transformation from $(q,\dot{q})$ to $(q,p)$ may not be well-defined, in which case some modification of the procedure is necessary (e.g. the introduction of Dirac brackets \cite{Dirac:1950pj}).
\ei
For all these reasons it would be very convenient to have a Lagrangian method for directly computing the Poisson bracket without reference to the Hamiltonian formalism, and indeed such a method was proposed long ago: the Peierls bracket \cite{Peierls:1952cb}.\footnote{The Peierls bracket has been surprisingly under-appreciated given its obvious theoretical value.  It seems to have mostly been the province of a few relativists and algebraic field theorists \cite{DeWitt:1962cg,DeWitt:2003pm,Marolf:1992rz,Marolf:1993zk,Marolf:1993af,Duetsch:2002yp,brennecke2008removal}. As far as we know this paper is the first since \cite{DeWitt:1962cg} back in 1962 to use it for concrete calculations.  We are indebted to Don Marolf, from whom we indirectly learned of its existence by way of Ahmed Almheiri and Netta Engelhardt.}

The basic idea of the Peierls bracket is to recast the Poisson bracket as the answer to a question in linear response theory: we perturb the action by $g$ and then ask how $f$ responds.  In the remainder of this section we will give an overview of this bracket, illustrating its equivalence to the Poisson bracket \eqref{poisson} in some simple examples and pointing out the kind of subtleties that can arise in the presence of spatial boundaries.  It is possible to show in general Lagrangian field theories (including those with gauge symmetries, higher derivatives, etc) that the Peierls bracket is always equivalent to the Poisson bracket constructed from the inverse of the symplectic form on covariant phase space \cite{Barnich:1991tc,Forger:2003jm,Khavkine:2014kya,Harlow:2019yfa}, and indeed covariant phase space \cite{Harlow:2019yfa,Crnkovic:1986ex,Lee:1990nz,Iyer:1994ys} is the most natural place to think about the Peierls bracket.  In this paper however we will restrict the use of covariant phase space to the appendices, as we are focused on practical computations and we hope to keep the main exposition as accessible as possible.

\subsection{Definition of the Peierls bracket}
Consider a Lagrangian field theory with action
\be
S_0=\int_M d^dx \sqrt{-g}\mathcal{L}(\phi,\partial\phi,\partial^2\phi,\ldots)+\int_{\Gamma} d^{d-1}x \sqrt{-\gamma}\ell\left(\phi,\partial \phi, \partial^2\phi,\ldots\right).
\ee
Here $\phi^a$ are some dynamical fields, $\Gamma$ is the spatial boundary, and we allow $\mathcal{L}$ and $\ell$ to depend on any finite number of their derivatives.  $\mathcal{L}$ and $\ell$ may also depend on some background fields, which we will not display explicitly.  We will assume that the boundary term $\ell$ is chosen such that any solution of the equations of motion which obeys the boundary conditions is a stationary point of this action up to terms at the future and past boundaries (see \cite{Harlow:2019yfa} for more on why this is right requirement and how a covariant symplectic form and Hamiltonian can be constructed from any such action).  Let $f[\phi]$ and $g[\phi]$ be two functionals of the dynamical fields that have support only during some finite window of time (for example we could integrate $\phi$ against some test function of compact support).  The \textit{Peierls bracket} of $f$ and $g$ is then defined as
\be\label{Pb}
\{f,g\}[\phi]\equiv \frac{d}{dk}f[\phi+k\delta_g\phi]\Big|_{k=0},
\ee
where $\phi$ is any solution of the equations of motion obeying the boundary conditions at $\Gamma$ and $\delta_g \phi$ is a certain solution of the linearized equations of motion about $\phi$ \cite{Peierls:1952cb}.  $\delta_g\phi$ is constructed in the following way: we introduce a deformed action
\be\label{Sdeform}
S\equiv S_0-k g,
\ee
and then look for a pair of configurations $\phi+k\delta \phi_R$ and $\phi+k\delta \phi_A$, called the ``retarded'' and ``advanced'' solutions, which obey the deformed equations of motion following from \eqref{Sdeform} at linear order in $k$ and also obey the original boundary conditions at $\Gamma$.  The retarded solution $\delta\phi_R$ is required to vanish at sufficiently early times, while the advanced solution $\delta\phi_A$ is required to vanish at sufficiently late times (this is where we use the compact support in time of $g$).\footnote{The existence of $\delta \phi_R, \delta\phi_A$ requires us to assume that the set of valid initial/final data for the deformed and undeformed equations of motion coincide on Cauchy slices which are outside of the support of $g$.  We expect this to be true in situations where the initial-value problem is well-posed, see e.g. \cite{Wald:1984rg}.}  We then have
\be
\delta_g\phi\equiv \delta \phi_R-\delta\phi_A,
\ee
which completes the definition \eqref{Pb}  \cite{Peierls:1952cb}.  In theories with gauge redundancies $\delta_g\phi$ can be non-unique, but the Peierls bracket will still be well-defined provided that $f$ and $g$ are both gauge-invariant.

The Peierls bracket \eqref{Pb} is manifestly covariant, and makes no reference to canonical momenta or a symplectic form.  On the other hand it is not manifest that it is antisymmetric in $f$ and $g$ or that it obeys the Jacobi identity: these properties follow indirectly from the demonstration that it is equivalent to the Poisson bracket on covariant phase space \cite{Forger:2003jm,Khavkine:2014kya,Harlow:2019yfa}.  In fact the non-manifest antisymmetry can be useful in practice: in what follows we will compute all Peierls brackets in two ways by deforming the action either by $f$ or $g$, and seeing that we get the same answer up to a sign is a convenient check of our calculations.

\subsection{Particle mechanics examples}
To develop some intuition we can study the Peierls bracket for the simple case of a single particle with action
\be
S_0=\frac{1}{2}\int dt \dot{x}^2.
\ee
Taking $g$ to be some function of $x(0)\equiv x_0$ and $\dot{x}(0)\equiv \dot{x}_0$ we have
\begin{align}\nonumber
S&=\int dt\frac{1}{2}\dot{x}^2-kg(x_0,\dot{x}_0)\\
&=\int dt\left(\frac{1}{2}\dot{x}^2-k\delta(t)g(x,\dot{x})\right)
\end{align}
and
\begin{align}\nonumber
\delta S&=\int dt\left[\frac{d}{dt}(\dot{x}\delta x)-\ddot{x}\delta x -k\delta(t)\left(\frac{\partial g}{\partial x_0}\delta x+\frac{\partial g}{\partial \dot{x}_0}\dot{\delta x}\right)\right]\\
&=\int dt \left[\frac{d}{dt}(\dot{x}\delta x)-\ddot{x}\delta x-k\left(\delta(t)\frac{\partial g}{\partial x_0}-\dot{\delta}(t)\frac{\partial g}{\partial \dot{x}_0}\right)\delta x\right],
\end{align}
so the deformed equation of motion is
\be\label{partdefEOM}
\ddot{x}=-k \delta(t)\frac{\partial g}{\partial x_0}+k\dot{\delta}(t)\frac{\partial g}{\partial \dot{x}_0}.
\ee
In these expressions the quantities $\frac{\partial g}{\partial x_0}$ and $\frac{\partial g}{\partial \dot{x}_0}$ should be understood as being evaluated on $x_0$ and $\dot{x}_0$, so that they both have vanishing time derivative.  Integrating \eqref{partdefEOM} across $t=0$ gives a jump
\be
\Delta \dot{x}=-k\frac{\partial g}{\partial x_0},
\ee
while multiplying it by $t$ and then integrating gives a jump
\be
\Delta x=k\frac{\partial g}{\partial \dot{x}_0}.
\ee
In expressions like these we will always work only to linear order in $k$, so the right hand side is well-defined even though $x_0$ and $\dot{x}_0$ both jump at $t=0$.  In manipulating the $\delta$ functions in these calculations we have used the identities
\begin{align}\nonumber
f(t)\delta(t)&=f(0)\delta(t)\\
t\dot{\delta}(t)&=-\delta(t).
\end{align}
Therefore about any unperturbed solution
\be
x(t)=x_0+\dot{x}_0t,
\ee
we have the advanced and retarded solutions
\begin{align}\nonumber
\delta x_R(t)&=\theta(t)\left(\frac{\partial g}{\partial \dot{x}_0}-\frac{\partial g}{\partial {x}_0}t\right)\\
\delta x_A(t)&=\theta(-t)\left(-\frac{\partial g}{\partial \dot{x}_0}+\frac{\partial g}{\partial {x}_0}t\right),
\end{align}
which combine to give the linearized solution
\be
\delta_g x(t)=\frac{\partial g}{\partial \dot{x}_0}-\frac{\partial g}{\partial {x}_0}t
\ee
of the (already linear) unperturbed equation of motion.  We thus have the Peierls brackets
\begin{align}\nonumber
\{x_0,g\}&=\frac{\partial g}{\partial \dot{x}_0}\\
\{\dot{x}_0,g\}&=-\frac{\partial g}{\partial {x}_0},
\end{align}
which are indeed equivalent to \eqref{poisson}.

More generally we can consider a set of particles $x^a(t)$ interacting via some action
\be
S_0=\int dt L(x,\dot{x}).
\ee
Taking again $g$ to depend only on $x^a(0)\equiv x^a_0$ and $\dot{x}^a(0)\equiv \dot{x}^a_0$, we have the deformed equations of motion
\be
\frac{d}{dt}\left(\frac{\partial L}{\partial \dot{x}^a}\right)-\frac{\partial L}{\partial x^a}=-k\delta(t)\frac{\partial g}{\partial x^a_0}+k\dot{\delta}(t)\frac{\partial g}{\partial \dot{x}^a_0}.
\ee
Integrating across $t=0$ as before one now finds the discontinuities
\begin{align}\nonumber
\Delta\left(\frac{\partial L}{\partial \dot{x}^a}\right)&=\frac{\partial^2 L}{\partial x^a\partial \dot{x}^b}\Delta x^b-k\frac{\partial g}{\partial x_0^a}\\
\frac{\partial^2L}{\partial \dot{x}^a\partial \dot{x}^b} \Delta x^b&=k\frac{\partial g}{\partial \dot{x}_0^a}.
\end{align}
Introducing the canonical momentum
\be
p_a\equiv \frac{\partial L}{\partial \dot{x}^a},
\ee
and assuming that the Hessian matrix $\frac{\partial^2L}{\partial \dot{x}^a\partial \dot{x}^b}=\frac{\partial p_b}{\partial\dot{x}^a}$ is invertible, we can rewrite these discontinuities as
\begin{align}\nonumber
\Delta x^a&=k\frac{\partial g}{\partial p_{a,0}}\\
\Delta p_a&=-k\frac{\partial g}{\partial x_0^a},
\end{align}
where the partial derivatives of $g$ are now taken with respect to $x_0^a$ and $p_{a,0}$ instead of $x_0^a$ and $\dot{x}_0^a$.  From these discontinuities we have the Peierls brackets
\begin{align}\nonumber
\{x^a_0,g\}&=\frac{\partial g}{\partial p_{a,0}}\\
\{p_{a,0},g\}&=-\frac{\partial g}{\partial x_0^a},
\end{align}
which again is equivalent to the Poisson bracket \eqref{poisson}.  This discussion is similar to one Peierls gave in his original paper, where he argued that his bracket was equivalent to the Poisson bracket in any two-derivative field theory.

\subsection{Boundary effects}\label{boundarysec}
\bfig
\includegraphics[height=5cm]{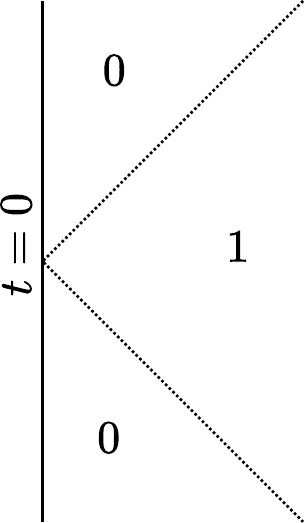}
\caption{The linearized solution $\delta_g \phi=\delta\phi_R-\delta\phi_A$ as a function of space and time.  The dotted lines indicate the locations of the shockwaves which restore the boundary conditions.}\label{boundaryfig}
\efig
Spatial boundaries can introduce interesting subtleties in the calculation of the Peierls bracket, here we discuss a simple example that illustrates the issue.  Indeed consider a free massless scalar field on a half-space in $1+1$ dimensions.  The action is
\be
S_0=\frac{1}{2}\int_{-\infty}^\infty dt \int_0^\infty dx \left(\dot{\phi}^2-\phi'^2\right),
\ee
and we will adopt the Dirichlet boundary condition
\be\label{modelbc}
\phi(t,0)=0.
\ee
We will now study the Peierls brackets of the observable
\be
g[\phi]=\int_{0}^\infty dx \dot{\phi}(0,x).
\ee
Including this observable in the action leads to the deformed equation of motion
\be
\ddot{\phi}-\phi''=k\dot{\delta}(t),
\ee
which naively suggests a $t=0$ discontinuity
\be\label{naivedis}
\Delta\phi=k
\ee
and retarded and advanced solutions
\begin{align}\nonumber
\delta \phi_R&=\theta(t)\\
\delta\phi_A&=-\theta(-t).
\end{align}
These however are not consistent with the boundary condition \eqref{modelbc}, and the same is true of the combination
\be
\delta_g\phi=\delta\phi_R-\delta\phi_A=1.
\ee
To fix this, we need to modify these solutions in such a way that the discontinuity \eqref{naivedis} is maintained away from the boundary but the boundary conditions continue to be obeyed.  We can do this by adding a linearized solution of the undeformed equations of motion which at $t=0$ has support only at $x=0$, but away from $t=0$ is chosen precisely to restore the boundary conditions:
\begin{align}\nonumber
\delta\phi_R&=\theta(t)-\theta(t-x)\\\nonumber
\delta\phi_A&=-\theta(-t)+\theta(-t-x)\\
\delta_g\phi&=1-\theta(t-x)-\theta(-t-x).
\end{align}
Near $t=0$ and away from the boundary these solutions coincide with the naive ones, but they also have ingoing and/or outgoing null shockwaves from the corner at $t=x=0$ which ensure the boundary conditions are respected at all times (see figure \ref{boundaryfig}).  We will encounter similar phenomena at several points in our discussion of JT gravity.

\section{Diffeomorphism-invariant observables in JT gravity}\label{JTsec}
We now turn to our topic of main interest: Jackiw-Teitelboim gravity coupled to matter.  The action is
\be\label{JTS}
S_0=\int_M d^2x\sqrt{-g}\left(\Phi_0 R+\Phi(R+2)\right)+2\int_\Gamma dt \sqrt{-\gamma} \left(\Phi_0 K+\Phi(K-1)\right)+S_{matter}(g,\psi),
\ee
where $\Phi_0$ is a parameter, $\Phi$ is a dynamical ``dilaton'' field, $\Gamma$ is the spatial boundary, and $\psi^i$ are matter fields  (here and throughout we set the $AdS_2$ curvature radius to one).  The terms involving $\Phi_0$ are topological and will play no role in our analysis, so from now on we will suppress them.  We will consider only the situation where the boundary $\Gamma$ has two connected components: a ``left'' component $\Gamma_-$ and a ``right'' component $\Gamma_+$, and at these boundaries we impose the boundary conditions
\begin{align}\nonumber
ds^2|_{\Gamma_\pm}&=-\frac{dt_{\pm}^2}{\epsilon^2}\\
\Phi|_{\Gamma_\pm}&=\frac{\phi_b}{\epsilon}.\label{BC}
\end{align}
We are particularly interested in the large-volume limit $\epsilon\to 0$, in which many expressions simplify and the theory is easier to understand, but to avoid delicate orders of limits we will do most of our calculations at finite $\epsilon$ and take $\epsilon\to 0$ only at the end.  We will assume that the matter theory also has spatial boundary conditions such that $S_{matter}$ is stationary under variations of the matter fields up to future/past boundary terms, but for the most part we will not need to discuss these boundary conditions explicitly.  The variation of the action \eqref{JTS} is
\begin{align}\nonumber
\delta S_0=\int_Md^2x\sqrt{-g}\Big[&(R+2)\delta\Phi+\left(\nabla^\mu\nabla^\nu\Phi+g^{\mu\nu}(\Phi-\nabla^2\Phi)+\frac{1}{2}T^{\mu\nu}\right)\delta g_{\mu\nu}\\
&+E_i\delta\psi^i\Big]+\mathrm{future/past\, terms},
\end{align}
where $E_i=0$ are the matter equations of motion and the dilaton and metric equations of motion are\footnote{In JT gravity the metric is determined by the equation of motion which is conjugate to the variation of the dilaton, while the dilaton is determined by the equation of motion which is conjugate to the variation of the metric.  We will always refer to the former as the dilaton equation of motion and the latter as the metric equation of motion.}
\begin{align}\nonumber
R+2&=0\\
\nabla_\mu\nabla_\nu \Phi+g_{\mu\nu}\left(\Phi-\nabla^2\Phi\right)+\frac{1}{2}T_{\mu\nu}&=0.\label{JTEOM}
\end{align}
Note in particular that the dilaton equation requires the spacetime to be locally $AdS_2$; this would not have been true had we allowed the matter fields to couple directly to the dilaton.

In this theory there are two kinds of diffeomorphisms which are of interest, one of which is a subset of the other.  The larger set is the set of diffeomorphisms whose actions on $\Gamma_\pm$ are time-translations of $t_{\pm}$ (more general diffeomorphisms than these do not preserve the boundary conditions and thus should not be considered).  The smaller set are those for which these time-translations vanish.  This latter set must be viewed as gauge redundancies in order for the theory to have sensible dynamics, while the former are physical symmetries which act nontrivially on phase space.  When we say that an observable in this theory is diffeomorphism-invariant, what we mean is that is invariant under the smaller set of diffemorphisms which act trivially at $\Gamma_{\pm}$.  Observables are allowed to transform nontrivially under boundary time translations, which are generated by a pair of Hamiltonians $H_\pm$.  In pure JT gravity without matter these are given by
\be\label{Hpm}
H_{\pm}=\frac{2}{\epsilon}\left(\Phi-n_\pm^\mu \nabla_\mu \Phi\right)|_{\Gamma_{\pm}},
\ee
where $n_\pm^\mu(t_\pm)$ are the outward unit normal vectors at the two boundaries.  $H_\pm$ are both conserved, and thus can be evaluated at any time on their respective boundaries.  $H_\pm$ continue to be the Hamiltonians for JT gravity coupled to some matter theories, but in general matter fields can contribute additional boundary terms to $H_\pm$.  For example this happens if there is a gauge field and we turn on a nonzero $A_0$ at the boundary.  The questions of which symmetries must be viewed as redundancies and how to construct the Hamiltonian in general Lagrangian field theories with spatial boundaries are naturally answered within the version of the covariant phase space formalism presented in \cite{Harlow:2019yfa}.  In appendix \ref{Tapp} we use this formalism study the question of when \eqref{Hpm} is the correct Hamiltonian for JT gravity coupled to a Lagrangian matter theory: it turns out that this is true if 1) the Hamiltonian  of the matter theory in any fixed background metric obeying the metric boundary conditions and possessing time-translation symmetry is the integral of its energy momentum tensor over a Cauchy slice without any additional boundary terms and 2) a certain additional boundary term potentially arising from the matter action vanishes.   Both requirements are true for example for a minimally-coupled or conformally-coupled free scalar with Dirichlet boundary conditions or a gauge field whose pullback to the boundary is required to vanish.  In this paper we will restrict to matter theories for which the boundary conditions are sufficiently restrictive that these conditions hold, in which case the full Hamiltonians are given by \eqref{Hpm} (this has been implicitly assumed without comment in all literature on the subject so far).

\bfig
\includegraphics[height=6cm]{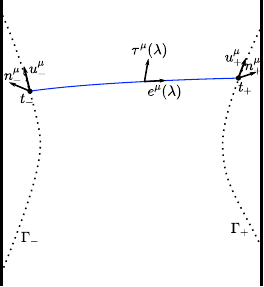}
\caption{The geometry of the two-sided geodesic $y^\mu_{t_-t_+}$, shown in blue, at finite $\epsilon$. The length $L(t_-,t_+)$ of this geodesic and the relative boosts $\eta_{\pm}(t_-,t_+)$ between its rest frame and those of the left and right boundaries $\Gamma_\pm$ define three diffeomorphism-invariant observables.}\label{twosidefig}
\efig
The Hamiltonians $H_{\pm}$ are our first examples of diffeomorphism-invariant observables in JT gravity plus matter.  To construct more it is convenient to first introduce two natural families of spacelike geodesics in this spacetime.  The first will be the two-sided geodesic $y_{t_- t_+}^\mu(\lambda)$, which we define to be the shortest geodesic from $\Gamma_-$ at time $t_-$ to $\Gamma_+$ at time $t_+$.  Moreover we will take $\lambda$ to be the unique affine parameter such that $y^\mu_{t_-t_+}$ intersects the left boundary when $\lambda=0$ and the right boundary when $\lambda=1$.   The length of this geodesic is
\be \label{Ldef}
L(t_-,t_+)=\int_0^1d\lambda \sqrt{\frac{dy^\mu_{t_-t_+}}{d\lambda}\frac{dy^\nu_{t_-t_+}}{d\lambda}g_{\mu\nu}(y_{t_-t_+}(\lambda))},
\ee
and it has unit tangent vector
\be\label{twoe}
e^\mu(\lambda)\equiv\frac{1}{L}\frac{dy_{t_-t_+}^{\mu}}{d\lambda}.
\ee
We will refer to its future-pointing unit normal vector as $\tau^\mu(\lambda)$.  Introducing the boundary unit tangent vectors
\be
u_{\pm}(t_\pm)\equiv \epsilon \partial_{t_{\pm}},
\ee
we can then define relative boosts $\eta_{\pm}(t_-,t_+)$ between the rest frame of the boundaries and the rest frame of the geodesic at its endpoints:
\begin{align}\nonumber
u^\mu_\pm(t_\pm)&\equiv\cosh \eta_{\pm} \tau^\mu(\lambda_\pm)\pm \sinh \eta_\pm e^\mu(\lambda_\pm)\\
n^\mu_\pm(t_\pm)&\equiv\sinh \eta_\pm \tau^\mu(\lambda_\pm)\pm \cosh \eta_\pm e^\mu(\lambda_\pm),\label{undef1}
\end{align}
where
\be
\lambda_\pm\equiv \frac{1}{2}\pm \frac{1}{2}.
\ee
The length $L(t_-,t_+)$ and relative boosts $\eta_{\pm}(t_-,t_+)$ are three diffeomorphism-invariant observables of this theory.  We will see that together with $H_\pm$ they generate an infinite-dimensional algebra which we will call the \textit{gravitational algebra}. We illustrate the various elements of this construction in figure \ref{twosidefig}.

\bfig
\includegraphics[height=5cm]{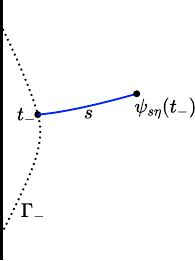}
\caption{Construction of a diffeomorphism-invariant ``one-sided'' dressed matter observable $\psi_{s\eta}(t_-)$: we fire a geodesic $y^\mu_{t_-\eta}(s)$, shown in blue, from the left boundary at time $t_-$ with relative boost $\eta$, and after a proper distance $s$ we evaluate the scalar matter field $\psi$.}\label{onesidefig}
\efig
We can also define diffeomorphism-invariant observables using the matter fields.  In particular if $\psi$ is some scalar observable from the matter theory, we can define a diffeomorphism-invariant ``two-sided'' matter observable
\be
\psi_{\lambda}(t_-,t_+)\equiv \psi(y_{t_-t_+}(\lambda)).
\ee
To define more we can use our second family of geodesics, $y_{t_-\eta}^\mu(s)$, which are fired from the left boundary at time $t_-$ with relative boost $\eta$ and parameterized by proper length $s$.  These have unit tangent vector
\be\label{onee}
e^\mu(s)\equiv\frac{dy_{t_-\eta}^\mu}{ds},
\ee
and at the left boundary obey
\begin{align}\nonumber
u^\mu_-(t_-)&=\cosh \eta \tau^\mu(0)-\sinh\eta e^\mu(0)\\
n^\mu_-(t_-)&=\sinh \eta \tau^\mu(0)-\cosh \eta e^\mu(0),\label{undef2}
\end{align}
where $\tau^\mu(s)$ is again the future-pointing unit normal vector to $y_{t_-\eta}^\mu(s)$.  We may then define a diffeomorphism-invariant ``one-sided'' matter observable
\be
\psi_{s,\eta}(t_-)\equiv \psi(y_{t_-\eta}(s)),
\ee
whose construction we illustrate in figure \ref{onesidefig}.

In the infinite-volume limit $\epsilon\to 0$ some of these observables diverge or vanish, so some renormalization is necessary.  We can identify the necessary subtractions by considering the behavior of our two geodesic families in the JT wormhole solution, which in global coordinates can be written as
\begin{align}\nonumber
ds^2&=-(x^2+1)d\tau^2+\frac{dx^2}{x^2+1}\\
\Phi&=\Phi_h \sqrt{x^2+1}\cos \tau,\label{nomattsol}
\end{align}
where $\Phi_h$ is the value of the dilaton on the bifurcate horizon.  The locations of the left and right boundaries in these coordinates are at
\begin{align}\nonumber
x_\pm(t_\pm)&=\pm\frac{\sqrt{\phi_b^2-\epsilon^2\Phi_h^2}}{\epsilon\Phi_h}\cosh\left(\frac{\Phi_h}{\sqrt{\phi_b^2-\epsilon^2\Phi_h^2}}t_\pm\right)\\
\tan (\tau_\pm(t_\pm))&=\sqrt{1-\frac{\epsilon^2\Phi_h^2}{\phi_b^2}}\sinh\left(\frac{\Phi_h}{\sqrt{\phi_b^2-\epsilon^2\Phi_h^2}}t_\pm\right).
\end{align}
It is also useful to consider the same solution in Schwarzschild coordinates, in which we have
\begin{align}\nonumber
ds^2&=-(r^2-r_s^2)d\hat{t}^2+\frac{dr^2}{r^2-r_s^2}\\
\Phi&=\phi_b r,\label{schwarzmetric}
\end{align}
with
\be\label{rsPhi}
r_s\equiv \frac{\Phi_h}{\phi_b},
\ee
and the boundary (these coordinates only cover one of the exterior regions) is now at
\begin{align}\nonumber
\hat{t}&=\frac{t}{\sqrt{1-\epsilon^2r_s^2}}\\
r&=\frac{1}{\epsilon}.
\end{align}
Thus the dilaton and boundary locations are simpler in Schwarzschild coordinates, for example the normal vector to the boundary is
\be
n^\mu=\sqrt{r^2-r_s^2}\delta^\mu_r,
\ee
so the Hamiltonians \eqref{Hpm} in this solution therefore evaluate to
\begin{align}\nonumber
H_\pm=&\frac{2\phi_b}{\epsilon^2}\left(1-\sqrt{1-r_s^2}\right)\\
=&\frac{\Phi_h^2}{\phi_b}+O(\epsilon^2),\label{Hnomatt}
\end{align}
but the full picture of the spacetime is more clear in global coordinates.  In particular by using the boost symmetry of this solution we can see that the length $L$ of the two-sided geodesic $y_{t_-t_+}^\mu$ depends only on $t_-+t_+$, and thus without loss of generality we can take $t_-=t_+$, in which case the geodesic is just a line of constant $\tau$ and we can compute its length via
\begin{align}\nonumber
L&=\int_{x_-((t_-+t_+)/2)}^{x_+((t_-+t_+)/2)}\frac{dx}{\sqrt{x^2+1}}\\
&=2\log\left(\frac{2\phi_b}{\epsilon}\right)+2\log\left(\frac{1}{\Phi_h}\cosh\left(\frac{\Phi_h}{\phi_b}\frac{t_-+t_+}{2}\right)\right)+O(\epsilon^2),\label{Lnomatt}
\end{align}
with the second line being written in a way that holds for any values of $t_-$, $t_+$. In the limit $\epsilon\to 0$ the renormalized length
\be\label{Ltdef}
\wt{L}\equiv L-2\log (2\Phi|_\Gamma)=L-2\log (2\phi_b/\epsilon)
\ee
is thus finite for any value of $\Phi_h$, and we will take \eqref{Ltdef} to be the definition of $\wt{L}$ also in configurations where the matter fields are nonvanishing.  Similarly we have
\begin{align}\nonumber
\sinh \eta_\pm&=\pm u_{\pm}^\mu(t_\pm) e_{\mu}(\lambda_\pm)\\\nonumber
&=\pm\frac{\epsilon}{\sqrt{x_\pm^2+1}}\frac{\partial x_{\pm}}{\partial t_{\pm}}\\
&=\frac{\epsilon}{2\phi_b}\left[2\Phi_h \tanh\left(\frac{\Phi_h}{\phi_b}\cdot\frac{t_-+t_+}{2}\right)+O(\epsilon^2)\right],\label{etanomatt}
\end{align}
where the second line assumes $t_-=t_+$ but the third line is true for any values of $t_\pm$.  Thus we can define a renormalized boost angle
\be\label{etatdef}
\wt{\eta}_\pm\equiv 2\Phi|_\Gamma\eta_\pm=\frac{2\phi_b\eta_\pm}{\epsilon},
\ee
which again we will adopt as the definition of $\wt{\eta}_\pm$ also when matter is present.

Renormalizations are also necessary for the dressed matter observables $\psi_{\lambda}$ and $\psi_{s\eta}$, since as $\epsilon \to 0$ any finite $\lambda\neq 1/2$ or $s\geq 0$ will correspond to a point which approaches the asymptotic boundary.  We can fix up the former by introducing the shifted and rescaled affine parameter
\be
\wt{\lambda}\equiv 2\left(\lambda-\frac{1}{2}\right)\log (2\Phi|_\Gamma)=2\left(\lambda-\frac{1}{2}\right)\log\frac{2\phi_b}{\epsilon},
\ee
which measures proper distance from the midpoint of $y_{t_-t_+}^\mu(\lambda)$, and then defining
\be
\psi_{\wt{\lambda}}(t_-,t_+)\equiv \psi(y_{t_-t_+}^\mu(\wt{\lambda})).
\ee
By $y_{t_-t_+}^\mu(\wt{\lambda})$ we mean the geodesic $y^\mu_{t_-t_+}$ parametrized by $\wt{\lambda}$, more explicitly $y_{t_-t_+}^\mu(\wt{\lambda})\equiv y_{t_-t_+}^\mu(\lambda(\wt{\lambda}))$.
Similarly we can define a shifted affine parameter
\be\label{sshift}
\wt{s}\equiv s-\log(2\Phi|_\Gamma)=s-\log \frac{2\phi_b}{\epsilon}
\ee
for our one-sided geodesic $y^\mu_{t_-\eta}(s)$,
which we also now specify using a renormalized boost
\be
\wt{\eta}\equiv 2\Phi|_\Gamma \eta,
\ee
and thus we have the renormalized one-sided matter observable
\be
\psi_{\wt{s},\wt{\eta}}(t_-)\equiv \psi(y_{t_-\wt{\eta}}(\wt{s})).
\ee

\bfig
\includegraphics[height=7cm]{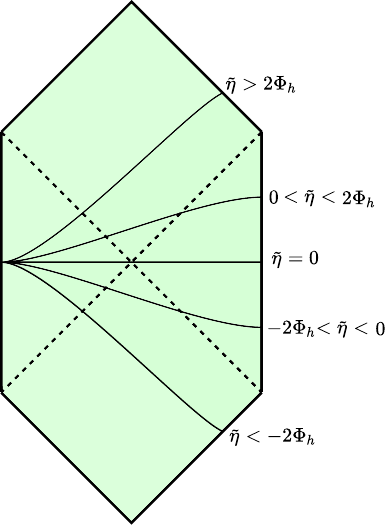}
\caption{Some examples of the one-sided geodesic $y^\mu_{t_-\wt{\eta}}(\wt{s})$ fired from the left boundary at $t_-=0$ in the pure JT solution \eqref{nomattsol}.  Geodesics for other starting times are obtained by acting with the boost isometry of this solution.  The dashed lines are horizons, and the spacetime region whose properties are determined by the initial data on any Cauchy slice connecting the two boundaries and the boundary conditions \eqref{BC} is shaded green.}\label{geodesicfig}
\efig
It is convenient to record here an explicit form for $y^\mu_{t_-\wt{\eta}}(\wt{s})$ in the solution \eqref{nomattsol}, \eqref{schwarzmetric}. In the $\epsilon\to 0$ limit in Schwarzschild coordinates we have
\begin{align}\nonumber
r(\wt{s})&=\frac{\phi_b(r_s^2-\alpha^2)}{2}e^{\wt{s}}+\frac{1}{2\phi_b}e^{-\wt{s}}\\
t(\wt{s})&=t_-\pm \frac{1}{2r_s}\log\left(\frac{1-e^{2\wt{s}}(r_s-\alpha)^2\phi_b^2}{1-e^{2\wt{s}}(r_s+\alpha)^2\phi_b^2}\right),\label{schgeod}
\end{align}
where
\be\label{alphaeta}
\alpha=\frac{|\wt{\eta}|}{2\phi_b},
\ee
and the sign choice in $t(\wt{s})$ is given by the sign of $\wt{\eta}$.  We have suppressed the distinction between $t$ and $\hat{t}$ since this vanishes in the limit $\epsilon\to 0$.  This geodesic has $r=\frac{1}{\epsilon}$ when
\be
\wt{s}=-\log \frac{2\phi_b}{\epsilon}+O(\epsilon^2),
\ee
which confirms the validity of the subtraction \eqref{sshift}.  Starting from the left boundary at $\wt{s}=-\infty$, it reaches the left future/past horizon when
\be
\wt{s}=-\log\big(\phi_b(r_s+\alpha)\big).\label{horizonst}
\ee
If $\alpha<r_s$, then it goes through the wormhole and exits via the right future/past horizon when
\be\label{horizonst2}
\wt{s}=-\log\big(\phi_b(r_s-\alpha)\big),
\ee
while if $\alpha>r_s$ it stays behind the horizon and leaves the regime of predictability of the system.  We show some examples in figure \ref{geodesicfig}.

\section{Computation of the algebra}\label{algsec}
We have now introduced our full set of diffeomorphism-invariant observables in JT gravity plus matter: the gravitational observables $L$, $\eta_\pm$, and $H_{\pm}$, the matter observables $\psi_{\lambda}$ and $\psi_{s,\eta}$, and the renormalized versions of each of these.  We now use the Peierls bracket to compute their algebra.  The method always follows the same steps:
\bi
\item Compute the variation $\delta g$ of some diffeomorphism-invariant observable $g$ with respect to the dynamical fields.  This often involves understanding how the dressing geodesic for $g$ (here we think of $y^\mu_{t_-t_+}$ as the dressing geodesic for $L$ and $\eta_\pm$) changes under a variation of the metric.
\item Deform the JT action $S_0$ by $-kg$, and use our expression for $\delta g$ to extract $\delta$-function and/or derivative of $\delta$-function contributions to the deformed equations of motion.
\item Integrate these deformed equations of motion to identify discontinuities in $\Phi$ and/or $\dot{\Phi}$ across the dressing geodesic for $g$.
\item Work out the discontinuities in the boundary tangent and normal vectors $u^\mu_\pm$ and $n^\mu_\pm$ which follow from these dilaton discontinuities, using the fact that these are related by the dilaton boundary conditions via the requirement that
\be\label{uphi}
u_\pm^\mu\nabla_\mu \Phi=0.
\ee
\item Compute the effects of these discontinuities on the other diffeomorphism-invariant observables we've introduced, from which we can read off the Peierls bracket of each observable with $g$.  This often boils down to identifying how the discontinuities affect the dressing geodesic for each observable.
\ei

The primary technical problem we encounter is thus understanding how geodesics vary under small changes of the metric or their boundary conditions, which is described by various versions of the equation of geodesic deviation (such problems are a special case of what \cite{Engelhardt:2019hmr} called ``surface theory''). We will do all calculations first at finite $\epsilon$, and then take the limit $\epsilon\to 0$ to compute the brackets of the renormalized observables.  As the tools are the same in each case, we will provide less detail as we go on.

Before we begin a word of warning: to avoid an explosive proliferation of notation,  many symbols in this section, such as $\alpha$, $\beta$, $e^\mu$, $\tau^\mu$, $\Delta y^\mu$, $\gamma$, $\sigma$, etc, will mean different (but closely analogous) things in different calculations.  They will be explicitly re-defined each time their meaning changes, but the reader should still be cautious and look back if at any point something is unsure.

\subsection[Brackets generated by two-sided length]{Brackets generated by $L$}\label{Lsec}
To compute Peierls brackets generated by the two-sided length observable $L(t_-,t_+)$ defined by \eqref{Ldef}, we first need to find the equations of motion for the deformed action
\be\label{LdefS}
S=S_0-kL.
\ee
Thus we need the variation
\be
\delta L=\frac{1}{2}\int_0^1d\lambda\frac{y^{\mu\prime}y^{\nu\prime}\delta g_{\mu\nu}(y(\lambda))}{\sqrt{y^{\alpha\prime}y^{\beta\prime} g_{\alpha\beta}(y(\lambda))}}=\frac{L}{2}\int_0^1 d\lambda e^\mu e^\nu\delta g_{\mu\nu}(y(\lambda)),\label{Lvar}
\ee
where here we have simplified our notation by dropping the explicit $t_-t_+$ on $y^\mu_{t_-t_+}(\lambda)$ and also defined
\be\label{ysimple}
y^{\mu\prime}\equiv \frac{d y^\mu}{d\lambda},
\ee
and we have also used that $\sqrt{y^{\alpha\prime}y^{\beta\prime} g_{\alpha\beta}(y(\lambda))}=L$ is constant along the geodesic since $\lambda$ is an affine parameter.  $e^\mu$ is the unit tangent vector given by \eqref{twoe}.  In \eqref{Lvar} we have included only the explicit variation of the metric, as the implicit metric-dependence in $y^\mu(\lambda)$ contributes a vanishing variation since by definition $y^\mu(\lambda)$ is a curve that extremizes $L$ (in the following sections we won't be so lucky).  To see how this variation modifies the equations of motion it is convenient to ``integrate in'' a $\delta$-function\footnote{Here we use the convention where $\delta^2(x)$ is a scalar, so written explicitly in coordinates it includes a factor of $1/\sqrt{-g}$.  For example in the metric \eqref{normalC} we have $\delta^2(x)=\frac{1}{L}\delta(\tau)\delta(\lambda)$.}
\begin{align}\nonumber
\delta L&=\frac{L}{2}\int d^2x\sqrt{-g}e^\mu(x)e^\nu(x)\delta g_{\mu\nu}(x)\int_0^1d\lambda' \delta^2(x-y(\lambda'))\\
&=\frac{1}{2}\int d^2 x \sqrt{-g}\delta(\tau)e^\mu e^\nu \delta g_{\mu\nu},
\end{align}
where in the second line we have used Gaussian normal coordinates adapted to the geodesic, in terms of which the metric is
\be\label{normalC}
ds^2=-d\tau^2+\ell(\tau,\lambda)^2d\lambda^2,
\ee
with $\ell(0,\lambda)=L$ and $\partial_\tau \ell(0,\lambda)=0$ since $\tau=0$ is a geodesic.  Here $\delta(\tau)$ indicates the ``usual'' $\delta$-function, obeying $\int_{-\epsilon}^\epsilon d\tau \delta(\tau)=1$, and its time derivative $\dot{\delta}(\tau)$ which appears below obeys $\int_{-\epsilon}^\epsilon d\tau \tau \dot{\delta}(\tau)=-1$.  $e^\mu(x)$ denotes some arbitrary extension of $e^\mu(\lambda)$ to the rest of the spacetime.  The equations of motion from the variation of the metric are thus modified to
\be\label{LEOM}
\nabla^\mu\nabla^\nu\Phi+g^{\mu\nu}\left(\Phi-\nabla^2\Phi\right)+\frac{1}{2}T^{\mu\nu}=\frac{k}{2}\delta(\tau)e^\mu e^\nu.
\ee

As in our discussion of particle mechanics, the source term on the right hand side of \eqref{LEOM} introduces discontinuities of the dynamical fields.   The equation $R+2=0$ is not modified, so the metric remains smooth and we can thus choose coordinates (such as Gaussian normal coordinates) where it does not jump.  The matter equations of motion are also unmodified, so there will be no discontinuity in the matter fields.  Contracting both sides of \eqref{LEOM} with $e_\mu e_\nu$, using that $\nabla^2\Phi=(e^\mu e^\nu-\tau^\mu \tau^\nu)\nabla_\mu\nabla_\nu \Phi$, and also noting that $\tau^\mu\nabla_\mu \tau^\nu=0$ since in Gaussian normal coordinates $\tau^\mu\equiv \delta^\mu_\tau$ is the unit tangent to a geodesic congruence fired orthogonally from $y^\mu_{t_-t_+}$, we have\footnote{Here we introduce a notation where for any scalar field $f$, $\dot{f}\equiv\tau^\mu\nabla_\mu f$, $\ddot{f}\equiv\tau^\mu\nabla_\mu(\tau^\nu\nabla_\nu f)$, and so on.}
\be
\ddot{\Phi}+\Phi+\frac{1}{2}e^\mu e^\nu T_{\mu\nu}=\frac{k}{2}\delta(\tau).
\ee
When integrating this equation across the surface $\tau=0$  we can ignore the energy-momentum tensor contribution since the matter fields are smooth, so the same manipulations as in our discussion of equation \eqref{partdefEOM} here tell us that dilaton field must have the discontinuities\footnote{By the diffeomorphism-invariance of $L$ these discontinuities are guaranteed to be consistent with the constraint equations obtained by contracting \eqref{LEOM} with $\tau^\mu\tau^\nu$ or $e^\mu \tau^\nu$, but one can also check it explicitly.  Indeed we will never need to use the constraint components of the equations of motion at any point in determine the algebra of our diffeomorphism-invariant observables.}
\begin{align}\nonumber
\Delta\Phi&=0\\
\Delta\dot{\Phi}&=\frac{k}{2}.\label{Ljump}
\end{align}

We can now use these discontinuities to compute some Peierls brackets.  Since the matter fields and metric do not jump, any observable which is constructed using only these will have a vanishing bracket with $L$.  In particular we have
\be\label{psilL}
\{\psi_\lambda(t_-,t_+),L(t_-,t_+)\}=0,
\ee
since the dilaton discontinuity \eqref{Ljump} has no effect on the two-sided geodesic $y^\mu(\lambda)$.  On the other hand since $\dot{\Phi}$ jumps at the boundary, the boundary tangent and normal vectors $u^\mu_\pm$ and $n^\mu_\pm$ will have discontinuities.  Indeed using the requirements that
\be
\Delta(u_\pm^\mu u_{\pm\mu})=\Delta(n_\pm^\mu n_{\pm\mu})=\Delta(u_\pm^\mu n_{\pm\mu})=\Delta(u^\mu_\pm\nabla_\mu \Phi)=0,
\ee
which follow from the definitions of $u_\pm^\mu$ and $n^\mu_\pm$ and \eqref{uphi}, and also using \eqref{undef1} and \eqref{Ljump}, we have the discontinuities
\begin{align}\nonumber
\Delta u^\mu_\pm=-\frac{k}{2}\frac{\cosh\eta_\pm}{n_\pm^\alpha\nabla_\alpha\Phi}n^\mu_\pm\\
\Delta n^\mu_\pm=-\frac{k}{2}\frac{\cosh\eta_\pm}{n_\pm^\alpha\nabla_\alpha\Phi}u^\mu_\pm.\label{Lun}
\end{align}
Thus we have discontinuities in the left and right Hamiltonians \eqref{Hpm}, given by
\begin{align}\nonumber
\Delta H_{\pm}&=-\frac{2}{\epsilon}\Delta(n^\mu_\pm\nabla_\mu\Phi)\\\nonumber
&=-\frac{2}{\epsilon}n^\mu_\pm \Delta \left(\nabla_\mu \Phi\right)\\\nonumber
&=-\frac{2}{\epsilon}\left(\sinh\eta_\pm \Delta \dot{\Phi}\pm \cosh \eta_\pm \Delta (e^\mu\nabla_\mu\Phi)\right)\\
&=-\frac{k}{\epsilon}\sinh\eta_\pm,
\end{align}
where in second equality we have used \eqref{Lun} and \eqref{uphi}, in the third we have used \eqref{undef1}, and in the fourth we have used \eqref{Ljump}. We therefore have the Peierls brackets
\be\label{HL}
\{H_\pm,L(t_-,t_+)\}=-\frac{1}{\epsilon}\sinh \eta_\pm(t_-,t_+).
\ee
Similarly noting that
\be\label{ueta}
u_\pm^\mu e_{\mu}(\lambda_\pm)=\pm \sinh \eta_\pm,
\ee
we have
\begin{align}\nonumber
\Delta\left(u_\pm^\mu e_{\mu}(\lambda_\pm)\right)&=-\frac{k}{2}\frac{\cosh\eta_\pm}{n_\pm^\alpha\nabla_\alpha\Phi}n^\mu_\pm e_{\mu}(\lambda_\pm)\\\nonumber
&=\mp \frac{k}{2}\frac{\cosh\eta_\pm}{n_\pm^\alpha\nabla_\alpha\Phi} \cosh \eta_\pm\\
&=\pm \cosh \eta_\pm \Delta \eta_\pm,
\end{align}
where we have used \eqref{Lun} and \eqref{undef1}, and thus
\be\label{etaLjump}
\Delta\eta_\pm=-\frac{k}{2}\frac{\cosh\eta_\pm(t_-,t_+)}{n_\pm^\alpha\nabla_\alpha\Phi(t_\pm)}
\ee
and
\be
\{\eta_\pm(t_-,t_+),L(t_-,t_+)\}=-\frac{1}{2}\frac{\cosh\eta_\pm(t_-,t_+)}{n_\pm^\alpha\nabla_\alpha\Phi(t_\pm)}.
\ee

Finally we can study the bracket of $L$ with the one-sided matter observable $\psi_{s,\eta}(t_-)$.  Unlike the two-sided geodesic $y^\mu_{t_-t_+}(\lambda)$, the one-sided geodesic $y_{t_-\eta}(s)$ is defined with a fixed boost relative to the boundary and thus will be modified by the discontinuity \eqref{Ljump} via the boundary kinks \eqref{Lun}.  The Peierls bracket of $L$ and $\psi_{s,\eta}(t_-)$ will therefore be nonzero.  We can parametrize the jump in $y_{t_-\eta}(s)$ as
\be
\Delta y^\mu_{t_-\eta}(s)=\alpha(s)e^\mu(s)+\beta(s)\tau^\mu(s),
\ee
where $e^\mu(s)$ and $\tau^\mu(s)$ are the unit tangent and future-pointing normal vectors to $y^\mu_{t_-\eta}(s)$.  We emphasize that $e^\mu$ is given by \eqref{onee}, in the one-sided case there is no factor of $L$ in the relation between $e^\mu$ and $y^{\mu \prime}$.  In what follows it is convenient to note that
\begin{align}\nonumber
e^\mu\nabla_\mu e^\nu&=0\\
e^\mu\nabla_\mu \tau^\nu&=0,\label{etdiv}
\end{align}
with the first equation being the geodesic equation and the second being a consequence of the first together with $e^\mu\tau_\mu=0$ and $\tau^\mu \tau_\mu=-1$.  The deviation $\Delta y^\mu_{t_-\eta}$ obeys the equation of geodesic deviation \cite{Wald:1984rg}
\be
e^\alpha\nabla_\alpha(e^\beta \nabla_\beta \Delta y^\mu_{t_-\eta})+R^\mu_{\phantom{\mu} \alpha\beta\gamma}e^\alpha \Delta y^\beta_{t_-\eta}e^\gamma=0,
\ee
which in terms of $\alpha$ and $\beta$ says that
\begin{align}\nonumber
\alpha''&=0\\
\beta''&=\beta.\label{onesidedev}
\end{align}
In showing this we use \eqref{etdiv}, and also that in any locally-$AdS_2$ space we have
\be\label{Riemann}
R_{\mu\nu\alpha\beta}=g_{\mu\beta}g_{\nu\alpha}-g_{\mu\alpha}g_{\nu\beta}.
\ee
To work out the boundary conditions on $\alpha$ and $\beta$, we first need to observe that the jump in the tangent vector $e^\mu$ is given by
\begin{align}\nonumber
\Delta e^\mu&\equiv\frac{d}{ds}(y^\mu_{t_-\eta}+\Delta y^\mu_{t_-\eta})-\frac{d y^\mu_{t_-\eta}}{ds}\\\nonumber
&=e^\lambda\partial_\lambda \Delta y^\mu_{t_-\eta}\\\nonumber
&=e^\lambda\nabla_\lambda \Delta y^\mu_{t_-\eta}-\Gamma^\mu_{\lambda\sigma}\Delta y^\sigma_{t_-\eta}e^\lambda\\
&=\alpha' e^\mu+\beta' \tau^\mu-\Gamma^\mu_{\lambda\sigma}\Delta y^\sigma_{t_-\eta}e^\lambda.
\end{align}
The geometric interpretation of this equation is as follows: the tangent vectors to $y_{t_-\eta}^\mu(s)$ and $y_{t_-\eta}^\mu(s)+\Delta y^\mu_{t_-\eta}(s)$ do not live in the same tangent spaces, so to compare them it is natural to first parallel transport $e^\mu(s)$ from $y_{t_-\eta}^\mu(s)$ to $y_{t_-\eta}^\mu(s)+\Delta y^\mu_{t_-\eta}(s)$.  Thus we can define a ``covariant jump''
\begin{align}\nonumber
\Delta^{(c)}e^\mu(s)&\equiv (e^\mu(s)+\Delta e^\mu(s))-(e^\mu(s)-\Gamma^\mu_{\alpha\beta}(y(s))e^\alpha(s)\Delta y_{t_-\eta}(s))\\\nonumber
&=e^\lambda\nabla_\lambda \Delta y^\mu_{t_-\eta}(s)\\
&=\alpha'(s)e^\mu(s)+\beta'(s) \tau^\mu(s),\label{covjump}
\end{align}
which is better-behaved than $\Delta e^\mu$.  Moreover since $e^\mu$ is a unit vector, we have
\be\label{alphap}
\frac{1}{2}\Delta^{(c)}(e^\mu e_\mu)=e_\mu\Delta^{(c)} e^\mu=\alpha'=0,
\ee
and thus\footnote{This conclusion is a bit more tedious to obtain if we use $\Delta$ instead of $\Delta^{(c)}$, as we then need to take into account the change in the metric:
\be
\frac{1}{2}\Delta(e^\mu e^\nu g_{\mu\nu})=e_\mu \Delta e^\mu+\frac{1}{2}e^\mu e^\nu \Delta y^\alpha_{t_-\eta}\partial_\alpha g_{\mu\nu}=0.
\ee
This derivative of $g_{\mu\nu}$ is canceled by the Christoffel symbols in $\Delta^{(c)}$ since $\Delta y^\alpha_{t_-\eta}\nabla_\alpha g_{\mu\nu}=0$.}
\be\label{ecovjump}
\Delta^{(c)}e^\mu(s)=\beta'(s)\tau^\mu(s).
\ee

Considering now the boundary conditions, the deformed geodesic still leaves the left boundary at time $t_-$ so we must have
\be\label{abbc}
\alpha(0)=\beta(0)=0.
\ee
In particular this means that at $s=0$ there is no distinction between $\Delta$ and $\Delta^{(c)}$, so we have
\be
\Delta e^\mu(0)=\beta'(0) \tau^\mu(0).
\ee
We still wish the relative boost of the deformed geodesic to be $\eta$ , so we have
\be
\Delta(u^\mu_- e_\mu)=\Delta u^\mu_- e_\mu+u_{-\mu} \Delta e^\mu=0,
\ee
which using \eqref{undef2} and \eqref{Lun} tells us that
\be
\beta'(0)=\frac{k}{2}\frac{\cosh\eta_-}{n_-^\alpha\nabla_\alpha\Phi}.
\ee
Since we have already seen that $\alpha'=0$ everywhere, these are enough boundary conditions to determine $\alpha$ and $\beta$ and indeed the unique solution is
\begin{align}\nonumber
\alpha(s)&=0\\
\beta(s)&=\frac{k}{2}\frac{\cosh\eta_-}{n_-^\alpha\nabla_\alpha\Phi}\sinh s.
\end{align}
Thus at last we have the discontinuity
\begin{align}\nonumber
\Delta \psi_{s,\eta}(t_-)&=\beta(s)\tau^\mu\nabla_\mu \psi (y_{t_-\eta}(s))\\
&=\frac{k}{2}\frac{\cosh\eta_-\sinh s}{n_-^\alpha\nabla_\alpha\Phi}\tau^\mu\nabla_\mu \psi (y_{t_-\eta}(s)),
\end{align}
and therefore the Peierls bracket
\be\label{psisL}
\{\psi_{s,\eta}(t_-),L(t_-,t_+)\}=\frac{1}{2}\frac{\cosh\eta_-(t_-,t_+)\sinh s}{n_-^\alpha\nabla_\alpha\Phi(t_-)}\tau^\mu\nabla_\mu \psi (y_{t_-\eta}(s)).
\ee

Altogether we have the following Peierls brackets involving $L$:
\begin{align}\nonumber
\{\psi_\lambda(t_-,t_+),L(t_-,t_+)\}&=0\\\nonumber
\{H_\pm,L(t_-,t_+)\}&=-\frac{1}{\epsilon}\sinh \eta_\pm(t_-,t_+)\\\nonumber
\{\eta_\pm(t_-,t_+),L(t_-,t_+)\}&=-\frac{1}{2}\frac{\cosh\eta_\pm(t_-,t_+)}{n_\pm^\alpha\nabla_\alpha\Phi(t_\pm)}\\
\{\psi_{s,\eta}(t_-),L(t_-,t_+)\}&=\frac{1}{2}\frac{\cosh\eta_-(t_-,t_+)\sinh s}{n_-^\alpha\nabla_\alpha\Phi(t_-)}\tau^\mu\nabla_\mu \psi (y_{t_-\eta}(s)).
\end{align}
In the large-volume limit $\epsilon\to 0$ we can rewrite these in terms of the renormalized versions of these observables:
\begin{align}\nonumber
\{\psi_{\wt{\lambda}}(t_-,t_+),\wt{L}(t_-,t_+)\}&=0\\\nonumber
\{H_\pm,\wt{L}(t_-,t_+)\}&=-\frac{\wt{\eta}_\pm(t_-,t_+)}{2\phi_b}\\\nonumber
\{\wt{\eta}_\pm(t_-,t_+),\wt{L}(t_-,t_+)\}&=-1\\
\{\psi_{\wt{s},\wt{\eta}}(t_-),\wt{L}(t_-,t_+)\}&=\frac{1}{2}e^{\wt{s}}\tau^\mu\nabla_\mu \psi (y_{t_-\wt{\eta}}(\wt{s})).
\end{align}
We will discuss the physics of these expressions in section \ref{appsec} once we have finished computing the rest of our brackets.

\subsection[Brackets generated by two-sided matter]{Brackets generated by $\psi_\lambda$}
We now consider Peierls brackets generated by the two-sided dressed matter observable $\psi_\lambda(t_-,t_+)$.  To compute the variation of the deformed action
\be
S=S_0-k\psi_{\lambda_0}(t_-,t_+),
\ee
where we indicate the affine location of the matter observable as $\lambda_0$ to avoid notational confusion later, we first need to understand how the two-sided geodesic $y_{t_-t_+}^\mu(\lambda)$ changes under a variation of the metric.  Its variation $\delta y^\mu_{t_-t_+}(\lambda)$ is controlled by what is called the ``sourced equation of geodesic deviation'' \cite{Engelhardt:2019hmr}, which says that\footnote{One way to think about this equation is the following: we consider a one-parameter family of metrics $g_{\mu\nu}+r\delta g_{\mu\nu}$, and we have a corresponding one-parameter family of geodesics $y_r^\mu(\lambda)$.  We can view this family as a map from some subset of $\mathbb{R}^2$ to spacetime, and $y^{\mu\prime}$ and $\delta y^\mu$ are the pushforwards of $\partial_\lambda$ and $\partial_r$ respectively.  We then have the modified geodesic equation $y^{\mu\prime}\nabla_\mu^{\{r\}}y^{\alpha\prime}=y^{\mu\prime}\nabla_\mu y^{\alpha\prime}+r\delta\Gamma^\alpha_{\mu\lambda}y^{\mu\prime}y^{\lambda\prime}+O(r^2)=0$, and substituting this into the appropriate step of the derivation of the equation of geodesic deviation (see e.g. \cite{Wald:1984rg}) gives \eqref{devmod}.}
\be\label{devmod}
y^{\mu\prime}\nabla_\mu(y^{\nu\prime} \nabla_\nu \delta y^\alpha)+R^\alpha_{\phantom \alpha\beta \mu\nu}y^{\beta\prime} y^{\nu\prime} \delta y^\mu+y^{\mu\prime}y^{\nu\prime}\delta \Gamma^{\alpha}_{\mu\nu}=0.
\ee
Here we have adopted the simplified notation \eqref{ysimple}, and also dropped the explicit $t_-t_+$ on $\delta y^\mu$.  We can parametrize any solution of \eqref{devmod} as
\be\label{twovar}
\delta y^\mu(\lambda)=\alpha(\lambda)e^\mu(\lambda)+\beta(\lambda)\tau^\mu(\lambda),
\ee
where as before $e^\mu=\frac{1}{L}y^{\mu\prime}$ is the unit tangent vector to $y^\mu_{t_-t_+}(\lambda)$ and $\tau^\mu$ is its future-pointing unit normal vector.  Making use of \eqref{etdiv}, which apply for $e^\mu(\lambda)$ as well as for $e^\mu(s)$, and also \eqref{Riemann}, we can rewrite \eqref{devmod} as
\begin{align}\nonumber
\alpha''+\frac{L}{2}\left(e^\alpha e^\beta \delta g_{\alpha\beta}\right)'&=0\\
\beta''-L^2 \beta-L\left(e^\alpha\tau^\beta\delta g_{\alpha\beta}\right)'+\frac{L^2}{2}e^\alpha e^\beta \tau^\lambda\nabla_\lambda \delta g_{\alpha\beta}&=0.\label{alphabeta2}
\end{align}
Since $y^\mu(\lambda)+\delta y^\mu(\lambda)$ is attached to the boundaries at the same times as $y^\mu(\lambda)$, we have the boundary conditions
\be
\alpha(0)=\alpha(1)=\beta(0)=\beta(1)=0.
\ee
The solutions of \eqref{alphabeta2} obeying these boundary conditions are
\begin{align}\nonumber
\alpha(\lambda)=&\frac{L}{2}\left[(\lambda-1)\int_0^\lambda d\lambda' e^\alpha(\lambda') e^\beta(\lambda') \delta g_{\alpha \beta}(y(\lambda'))+\lambda \int_\lambda^1d\lambda'e^\alpha(\lambda') e^\beta(\lambda') \delta g_{\alpha \beta}(y(\lambda'))\right]\\\nonumber
\beta(\lambda)=&\frac{L}{\sinh L}\Bigg[\sinh(L(1-\lambda))\int_0^\lambda d\lambda'\cosh (L \lambda ')\tau^\alpha(\lambda') e^\beta(\lambda') \delta g_{\alpha \beta}(y(\lambda'))\\\nonumber
&+\frac{\sinh(L(1-\lambda))}{2}\int_0^\lambda d\lambda'\sinh(L\lambda')e^\alpha(\lambda')e^\beta(\lambda')\tau^\mu \nabla_\mu \delta g_{\alpha\beta}(y(\lambda'))\\\nonumber
&-\sinh[L \lambda]\int_{\lambda}^1d\lambda'\cosh(L(1-\lambda'))\tau^\alpha(\lambda') e^\beta(\lambda') \delta g_{\alpha \beta}(y(\lambda'))\\
&+\frac{\sinh(L\lambda)}{2}\int_\lambda^1 d\lambda'\sinh(L(1-\lambda'))e^\alpha(\lambda')e^\beta(\lambda')\tau^\mu \nabla_\mu \delta g_{\alpha\beta}(y(\lambda'))\Bigg],\label{alphabeta}
\end{align}
in terms of which we have the variation
\be
\delta\psi_{\lambda_0}=\delta \psi(y(\lambda_0))+\alpha(\lambda_0)e^\mu(\lambda_0)\nabla_\mu \psi(y(\lambda_0))+\beta(\lambda_0)\tau^\mu(\lambda_0)\nabla_\mu\psi(y(\lambda_0)).
\ee
To facilitate finding the deformed equations of motion we can insert $\delta$ functions and use Gaussian normal coordinates (as in our discussion of $\delta L$), obtaining
\be
\alpha(\lambda_0)=\frac{1}{2}\int d^2 x \sqrt{-g}e^\alpha(x)e^\beta(x)\delta g_{\alpha\beta}(x)\delta(\tau)\Big((\lambda_0-1)\theta(\lambda_0-\lambda)+\lambda_0\theta(\lambda-\lambda_0)\Big)
\ee
and
{\small\begin{align}\nonumber
\beta(\lambda_0)=\frac{1}{\sinh L}\int d^2 x \sqrt{-g}\Bigg[&\Bigg(\sinh(L(1-\lambda_0))\cosh(\lambda L)\theta(\lambda_0-\lambda)\\\nonumber
&-\sinh(L\lambda_0)\cosh(L(1-\lambda))\theta(\lambda-\lambda_0)\Bigg)\tau^\alpha(x)e^\beta(x)\delta g_{\alpha \beta}(x)\delta(\tau)\\\nonumber
&-\frac{1}{2}\Bigg(\sinh(L(1-\lambda_0))\sinh(L\lambda)\theta(\lambda_0-\lambda)\\\nonumber
&+\sinh(L\lambda_0)\sinh(L(1-\lambda))\theta(\lambda-\lambda_0)\Bigg)e^\alpha(x)e^\beta(x)\delta g_{\alpha\beta}(x)\dot{\delta}(\tau)\Bigg],
\end{align}}
where in integrating by parts in the last line $e^\alpha(x)$ and $e^\beta(x)$ are understood to be extended away from the geodesic in such a way that $\tau^\mu\nabla_\mu e^\alpha=0$ at $\tau=0$ (this is not necessary, but simplifies equations), and we have used that $\partial_\tau\ell=0$. Thus the deformed equations of motion are
\begin{align}\nonumber
\ddot{\Phi}+\Phi+\frac{1}{2}e^\mu e^\nu T_{\mu\nu}&=\frac{k}{2}e^\mu(\lambda_0)\nabla_\mu \psi(y(\lambda_0))\Bigg((\lambda_0-1)\theta(\lambda_0-\lambda)+\lambda_0\theta(\lambda-\lambda_0)\Bigg)\delta(\tau)\\\nonumber
&-\frac{k}{2}\tau^\mu(\lambda_0)\nabla_\mu\psi(y(\lambda_0))\Bigg(\frac{\sinh(L(1-\lambda_0))\sinh(L\lambda)}{\sinh L}\theta(\lambda_0-\lambda)\\
&+\frac{\sinh(L\lambda_0)\sinh(L(1-\lambda))}{\sinh L}\theta(\lambda-\lambda_0)\Bigg)\dot{\delta}(\tau).
\end{align}
The matter fields will now also be discontinuous at $y^\mu(\lambda_0)$, in a way that depends in detail on how $\psi$ is built from the fundamental fields $\psi_i$.  This discontinuity however only affects the energy momentum tensor at $y(\lambda_0)$, so away from this point we can again ignore the contribution of the energy momentum tensor when we integrate this equation or this equation times $\tau$ across $\tau=0$ to extract the discontinuities of $\Phi$.  This at last leads to
\begin{align}\nonumber
\Delta\Phi&=-\frac{k}{2}\tau^\mu(\lambda_0)\nabla_\mu\psi(y(\lambda_0))\Big(\frac{\sinh(L(1-\lambda_0))\sinh(L\lambda)}{\sinh L}\theta(\lambda_0-\lambda)+\frac{\sinh(L\lambda_0)\sinh(L(1-\lambda))}{\sinh L}\theta(\lambda-\lambda_0)\Big)\\
\Delta\dot{\Phi}&=\frac{k}{2}e^\mu(\lambda_0)\nabla_\mu \psi(y(\lambda_0))\Big((\lambda_0-1)\theta(\lambda_0-\lambda)+\lambda_0\theta(\lambda-\lambda_0)\Big),\label{psijump}
\end{align}
while as before there is no jump in the metric or the matter fields away from $y^\mu(\lambda_0)$.

We can now use these discontinuities to compute some Peierls brackets.  As before \eqref{psijump} has no effect on the location of the two-sided geodesic $y^\mu(\lambda)$ or the metric in its vicinity, so we find
\be
\{L(t_-,t_+),\psi_\lambda(t_-,t_+)\}=0,
\ee
which is compatible with \eqref{psilL} and thus gives our first confirmation of the antisymmetry of the Peierls bracket.  We can work out the discontinuities in $u_\pm^\mu$ and $n_\pm^\mu$ in the same way we did for \eqref{Lun}, now finding
\begin{align}\nonumber
\Delta u_\pm^\mu&=\sigma_\pm n^\mu_\pm\\
\Delta n_\pm^\mu&=\sigma_\pm u^\mu_\pm
\end{align}
with
\begin{align}\nonumber
\sigma_+&=-\frac{k}{2}\frac{1}{n_+^\alpha\nabla_\alpha \Phi(t_+)}\left(\lambda_0\cosh \eta_+ e^\mu\nabla_\mu\psi(y(\lambda_0))+\frac{\sinh(L\lambda_0)}{\sinh L}\sinh\eta_+\tau^\mu\nabla_\mu\psi(y(\lambda_0))\right)\\
\sigma_-&=-\frac{k}{2}\frac{1}{n_-^\alpha\nabla_\alpha \Phi(t_-)}\left((\lambda_0-1)\cosh \eta_- e^\mu\nabla_\mu\psi(y(\lambda_0))+\frac{\sinh(L(1-\lambda_0))}{\sinh L}\sinh\eta_-\tau^\mu\nabla_\mu\psi(y(\lambda_0))\right).
\end{align}
In showing this we make use of the fact that \eqref{psijump} gives the following boundary discontinuities of $\Phi$:
\begin{align}\nonumber
\Delta\dot{\Phi}(y(0))&=\frac{k}{2}(\lambda_0-1)e^\mu\nabla_\mu\psi(y(\lambda_0))\\\nonumber
\Delta\dot{\Phi}(y(1))&=\frac{k}{2}\lambda_0e^\mu\nabla_\mu\psi(y(\lambda_0))\\\nonumber
\Delta
\left(e^\mu\nabla_\mu \Phi(y(0))\right)&=-\frac{k}{2}\frac{\sinh(L(1-\lambda_0))}{\sinh L}\tau^\mu\nabla_\mu\psi(y(\lambda_0))\\
\Delta \left(e^\mu\nabla_\mu \Phi(y(1))\right)&=\frac{k}{2}\frac{\sinh(L\lambda_0))}{\sinh L}\tau^\mu\nabla_\mu\psi(y(\lambda_0)).
\end{align}
Using these results we can also determine the discontinuities of $\eta_\pm$ and $H_\pm$ in the same way as the previous section, giving
\be
\Delta \eta_\pm=\sigma_\pm
\ee
and
\begin{align}\nonumber
\Delta H_+&=-\frac{k}{\epsilon}\left(\lambda_0\sinh \eta_+e^\mu\nabla_\mu\psi(y(\lambda_0))+\frac{\sinh(L\lambda_0)}{\sinh L}\cosh\eta_+\tau^\mu\nabla_\mu\psi(y(\lambda_0))\right)\\
\Delta H_-&=-\frac{k}{\epsilon}\left((\lambda_0-1)\sinh \eta_-e^\mu\nabla_\mu\psi(y(\lambda_0))+\frac{\sinh(L(1-\lambda_0))}{\sinh L}\cosh\eta_-\tau^\mu\nabla_\mu\psi(y(\lambda_0))\right).
\end{align}
Thus we at last have the Peierls brackets
\begin{align}\nonumber
\{\eta_+,\psi_{\lambda}\}&=-\frac{1}{2n^\alpha_+\nabla_\alpha \Phi(t_+)}\left(\lambda\cosh\eta_+e^\mu\nabla_\mu\psi(y(\lambda))+\sinh\eta_+\frac{\sinh(L\lambda)}{\sinh L}\tau^\mu\nabla_\mu\psi(y(\lambda))\right)\\\nonumber
\{\eta_-,\psi_{\lambda}\}&=-\frac{1}{2n^\alpha_-\nabla_\alpha \Phi(t_-)}\left((\lambda-1)\cosh\eta_-e^\mu\nabla_\mu\psi(y(\lambda))+\sinh\eta_-\frac{\sinh(L(1-\lambda))}{\sinh L}\tau^\mu\nabla_\mu\psi(y(\lambda))\right)\\\nonumber
\{H_+,\psi_{\lambda}\}&=-\frac{1}{\epsilon}\left(\lambda\sinh\eta_+e^\mu\nabla_\mu\psi(y(\lambda))+\cosh\eta_+\frac{\sinh(L\lambda)}{\sinh L}\tau^\mu\nabla_\mu\psi(y(\lambda))\right)\\
\{H_-,\psi_{\lambda}\}&=-\frac{1}{\epsilon}\left((\lambda-1)\sinh\eta_-e^\mu\nabla_\mu\psi(y(\lambda))+\cosh\eta_-\frac{\sinh(L(1-\lambda))}{\sinh L}\tau^\mu\nabla_\mu\psi(y(\lambda))\right),\label{twosidepb}
\end{align}
where to save space we have suppressed the explicit $t_\pm$ dependence of the observables but $\eta_\pm$ and $\psi_\lambda$ should be evaluated at the same times.  These expressions simplify considerably in the $\epsilon\to 0$ limit, there we have
\begin{align}\nonumber
\{\wt{\eta}_{\pm},\psi_{\wt{\lambda}}\}&=\mp\frac{1}{2}e^\mu(\wt{\lambda})\nabla_\mu\psi(y(\wt{\lambda}))\\
\{H_{\pm},\psi_{\wt{\lambda}}\}&=\mp\frac{\wt{\eta}_{\pm}}{4\phi_b}e^\mu(\wt{\lambda})\nabla_\mu\psi(y(\wt{\lambda}))-\frac{1}{2\phi_b}e^{-\frac{\wt{L}}{2}\pm \wt{\lambda}}\tau^\mu(\wt{\lambda})\nabla_\mu\psi(y(\wt{\lambda})).\label{comms}
\end{align}

\subsection[Brackets generated by one-sided matter]{Brackets generated by $\psi_{s,\eta}$}\label{onesidebracketsec}
We now consider Peierls brackets generated by the one-sided dressed matter observable $\psi_{s,\eta}(t_-)$.  We need to compute the variation of the deformed action
\be
S=S_0-k\psi_{s_0\eta}(t_-),
\ee
where to simplify notation later we denote by $s_0$ the proper length along the one-sided geodesic $y_{t_-\eta}^\mu(s)$ at which we locate the observable.  The initial steps are the same as for $\psi_\lambda$: the variation $\delta y_{t_-\eta}^\mu(s)$ of this geodesic again obeys the sourced equation of geodesic deviation
\be
e^\mu\nabla_\mu(e^\nu\nabla_\nu \delta y^\alpha)+R^\alpha_{\phantom{\alpha}\beta\mu\nu}e^\beta e^\nu \delta y^\mu+e^\mu e^\nu \delta \Gamma^\alpha_{\mu\nu}=0,\label{devmod2}
\ee
where we have dropped the explicit $t_-\eta$ on $\delta y^\mu$ and also used the relationship \eqref{onee} between $y^{\mu\prime}$ and $e^\mu$.  We can again express this variation as
\be
\delta y^\mu(s)=\alpha(s)e^\mu(s)+\beta(s)\tau^\mu(s),
\ee
in terms of which \eqref{devmod2} is equivalent to requiring that
\begin{align}\nonumber
\alpha''+\frac{1}{2}(e^\mu e^\nu\delta g_{\mu\nu})'&=0\\
\beta''-\beta-(e^\mu \tau^\nu \delta g_{\mu\nu})'+\frac{1}{2}e^\mu e^\nu \tau^\alpha\nabla_\alpha \delta g_{\mu\nu}&=0.\label{devmodab2}
\end{align}
The variation of $e^\mu$ is given by
\begin{align}\nonumber
\delta^{(c)} e^\mu &\equiv \delta e^\mu+\Gamma^\mu_{\lambda \nu}\delta y^\lambda e^\nu\\\nonumber
&=e^\nu \nabla_\nu \delta y^\mu\\
&=\alpha' e^\mu+\beta' \tau^\mu,
\end{align}
and requiring that $\delta^{(c)} (e^\mu e^\nu g_{\mu\nu})=0$ we see that
\be\label{alphapeqvar}
\alpha'=-\frac{1}{2}e^\alpha e^\beta\delta g_{\alpha\beta},
\ee
which is compatible with the first line of \eqref{devmodab2}, and thus
\be
\delta^{(c)}e^\mu=\beta'\tau^\mu-\frac{1}{2}e^\alpha e^\beta\delta g_{\alpha\beta} e^\mu.\label{ecovjumpvar}
\ee
The boundary conditions for $\alpha$ and $\beta$ are determined by the requirements that at $s=0$ we have
\begin{align}\nonumber
\delta y^\mu&=0\\
u^\mu_- e_\mu&=-\sinh \eta.\label{onesidebc}
\end{align}
The first of these tells us that
\be
\alpha(0)=\beta(0)=0,
\ee
which is already sufficient to determine $\alpha$ from \eqref{alphapeqvar}. Moreover at $s=0$ we have
\be
\delta u^\mu=0
\ee
by the boundary conditions \eqref{BC}, so the variation of the second line of \eqref{onesidebc}
tells us that
\begin{align}
\beta'(0)&=\tau^\alpha e^\beta\delta g_{\alpha\beta}(y(0))-\frac{1}{2}\tanh\eta e^\alpha e^\beta \delta g_{\alpha\beta}(y(0)).
\end{align}
The solutions for $\alpha$ and $\beta$ obeying these boundary conditions are
\begin{align}\nonumber
\alpha(s)=&-\frac{1}{2}\int_0^sds'e^\alpha e^\beta \delta g_{\alpha\beta}(y(s'))\\\nonumber
\beta(s)=&\int_0^sds'\left[\cosh(s-s')\tau^\alpha e^\beta \delta g_{\alpha\beta}(y(s'))-\frac{1}{2}\sinh(s-s')e^\alpha e^\beta \tau^\mu \nabla_\mu \delta g_{\alpha\beta}(y(s'))\right]\\
&-\frac{1}{2}\sinh (s)\tanh \eta e^\alpha e^\beta \delta g_{\alpha \beta}(y(0)).\label{oneabsol}
\end{align}
We then have the variation
\be\label{psi1var}
\delta \psi_{s_0,\eta}(t_-)=\delta\psi(y(s_0))+\alpha(s_0)e^\mu \nabla_\mu \psi(y(s_0))+\beta(s_0)\tau^\mu \nabla_\mu \psi(y(s_0)).
\ee
To see how this modifies the equations of motion, it is again convenient to integrate in $\delta$ functions and use Gaussian normal coordinates, in terms of which the metric in the vicinity of the dressing geodesic is
\be
ds^2=-d\tau^2+\ell(\tau,s)^2ds^2,
\ee
with $\ell(0,s)=1$ and $\partial_\tau \ell(0,s)=0$, and we have \be
\alpha(s_0)=-\frac{1}{2}\int d^2 x \sqrt{-g}e^\alpha(x) e^\beta(x) \delta g_{\alpha \beta}(x)\delta(\tau)\theta(s_0-s)
\ee
and
\begin{align}\nonumber
\beta(s_0)=\int d^2 x \sqrt{-g}\theta(s_0-s)\Bigg[&\cosh(s_0-s)e^\alpha(x)\tau^\beta(x)\delta g_{\alpha \beta}(x)\delta(\tau)\\\nonumber
&+\frac{1}{2}\sinh(s_0-s)e^\alpha(x)e^\beta(x)\delta g_{\alpha\beta}(x)\dot{\delta}(\tau)\Bigg]\\
&-\frac{1}{2}\sinh s_0\tanh\eta e^\alpha(y(0)) e^\beta(y(0))\delta g_{\alpha\beta}(y(0)).
\end{align}
The boundary term in $\beta(s_0)$ does not affect the equations of motion for $s>0$, so away from the boundary the $e^\mu e^\nu$ component of the deformed metric equation of motion is
\be
\ddot{\Phi}+\Phi+\frac{1}{2}e^\mu e^\nu T_{\mu\nu}=\frac{k}{2}\theta(s_0-s)\left[-e^\alpha\nabla_\alpha \psi(y(s_0))\delta(\tau)+\tau^\alpha\nabla_\alpha\psi(y(s_0))\sinh(s_0-s)\dot{\delta}(\tau)\right].
\ee
Integrating as usual, we find (for $s>0$) the discontinuities
\begin{align}\nonumber
\Delta\Phi&=\frac{k}{2}\theta(s_0-s)\sinh(s_0-s)\tau^\alpha\nabla_\alpha\psi(y(s_0))\\
\Delta\dot{\Phi}&=-\frac{k}{2}\theta(s_0-s)e^\alpha\nabla_\alpha \psi(y(s_0)),\label{psisjump}
\end{align}
as well as a theory-dependent matter field discontinuity at $y^\mu(s_0)$.

\bfig
\includegraphics[height=4.5cm]{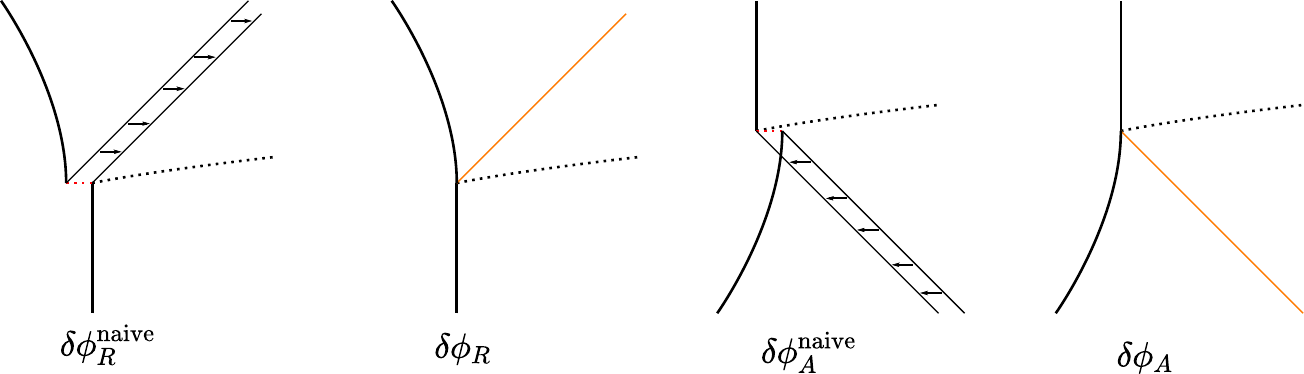}
\caption{Boundary jumping in the naive retarded and advanced solutions for the action deformed by $\psi_{s,\eta}$.  The geodesic $y^\mu_{t_-\eta}(s)$ is shown as a dotted black line.  In the naive retarded solution shown here the boundary jumps outwards to the future, so we need to act with a diffeomorphism to move it back.  Similarly in the naive advanced solution the boundary jumps inward to the past, so we need to act with a diffeomorphism to move it back.  If the matter fields are nonzero in the vicinity of the boundary, then we need to also modify them on the red dashed line in such a way that in the retarded/advanced solution they continue to obey the boundary conditions in the future/past.  The diffeomorphism has no physical effect in the interior of the spacetime, but the matter modification can lead to shockwaves shown in orange in the retarded/advanced solutions $\delta \phi_R$/$\delta\phi_A$.}\label{jumpfig}
\efig
So far this has all proceeded as in the discussion for $\psi_\lambda$, but now a new complication arises.  Our derivation of \eqref{psisjump} held for $s>0$, but if we try extending this discontinuity to $s=0$ then $\Delta\Phi$ is nonvanishing there and thus does not respect the boundary conditions \eqref{BC}.  Relatedly at $s=0$ there seems to be an additional term in the deformed equations motion, arising from the boundary term in $\beta(s)$.  In fact this is the same situation we found in the simple example in section \ref{boundarysec}: the naive retarded and advanced solutions obtained from the ``bulk'' equations of motion do not obey the boundary conditions, and thus need to be modified at the boundary to ensure that the boundary conditions continue to be respected.  In the current situation the needed modification has two parts.  The first part arises from the observation that in the naive retarded and advanced solutions sourced by the discontinuity \eqref{psisjump}, the surface at which the dilaton boundary condition \eqref{BC} is obeyed jumps discontinuously.   We can therefore undo this jump by acting on the retarded and advanced solutions with diffeomorphisms which move the ``new'' boundary locations back to the ``old'' boundary locations.  These diffeomorphisms will necessarily be discontinuous at $y^\mu_{t_-\eta}(0)$, as in the retarded solution the naive boundary jumps only to the future of this point while in the advanced solution it jumps only to its past. These diffeomorphisms have no physical effects away from the boundary, so they can be extended into the bulk in an arbitrary manner that we will not need to specify.  The second part arises because although the matter fields obey their boundary conditions at the ``old'' boundary, they don't necessarily do so at the ``new'' boundary and thus we need to modify them to ensure the boundary conditions are respected.  We illustrate the situation in figure \ref{jumpfig}.  In this subsection and the following one we will assume that the matter fields vanish in the vicinity $y^\mu_{t_-\eta}(0)$, in which case only the first modification is necessary.  In the $\epsilon\to 0$ limit this is true automatically for most kinds of matter,\footnote{\label{stressfootnote}We will see below (equations \eqref{etaH2}, \eqref{Honepsi2}) that nonvanishing boundary matter leads to extra contributions to the Peierls brackets of $H_-$ with $\psi_{s,\eta}$ and $H_\pm$ with $\wt{\eta}_\pm$ which at small $\epsilon$ are proportional to $\frac{1}{\epsilon} n_-^\mu n_-^\nu T_{\mu\nu}$.  For a weakly interacting scalar field dual to a boundary operator of dimension $\Delta$, this is of order $\epsilon^{2\Delta-1}$ and thus vanishes for $\Delta>\frac{1}{2}$.  If the scalar field has mass $m$ and Dirichlet boundary conditions, then we have the usual formula $\Delta=\frac{1}{2}+\frac{1}{2}\sqrt{1+4m^2}$ and thus this contribution vanishes for any $m^2>-\frac{1}{4}$.} but for completeness in appendix \ref{mattapp} we show how to generalize the analysis to the situation where they do not vanish for the special case of a free scalar field.  It is also interesting to consider the physical meaning of the boundary term in $\beta(s)$; in appendix \ref{boundaryapp} we show that this term is necessary to ensure that Peierls brackets we compute here genuinely coincide with the Poisson bracket on covariant phase space, or equivalently that the retarded and advanced solutions we construct here are geniune stationary points of the deformed action including boundary terms.

To work out the diffeomorphism which is necessary to fix the naive retarded solution (we will not need to discussed the advanced solution), we can parametrize the left boundary location in the naive solution as
\be
\hat{y}^\mu(t)=\hat{y}^\mu_0(t)+\theta(t-t_-)\Delta \hat{y}^\mu(t),
\ee
where $\hat{y}_0^\mu(t)$ is the unperturbed boundary trajectory.  The boundary conditions \eqref{BC} require that
\be
u_-^\mu(t)\equiv \epsilon \frac{d\hat{y}^\mu}{dt}
\ee
be a unit vector, so to avoid a $\delta$-function contribution to $u^\mu u_\mu$ at linear order in $k$ we must have
\be
u^\mu_{-}(t_-)\Delta \hat{y}_\mu(t_-)=0.
\ee
This then implies that
\be\label{dely}
\Delta \hat{y}^\mu(t_-)=\gamma n_-^\mu(t_-)
\ee
for some quantity $\gamma$ which we will now determine.  To simplify our expressions, for the remainder of this subsection unless otherwise indicated all quantities can be assumed to be evaluated at $t_-$ or $\hat{y}_0^\mu(t_-)=y^\mu_{t_-\eta}(0)$ as appropriate.  For the boundary condition $\Phi=\phi_b/\epsilon$ to be obeyed, the jump in the boundary location must cancel the the jump in the value of the dilaton:
\be
\Delta \Phi+\Delta \hat{y}^\mu\nabla_\mu \Phi=0.
\ee
From \eqref{dely} and \eqref{psisjump}, we thus have
\be
\gamma=-\frac{k}{2}\sinh s_0 \frac{\tau^\alpha\nabla_\alpha \psi(y(s_0))}{n_-^\mu\nabla_\mu \Phi}.\label{gamma}
\ee

We can also work out the discontinuities in $u^\mu_-$ and $n^\mu_-$ in the following way. We can describe the jumps as
\begin{align}\nonumber
u^\mu_-(t)&=u^\mu_{0-}(t)+\epsilon\gamma \delta(t-t_-)n^\mu_-(t)+\theta(t-t_-)\Delta u^\mu_-(t)\\
n^\mu_-(t)&=n^\mu_{0-}(t)+\epsilon\gamma \delta(t-t_-)u^\mu_-(t)+\theta(t-t_-)\Delta n^\mu_-(t),
\end{align}
where
\be
\Delta u^\mu_-\equiv \epsilon\frac{d\Delta \hat{y}^\mu}{dt},
\ee
$u^\mu_{0-}$ and $n^\mu_{0-}$ are the unperturbed versions of $u^\mu_{-}$ and $n^\mu_{-}$, and the $\delta$-function term in $n^\mu(t)$ is determined by the requirements that $n_\mu n^\mu=1$ and $n^\mu u_\mu=0$.  Applying these requirements (and also $u^\mu u_\mu=-1$) to the step function contributions, we have
\begin{align}\nonumber
\left(u_{0-}^\mu+\Delta u^\mu_-\right)\left(u_{0-}^\nu+\Delta u_-^\nu\right)g_{\mu\nu}\left(\hat{y}+\Delta\hat{y}\right)&=-1\\\nonumber
\left(n_{0-}^\mu+\Delta n^\mu_-\right)\left(n_{0-}^\nu+\Delta n_-^\nu\right)g_{\mu\nu}\left(\hat{y}+\Delta\hat{y}\right)&=1\\
\left(u_{0-}^\mu+\Delta u^\mu_-\right)\left(n_{0-}^\nu+\Delta n_-^\nu\right)g_{\mu\nu}\left(\hat{y}+\Delta\hat{y}\right)&=0.
\end{align}
At linear order in $k$ these imply that
\begin{align}\nonumber
\Delta^{(c)}u_-^\mu\equiv\Delta u^\mu_-+\Gamma^\mu_{\alpha\beta}u_-^\alpha\Delta \hat{y}^\beta&=\sigma n^\mu_-\\
\Delta^{(c)}n_-^\mu\equiv\Delta n^\mu_-+\Gamma^\mu_{\alpha\beta}n_-^\alpha\Delta \hat{y}^\beta&=\sigma u^\mu_-,\label{vecjump}
\end{align}
where $\sigma$ will be determined in a moment and $\Delta^{(c)}$ is the covariant operation we met in equation \eqref{covjump} which parallel-transports $u^\mu_-$ and $n^\mu_-$ from $\hat{y}$ to $\hat{y}+\Delta\hat{y}$ before comparing them with $u^\mu_-+\Delta u^\mu_-$ and $n^\mu_-+\Delta n^\mu_-$ respectively.  In fact \eqref{vecjump} is easily obtained by acting on $u^\mu u_\mu=-n_\mu n^\mu=-1$ and $u_\mu n^\mu=0$ with $\Delta^{(c)}$ and making use of the fact that
\be
\Delta^{(c)}g_{\mu\nu}=0+O(k^2).
\ee
Similarly we can determine $\sigma$ by acting with $\Delta^{(c)}$ on \eqref{uphi}, which at first order in $k$ tells us that\footnote{Note that we are defining the action of $\Delta^{(c)}$ on $\Phi$ to include both the change in $\Phi$ due to the jump $\Delta\hat{y}^\mu$ and also the change due to the discontinuity \eqref{psisjump}.}
\be
(\Delta^{(c)} u^\mu_-)\nabla_\mu\Phi+u_-^\mu\left(\Delta (\nabla_\mu\Phi)+\Delta\hat{y}^\nu\nabla_\nu\nabla_\mu\Phi\right)=0.
\ee
Using \eqref{undef2} and \eqref{vecjump} we can rewrite this as
\be
\sigma n_-^\mu\nabla_\mu \Phi+\cosh\eta \Delta\dot{\Phi}-\sinh\eta e^\mu\nabla_\mu\Delta\Phi+u^\mu_-\Delta\hat{y}^\nu \nabla_\mu\nabla_\nu \Phi=0,
\ee
and using the discontinuities \eqref{psisjump}, the unperturbed equations of motion \eqref{JTEOM}, and the fact that energy conservation requires the matter energy-momentum tensor to obey the boundary condition $T_{\mu\nu}u^\mu n^\nu=0$ (see \eqref{boundaryT}, but anyways we are assuming that the $T_{\mu\nu}=0$ near the boundary), we find that
\be
\sigma=\frac{k}{2}\frac{\cosh\eta \,e^\alpha\nabla_\alpha \psi(y(s_0))-\sinh\eta \cosh s_0\,\tau^\alpha \nabla_\alpha\psi(y(s_0))}{n_-^\beta\nabla_\beta \Phi}.\label{sigma}
\ee

We are now in a position to compute Peierls brackets involving $\psi_{s,\eta}(t_-)$.  In doing so we need to be sure to include both the naive discontinuities \eqref{psisjump} and also the additional discontinuities arising from the diffeomorphism which undoes the boundary jump \eqref{dely}.  On scalar quantities both effects are incorporated in our definition of $\Delta^{(c)}$, so for example we have
\begin{align}\nonumber
\Delta^{(c)}H_-&=-\frac{2}{\epsilon}\Delta^{(c)}\left(n_-^\mu\nabla_\mu \Phi\right)\\\nonumber
&=-\frac{2}{\epsilon}\left(\sigma u^\mu_-\nabla_\mu \Phi+n^\mu_- \Delta^{(c)}\nabla_\mu\Phi\right)\\\nonumber
&=-\frac{2}{\epsilon}\left(\gamma n^\mu_-n^\nu_-\nabla_\mu\nabla_\nu \Phi+\sinh \eta \Delta \dot{\Phi}-\cosh \eta e^\mu\nabla_\mu\Delta\Phi\right)\\
&=\frac{k}{\epsilon}\left(\sinh s_0 \frac{\tau^\alpha \nabla_\alpha\psi(y(s_0))}{n_-^\beta\nabla_\beta\Phi}\cdot \frac{\phi_b}{\epsilon}+\sinh \eta e^\alpha \nabla_\alpha \psi(y(s_0))-\cosh \eta \cosh s_0 \tau^\alpha\nabla_\alpha \psi(y(s_0))\right),\label{Hcalc}
\end{align}
where we have used \eqref{vecjump}, \eqref{dely}, \eqref{undef2}, the unperturbed equations of motion \eqref{JTEOM} with $T_{\mu\nu}=0$, which can be written as
\be\label{vacEOM}
\nabla_\mu\nabla_\nu \Phi=g_{\mu\nu}\Phi,
\ee
the boundary conditions \eqref{BC} (through \eqref{uphi}), and \eqref{psisjump}.  Thus we have the Peierls bracket
\be
\{H_-,\psi_{s,\eta}(t_-)\}=\frac{\phi_b}{\epsilon^2}\sinh s \frac{\tau^\alpha \nabla_\alpha\psi(y(s))}{n_-^\beta\nabla_\beta\Phi(t_-)}+\frac{1}{\epsilon}\Big(\sinh \eta e^\alpha\nabla_\alpha \psi(y(s))-\cosh \eta \cosh s\tau^\alpha\nabla_\alpha \psi(y(s))\Big).\label{Honepsi}
\ee
There are no discontinuities of any kind in the vicinity of the right boundary, so we trivially have
\be
\{H_+,\psi_{s,\eta}(t_-)\}=0.
\ee

We can also compute the jump in the two-sided length observable $L(t_-,t_+)$ and the relative boosts $\eta_\pm(t_-,t_+)$.  Indeed viewing $\Delta \hat{y}^\mu$ as the initial displacement for a deviation $\Delta y^\mu(\lambda)$ of the two-sided geodesic $y^\mu_{t_-t_+}(\lambda)$, we can decompose
\be
\Delta y^\mu(\lambda)=\alpha(\lambda)e^\mu(\lambda)+\beta(\lambda)\tau^\mu(\lambda),
\ee
in terms of which the (unsourced) geodesic deviation equation
\be
y^{\mu\prime}\nabla_\mu(y^{\nu\prime} \nabla_\nu \Delta y^\alpha)+R^\alpha_{\phantom \alpha\beta \mu\nu}y^{\beta\prime} y^{\nu\prime} \Delta y^\mu=0
\ee
tells us that
\begin{align}\nonumber
\alpha''&=0\\
\beta''&=L^2\beta.\label{twodev}
\end{align}
The relevant boundary conditions are that $\Delta y^\mu(0)=\Delta\hat{y}^\mu(t_-)$ and $\Delta y^\mu(1)=0$, so we therefore have
\begin{align}\nonumber
\alpha(\lambda)&=-\gamma (1-\lambda)\cosh \eta_-\\
\beta(\lambda)&=\gamma \frac{\sinh (L(1-\lambda))}{\sinh L}\sinh \eta_-.\label{ab2}
\end{align}
We then have
\begin{align}\nonumber
\Delta L&=\Delta\sqrt{y^{\mu\prime} y^{\nu\prime}g_{\mu\nu}}\\\nonumber
&=\frac{1}{2L}2y^{\mu\prime}y^{\alpha\prime}\nabla_\alpha\Delta y^{\nu}g_{\mu\nu}\\\nonumber
&=\alpha'(\lambda)\\
&=\gamma\cosh\eta_-,
\end{align}
so we have the Peierls bracket
\be
\{L(t_-,t_+),\psi_{s,\eta}(t_-)\}=-\frac{1}{2}\frac{\cosh \eta_-(t_-,t_+)\sinh s}{n_-^\alpha\nabla_\alpha\Phi(t_-)}\tau^\mu\nabla_\mu\psi(y(s)),
\ee
which is indeed compatible with \eqref{psisL}.  The jumps in $\eta_\pm$ are most easily computed by noting (see also \eqref{ecovjump}) that
\begin{align}\nonumber
\Delta^{(c)}e^\mu(\lambda)&\equiv \Delta e^\mu(\lambda)+\Delta y^\nu \Gamma^\mu_{\nu\lambda}e^\lambda\\\nonumber
&=\frac{1}{L}y^{\lambda\prime}\nabla_\lambda \Delta y^\mu-\frac{\Delta L}{L^2}y^{\mu\prime}\\
&=\frac{\beta'(\lambda)}{L}\tau^\mu(\lambda)\label{e2jump},
\end{align}
and then applying $\Delta^{(c)}$ to
\be
\cosh\eta_\pm=\pm e_\mu(\lambda_\pm) n_\pm^\mu,
\ee
which by way of \eqref{vecjump}, \eqref{ab2}, and \eqref{e2jump} gives
\begin{align}\nonumber
\Delta \eta_+&=\frac{\gamma\sinh \eta_-}{\sinh L}\\
\Delta \eta_-&=-\frac{\gamma \sinh\eta_-\cosh L}{\sinh L}+\sigma.
\end{align}
Thus using \eqref{gamma} and \eqref{sigma} we have the Peierls brackets
\begin{align}\nonumber
\{\eta_+(t_-,t_+),\psi_{s,\eta}(t_-)\}=&-\frac{1}{2}\frac{\sinh s\, \sinh \eta_-\,\tau^\alpha \nabla_\alpha \psi(y(s))}{\sinh L \,n_-^\mu \nabla_\mu \Phi(t_-)}\\\nonumber
\{\eta_-(t_-,t_+),\psi_{s,\eta }(t_-)\}=&\frac{1}{2 n_-^\mu\nabla_\mu\Phi(t_-)}\Big[\cosh\eta e^\alpha\nabla_\alpha \psi(y(s))\\
&+\big(\sinh s \sinh \eta_- \coth L-\sinh \eta \cosh s\big)\tau^\alpha\nabla_\alpha \psi(y(s))\Big].
\end{align}

Finally we can again study the renormalized versions of these brackets, which leads to
\begin{align}\nonumber
\{H_+,\psi_{\wt{s},\wt{\eta}}(t_-)\}&=0\\\nonumber
\{H_-,\psi_{\wt{s},\wt{\eta}}(t_-)\}&=\frac{1}{2\phi_b}\left[\left(-e^{-\wt{s}}+\phi_b e^{\wt{s}}H_--\frac{1}{4}e^{\wt{s}}\wt{\eta}^2\right)\tau^\alpha \nabla_\alpha \psi(y(\wt{s}))+\wt{\eta}e^\alpha \nabla_\alpha\psi(y(\wt{s}))\right]\\\nonumber
\{\wt{L}(t_-,t_+),\psi_{\wt{s},\wt{\eta}}(t_-)\}&=-\frac{1}{2}e^{\wt{s}}\tau^\alpha \nabla_\alpha \psi(y(\wt{s}))\\\nonumber
\{\wt{\eta}_+(t_-,t_+),\psi_{\wt{s},\wt{\eta}}(t_-)\}&=0\\
\{\wt{\eta}_-(t_-,t_+),\psi_{\wt{s},\wt{\eta}}(t_-)\}&=e^\alpha\nabla_\alpha \psi(y(\wt{s}))+\frac{1}{2}\left(\wt{\eta}_--\wt{\eta}\right)e^{\wt{s}}\tau^\alpha \nabla_\alpha \psi(y(\wt{s})).
\end{align}
In computing the renormalized bracket with the left Hamiltonian, we need the higher-order expressions
\begin{align}\nonumber
\sinh s&=\frac{\phi_b}{\epsilon}\left(e^{\wt{s}}-\frac{\epsilon^2}{4\phi_b^2}e^{-\wt{s}}\right)\\\nonumber
\cosh s&=\frac{\phi_b}{\epsilon}\left(e^{\wt{s}}+\frac{\epsilon^2}{4\phi_b^2}e^{-\wt{s}}\right)\\\nonumber
\frac{1}{n^\mu_- \nabla_\mu \Phi}&=\frac{\epsilon}{\phi_b}\left(1+\frac{\epsilon^2}{2\phi_b}H_-+\ldots\right)\\
\cosh \eta&=1+\frac{\epsilon^2}{8\phi_b^2}\wt{\eta}^2+\ldots.
\end{align}

\subsection[Brackets generated by the relative boosts]{Brackets generated by $\eta_\pm$}\label{etabracketsec}
We now consider Peierls brackets generated by the two-sided boost observables $\eta_\pm(t_-,t_+)$.  In computing the variation of the deformed action
\be
S=S_0-k\eta_\pm,
\ee
we can use the same sourced geodesic variation
\be
\delta y^\mu(\lambda)=\alpha(\lambda)e^\mu(\lambda)+\beta(\lambda)\tau^\mu(\lambda)
\ee
as we did in computing the variation of $\psi_\lambda$, with $\alpha$ and $\beta$ given by \eqref{alphabeta}.  Defining again
\be
\lambda_\pm\equiv \frac{1}{2}\pm \frac{1}{2},
\ee
we have
\be
u_\pm^\mu(t_\pm)e_\mu(\lambda_\pm)=\pm \sinh \eta_\pm.
\ee
Computing the variation of both sides of this expression, and using that $\delta u_\pm^\mu=0$ by the boundary conditions \eqref{BC}, we have
\be\label{etavar0}
u_{\pm\mu}(t_\pm)\delta e^\mu(\lambda_\pm)+u_\pm^\mu(t_\pm) e^\nu(\lambda_\pm)\delta g_{\mu\nu}(y(\lambda_\pm))=\pm \cosh \eta_\pm \delta \eta_\pm.
\ee
We can compute the variation of $e^\mu(\lambda_\pm)$ via
\begin{align}\nonumber\nonumber
\delta e^\mu(\lambda_\pm)&=\delta\left(\frac{y^{\mu\prime}(\lambda_\pm)}{L}\right)\\\nonumber
&=-\frac{\delta L}{L^2}y^{\mu\prime}(\lambda_\pm)+\frac{1}{L} y^{\lambda\prime}\partial_\lambda \delta y^\mu(\lambda_\pm)\\\nonumber
&=-\frac{e^\mu(\lambda_\pm)}{2}\int_0^1d\lambda e^\alpha e^\beta\delta g_{\alpha\beta}(y(\lambda))+\frac{1}{L} y^{\lambda\prime}\left(\nabla_\lambda \delta y^\mu(\lambda_\pm)-\Gamma^\mu_{\lambda\sigma}\delta y^\sigma(\lambda_\pm)\right)\\\nonumber
&=-\frac{e^\mu(\lambda_\pm)}{2}\int_0^1d\lambda e^\alpha e^\beta\delta g_{\alpha\beta}(y(\lambda))  +\frac{1}{L}\left(\alpha'(\lambda_\pm)e^\mu(\lambda_\pm)+\beta'(\lambda_\pm)\tau^\mu(\lambda_\pm)\right)\\
&=\frac{1}{L}\beta'(\lambda_\pm)\tau^\mu(\lambda_\pm)-\frac{1}{2}e^\alpha e^\beta\delta g_{\alpha\beta}(y(\lambda_\pm))e^\mu(\lambda_\pm),\label{edelt2}
\end{align}
where we have used \eqref{alphabeta}, \eqref{Lvar}, \eqref{etdiv}, and the vanishing of $\delta y^\mu(\lambda_\pm)$.  Combining this with \eqref{etavar0} and \eqref{alphabeta} we have the variations
\begin{align}\nonumber
\delta \eta_+=&\frac{L}{\sinh L}\int_0^1 d\lambda\left(\cosh(L\lambda)\tau^\alpha e^\beta \delta g_{\alpha\beta}(y(\lambda))+\frac{1}{2}\sinh(L\lambda)e^\alpha e^\beta\tau^\gamma\nabla_\gamma \delta g_{\alpha\beta}(y(\lambda))\right)\\\nonumber
&+\frac{1}{2}\tanh \eta_+e^\alpha e^\beta \delta g_{\alpha\beta}(y(\lambda_+))\\\nonumber
\delta \eta_-=&-\frac{L}{\sinh L}\int_0^1 d\lambda\left(\cosh(L(1-\lambda))\tau^\alpha e^\beta \delta g_{\alpha\beta}(y(\lambda))-\frac{1}{2}\sinh(L(1-\lambda))e^\alpha e^\beta\tau^\gamma\nabla_\gamma \delta g_{\alpha\beta}(y(\lambda))\right)\\
&+\frac{1}{2}\tanh \eta_-e^\alpha e^\beta \delta g_{\alpha\beta}(y(\lambda_-)).
\end{align}
Inserting $\delta$ functions and adopting Gaussian normal coordinates as usual, the deformed equations of motion are
\be
\ddot{\Phi}+\Phi+\frac{1}{2}e^\mu e^\nu T_{\mu\nu}=-\frac{k}{2}\frac{\sinh(L\lambda)}{\sinh L}\dot{\delta}(\tau)
\ee
for the $\eta_+$ deformation and
\be
\ddot{\Phi}+\Phi+\frac{1}{2}e^\mu e^\nu T_{\mu\nu}=-\frac{k}{2}\frac{\sinh(L(1-\lambda))}{\sinh L}\dot{\delta}(\tau)
\ee
for the $\eta_-$ deformation.  Integrating as usual, for $0<\lambda<1$ we get the dilaton discontinuities
\begin{align}\nonumber
\Delta \dot{\Phi}&=0\\\nonumber
\Delta \Phi&=-\frac{k}{2}\frac{\sinh(L\lambda)}{\sinh L}\\
e^\mu\nabla_\mu \Delta\Phi&=-\frac{k}{2}\frac{\cosh(L\lambda)}{\sinh L}
\end{align}
for the $\eta_+$ deformation and
\begin{align}\nonumber
\Delta \dot{\Phi}&=0\\\nonumber
\Delta \Phi&=-\frac{k}{2}\frac{\sinh(L(1-\lambda))}{\sinh L}\\
e^\mu\nabla_\mu \Delta\Phi&=\frac{k}{2}\frac{\cosh(L(1-\lambda))}{\sinh L}
\end{align}
for the $\eta_-$ deformation.

As in the previous subsection these discontinuities do not respect the boundary conditions, and the deformation by $\eta_\pm$ leads to a boundary jump
\begin{align}\nonumber
\Delta_\pm\hat{y}^\mu_\pm&=\frac{k}{2}\frac{1}{n_\pm^\alpha\nabla_\alpha\Phi(t_\pm)}n_\pm^\mu\\
\Delta_\mp\hat{y}^\mu_\pm&=0.
\end{align}
Here the $\pm$ on $\Delta_\pm$ indicates whether we have deformed the action by $\eta_+$ or $\eta_-$, while as usual the $\pm$ on the other quantities such as $\hat{y}^\mu_\pm$ indicate which boundary we are discussing.  These jumps need to be fixed by a discontinuous diffeomorphism, as in figure \ref{jumpfig}, to give the true advanced/retarded solutions.  As before we can define a discretized differential $\Delta_{\pm}^{(c)}$, where the $\pm$ again indicates whether we have deformed the action by $\eta_+$ or $\eta_-$, and via a similar argument as in the previous section we have
\begin{align}\nonumber
\Delta^{(c)}_\pm u_\pm^\mu&=\sigma_\pm n_\pm^\mu\\\nonumber
\Delta^{(c)}_\pm n_\pm^\mu&=\sigma_\pm u_\pm^\mu\\\nonumber
\Delta^{(c)}_\mp u_\pm^\mu&=\wt{\sigma}_\pm n_\pm^\mu\\
\Delta^{(c)}_\mp n_\pm^\mu&=\wt{\sigma}_\pm u_\pm^\mu,
\end{align}
with
\begin{align}\nonumber
\sigma_\pm&=\frac{k}{2}\frac{\cosh L \sinh \eta_\pm}{\sinh L n_\pm^\mu\nabla_\mu \Phi(t_\pm)}\\
\wt{\sigma}_\pm&=-\frac{k}{2}\frac{\sinh \eta_\pm}{\sinh L n_\pm^\mu\nabla_\mu\Phi(t_\pm)}.
\end{align}
These lead to discontinuities
\be
\Delta_\pm y^\mu(\lambda)=\alpha_\pm(\lambda)e^\mu(\lambda)+\beta_\pm(\lambda)\tau^\mu(\lambda)
\ee
of the two-sided geodesic $y_{t_-t_+}^\mu(\lambda)$, with $\alpha_\pm$ and $\beta_\pm$ obeying \eqref{twodev} and given by
\begin{align}\nonumber
\alpha_+(\lambda)&=\frac{k}{2}\frac{\cosh\eta_+}{n_+^\mu\nabla_\mu\Phi(t_+)}\lambda\\\nonumber
\beta_+(\lambda)&=\frac{k}{2}\frac{\sinh \eta_+\sinh(L\lambda)}{ n_+^\mu\nabla_\mu \Phi(t_+)\sinh L}\\\nonumber
\alpha_-(\lambda)&=-\frac{k}{2}\frac{\cosh\eta_-}{n_-^\mu\nabla_\mu\Phi(t_-)}(1-\lambda)\\
\beta_-(\lambda)&=\frac{k}{2}\frac{\sinh \eta_-\sinh(L(1-\lambda))}{ n_-^\mu\nabla_\mu \Phi(t_-)\sinh L},
\end{align}
and discontinuities
\be
\Delta_\pm y^\mu(s)=\alpha_\pm(s)e^\mu(s)+\beta_\pm(s)\tau^\mu(s)
\ee
of the one-sided geodesic $y_{t_-\eta}^\mu(s)$, with $\alpha_\pm$ and $\beta_\pm$ obeying \eqref{onesidedev} and given by
\begin{align}\nonumber
\alpha_+&=0\\\nonumber
\beta_+&=-\wt{\sigma}_-\sinh s\\\nonumber
\alpha_-&=-\frac{k}{2}\frac{\cosh \eta}{n_-^\mu\nabla_\mu \Phi(t_-)}\\
\beta_-&=\frac{k}{2}\frac{\sinh \eta}{n_-^\mu\nabla_\mu \Phi(t_-)}\cosh s-\sigma_- \sinh s.
\end{align}
We may then use these discontinuities to work out the discontinuities of our various observables as in the previous subsections, leading to the Peierls brackets
\begin{align}\nonumber
\{L,\eta_\pm\}&=\frac{1}{2} \frac{\cosh \eta_\pm}{n_\pm^\mu\nabla_\mu \Phi(t_\pm)}\\\nonumber
\{H_\mp,\eta_\pm\}&=-\frac{1}{\epsilon}\frac{\cosh\eta_{\mp}}{\sinh L}\\\nonumber
\{H_\pm,\eta_\pm\}&=\frac{1}{\epsilon}\left(\cosh \eta_\pm\coth L-\frac{\phi_b}{\epsilon}\frac{1}{n_\pm^\mu\nabla_\mu \Phi(t_\pm)}\right)\\\nonumber
\{\psi_\lambda,\eta_+\}&=\frac{1}{2n^\alpha_+\nabla_\alpha \Phi(t_+)}\left(\lambda\cosh\eta_+e^\mu\nabla_\mu\psi(y(\lambda))+\sinh\eta_+\frac{\sinh(L\lambda)}{\sinh L}\tau^\mu\nabla_\mu\psi(y(\lambda))\right)\\\nonumber
\{\psi_\lambda,\eta_-\}&=\frac{1}{2n^\alpha_-\nabla_\alpha \Phi(t_-)}\left((\lambda-1)\cosh\eta_-e^\mu\nabla_\mu\psi(y(\lambda))+\sinh\eta_-\frac{\sinh(L(1-\lambda))}{\sinh L}\tau^\mu\nabla_\mu\psi(y(\lambda))\right)\\\nonumber
\{\psi_{s,\eta},\eta_+\}&=\frac{1}{2}\frac{\sinh s\, \sinh \eta_-\,\tau^\alpha \nabla_\alpha \psi(y(s))}{\sinh L \,n_-^\mu \nabla_\mu \Phi(t_-)}\\\nonumber
\{\psi_{s,\eta},\eta_-\}&=-\frac{1}{2 n_-^\mu\nabla_\mu\Phi(t_-)}\Big[\cosh\eta e^\alpha\nabla_\alpha \psi(y(s))+\big(\sinh s \sinh \eta_- \coth L-\sinh \eta \cosh s\big)\tau^\alpha\nabla_\alpha \psi(y(s))\Big]\\
\{\eta_-,\eta_+\}&=\frac{1}{2\sinh L}\left[\frac{\sinh \eta_+}{n_+^\mu\nabla_\mu(t_+)}-\frac{\sinh \eta_-}{n_-^\mu\nabla_\mu(t_-)}\right].\label{etaresults}
\end{align}
For brevity we have again suppressed the time-dependence of the obserables on the left-hand side, but $\eta_\pm$, $\psi_\lambda$, and $L$ should all be evaluated at the same times $t_\pm$ and $\psi_{s,\eta}$ is evaluted at $t_-$. In computing $\{H_\pm,\eta_\pm\}$  we have again assumed that the energy momentum tensor vanishes at the boundaries, allowing the use of \eqref{vacEOM}, with the same justifications as in the previous subsection.  The first, fourth, fifth, sixth, and seventh of these brackets are compatible with our previous results computed using the other observable as the deformation of the action, and the last is manifestly antisymmetric and thus also compatible with the antisymmetry of the Peierls bracket.  We have already computed the renormalized versions of most of these brackets, but for the ones which are new in the $\epsilon\to 0$ limit we have
\begin{align}\nonumber
\{H_\pm,\wt{\eta}_\pm\}&=\frac{\wt{\eta}_\pm^2}{4\phi_b}-H_\pm\\\nonumber
\{H_\mp,\wt{\eta}_\pm\}&=-\frac{1}{\phi_b}e^{-\wt{L}}\\
\{\wt{\eta}_-,\wt{\eta}_+\}&=0.
\end{align}

\subsection[Brackets generated by the Hamiltonians]{Brackets generated by $H_\pm$}\label{Hsec}
We finally consider brackets generated by the Hamiltonians $H_\pm$.  We have already computed all of these brackets using our other observables as sources, but in those calculations we paid no attention to the special role of $H_\pm$ as generators of boundary time translations.  Here we will use this directly, which simplifies our calculations substantially.  Indeed we know that for any observable $f(t_-,t_+)$ defined relative to the left and/or right boundaries at times $t_\pm$, we must have
\be
\{f,H_\pm\}=\partial_\pm f.
\ee
We thus merely need to see how our various dressing geodesics behave as we perturb the times $t_\pm$ at which they intersect the right and left boundaries by displacements $\Delta t_\pm$, which once again is an exercise in solving the geodesic deviation equation.

We will first consider the deviation $\Delta y^\mu$ of the two-sided geodesic $y_{t_-t_+}^\mu(\lambda)$, which as usual we parametrize as
\be
\Delta y^\mu(\lambda)=\alpha(\lambda)e^\mu(\lambda)+\beta(\lambda)\tau^\mu(\lambda).
\ee
$\alpha$ and $\beta$ must obey the geodesic deviation equation \eqref{twodev}, with their boundary conditions determined by the requirement that
\be
\Delta y^\mu(\lambda_\pm)=\frac{\Delta t_\pm}{\epsilon}u_\pm^\mu(t_\pm).
\ee
We thus have
\begin{align}\nonumber
\alpha(\lambda)&=\frac{\Delta t_+}{\epsilon}\sinh \eta_+ \lambda-\frac{\Delta t_-}{\epsilon}\sinh \eta_- (1-\lambda)\\
\beta(\lambda)&=\frac{\Delta t_+}{\epsilon}\cosh \eta_+\frac{\sinh(L\lambda)}{\sinh L}+\frac{\Delta t_-}{\epsilon}\cosh \eta_- \frac{\sinh(L(1-\lambda))}{\sinh L}.
\end{align}
We can immediately compute the deviations in $L$ and $\psi_\lambda$:
\begin{align}\nonumber
\Delta L&=\alpha'(0)\\
&=\frac{\Delta t_+}{\epsilon}\sinh\eta_++\frac{\Delta t_-}{\epsilon}\sinh \eta_-,
\end{align}
and
\be
\Delta \psi_\lambda=\alpha(\lambda)e^\mu\nabla_\mu \psi(y(\lambda))+\beta(\lambda)\tau^\mu\nabla_\mu \psi(y(\lambda)),
\ee
which give brackets
\begin{align}\nonumber
\{L,H_\pm\}&=\frac{\sinh \eta_\pm}{\epsilon}\\\nonumber
\{\psi_\lambda,H_+\}&=\frac{1}{\epsilon}\left(\lambda\sinh\eta_+e^\mu\nabla_\mu\psi(y(\lambda))+\cosh\eta_+\frac{\sinh(L\lambda)}{\sinh L}\tau^\mu\nabla_\mu\psi(y(\lambda))\right)\\
\{\psi_\lambda,H_-\}&=\frac{1}{\epsilon}\left((\lambda-1)\sinh\eta_-e^\mu\nabla_\mu\psi(y(\lambda))+\cosh\eta_-\frac{\sinh(L(1-\lambda))}{\sinh L}\tau^\mu\nabla_\mu\psi(y(\lambda))\right),
\end{align}
which are compatible with \eqref{HL} and \eqref{twosidepb} above.  Computing the deviations of $\eta_\pm$ requires a bit more work, as we need the deviations
\be
\Delta^{(c)} e^\mu(\lambda)=\frac{\beta'(\lambda)}{L}\tau^\mu(\lambda)
\ee
(see \eqref{e2jump}) and
\be
\Delta^{(c)}u_\pm^\mu\equiv \Delta y^\nu(\lambda_\pm)\nabla_\nu u_\pm^\mu.
\ee
Since $\Delta^{(c)}(u_\pm^\mu u_{\pm\mu})=0$ we must have
\be
\Delta^{(c)}u^\mu_\pm=\sigma_\pm n_\pm^\mu,
\ee
and acting with $\Delta^{(c)}$ on equation \eqref{uphi} we find
\begin{align}\nonumber
\Delta^{(c)}(u^\mu_\pm \nabla_\mu \Phi)&=\sigma_\pm n^\mu_\pm \nabla_\mu\Phi(t_\pm)+\frac{\Delta t_\pm}{\epsilon}u_\pm^\mu u_\pm^\nu \nabla_\mu\nabla_\nu\Phi\\\nonumber
&=\sigma_\pm n^\mu_\pm \nabla_\mu\Phi(t_\pm)-\frac{\Delta t_\pm}{\epsilon}\left(\frac{\phi_b}{\epsilon}+\frac{1}{2}n_\pm^\mu n_\pm^\nu T_{\mu\nu}(y(\lambda_\pm))\right)\\
&=0,
\end{align}
where in the second equality we have used the equations of motion \eqref{JTEOM} in the form
\be
\nabla_\mu\nabla_\nu \Phi=g_{\mu\nu}\Phi+\frac{1}{2}Tg_{\mu\nu}-\frac{1}{2}T_{\mu\nu}
\ee
and the boundary conditions \eqref{BC}, and thus
\be
\sigma_\pm=\frac{\Delta t_\pm}{\epsilon n^\mu_\pm \nabla_\mu\Phi(t_\pm)}\left(\frac{\phi_b}{\epsilon}+\frac{1}{2}n_\pm^\alpha n_\pm^\beta T_{\alpha\beta}(y(\lambda_\pm))\right).
\ee
Now applying $\Delta^{(c)}$ to equation \eqref{ueta}, we then find
\begin{align}\nonumber
\Delta \eta_\pm&=\sigma_\pm\mp \frac{\beta'(\lambda_\pm)}{L}\\
&=\frac{1}{\epsilon}\left[\frac{\Delta t_\pm}{n_\pm^\mu\nabla_\mu \Phi(t_\pm)}\left(\frac{\phi_b}{\epsilon}+\frac{1}{2}n_\pm^\alpha n_\pm^\beta T_{\alpha\beta}(y(\lambda_\pm))\right)+\Delta t_\mp\frac{\cosh \eta_\mp}{\sinh L}-\coth L \cosh \eta_\pm \Delta t_\pm\right],\label{etaHdev}
\end{align}
which gives
\begin{align}\nonumber
\{\eta_\pm,H_\pm\}&=\frac{1}{\epsilon}\left[\frac{1}{n_\pm^\mu\nabla_\mu \Phi(t_\pm)}\left(\frac{\phi_b}{\epsilon}+\frac{1}{2}n_\pm^\alpha n_\pm^\beta T_{\alpha\beta}(y(\lambda_\pm))\right)-\coth L \cosh \eta_\pm\right]\\
\{\eta_\pm,H_\mp\}&=\frac{1}{\epsilon}\frac{\cosh \eta_\mp}{\sinh L}.\label{etaH2}
\end{align}
These are compatible with the second and third lines of \eqref{etaresults}, and in fact generalize them to the situation where the energy-momentum tensor does not necessarily vanish at the boundary.

We can also study the deviation $\Delta y^\mu$ of the one-sided geodesic $y_{t_-\eta}^\mu(s)$, which we can parametrize as
\be
\Delta y^\mu(s)=\alpha(s)e^\mu(s)+\beta(s)\tau^\mu(s).
\ee
$\alpha$ and $\beta$ now obey \eqref{onesidedev}, as well as the boundary conditions
\begin{align}\nonumber
\Delta y^\mu(0)&=\frac{\Delta t_-}{\epsilon}u_-^\mu(t_-)\\\nonumber
\alpha'(0)&=0\\
\beta'(0)&=-\sigma_-,
\end{align}
where the second of these arises as in \eqref{alphap} and the third arises from an analogous argument to the one leading to the first line of \eqref{etaHdev}.  We thus have
\begin{align}\nonumber
\alpha(s)&=-\frac{\Delta t_-}{\epsilon}\sinh \eta\\
\beta(s)&=\frac{\Delta t_-}{\epsilon}\left[\cosh \eta \cosh s-\frac{\sinh s}{n_-^\mu\nabla_\mu \Phi(t_-)}\left(\frac{\phi_b}{\epsilon}+\frac{1}{2}n_-^\alpha n_-^\beta T_{\alpha\beta}(y(0))\right)\right],
\end{align}
which gives
\begin{align}\nonumber
\{\psi_{s,\eta},H_-\}&=-\frac{1}{\epsilon}\Bigg[\sinh \eta e^\mu\nabla_\mu\psi(y(s))\\
&+\left\{\frac{\sinh s}{n_-^\alpha\nabla_\alpha \Phi(t_-)}\left(\frac{\phi_b}{\epsilon}+\frac{1}{2}n_-^\alpha n_-^\beta T_{\alpha\beta}(y(0))\right)
-\cosh \eta \cosh s\right\}\tau^\mu\nabla_\mu\psi(y(s))\Bigg].\label{Honepsi2}
\end{align}
This is compatible with \eqref{Honepsi}, and in fact generalizes it to the situation where the energy-momentum tensor is possibly nonzero at the boundary.  We also trivially have
\be
\{\psi_{s,\eta},H_+\}=0,
\ee
since $\psi_{s,\eta}$ does not depend on $t_+$.

This completes the evaluation of the Peierls brackets of our set of diffeomorphism-invariant observables in JT gravity coupled to matter. We have now computed each bracket in two distinct ways, using the non-manifest antisymmetry of the Peierls bracket to check our calculations.  In fact another check is also possible: we can compute the Peierls bracket of each observable with \textit{itself} and see that it vanishes.  We have confirmed this in a few cases, but will spare the reader the details as this section is already long enough.

\section{First Applications}\label{appsec}
We now gather together the various renormalized brackets we have computed, starting with those that involve only the gravitational variables:
\begin{align}\nonumber
\{\wt{L},\wt{\eta}_{\pm}\}&=1\\\nonumber
\{\wt{L},H_{\pm}\}&=\frac{\wt{\eta}_{\pm}}{2\phi_b}\\\nonumber
\{\wt{\eta}_{\pm},H_{\pm}\}&=H_{\pm}-\frac{\wt{\eta}_{\pm}^2}{4\phi_b}\\\nonumber
\{\wt{\eta}_{\pm},H_{\mp}\}&=\frac{1}{\phi_b}e^{-\wt{L}}\\\nonumber
\{H_+,H_-\}&=0\\
\{\wt{\eta}_+,\wt{\eta}_-\}&=0.\label{gravalg}
\end{align}
Here $\wt{L}$ and $\wt{\eta}_{\pm}$ should be understood as being evaluated at the same boundary times $t_-,t_+$, while the Hamiltonians can be evaluated at any time.  We refer to \eqref{gravalg} as the \textit{gravitational algebra}.   We also have  brackets of these renormalized gravitational observables with the two-sided matter observable $\psi_{\wt{\lambda}}(t_-,t_+)$:
\begin{align}\nonumber
\{\wt{L},\psi_{\wt{\lambda}}\}&=0\\\nonumber
\{\wt{\eta}_{\pm},\psi_{\wt{\lambda}}\}&=\mp \frac{1}{2}e^\mu\nabla_\mu\psi(y(\wt{\lambda}))\\
\{H_{\pm},\psi_{\wt{\lambda}}\}&=\mp\frac{\wt{\eta}_{\pm}}{4\phi_b}e^\mu\nabla_\mu\psi(y(\wt{\lambda}))-\frac{1}{2\phi_b}e^{-\frac{\wt{L}}{2}\pm \wt{\lambda}}\tau^\mu\nabla_\mu\psi(y(\wt{\lambda})),\label{twosidealg}
\end{align}
where again all observables are defined at boundary times $t_-,t_+$, as well as the brackets with the one-sided matter observable $\psi_{\wt{s},\wt{\eta}}$:
\begin{align}\nonumber
\{\wt{L},\psi_{\wt{s},\wt{\eta}}\}&=-\frac{1}{2}e^{\wt{s}}\tau^\alpha \nabla_\alpha \psi(y(\wt{s}))\\\nonumber
\{\wt{\eta}_+,\psi_{\wt{s},\wt{\eta}}\}&=0\\\nonumber
\{\wt{\eta}_-,\psi_{\wt{s},\wt{\eta}}\}&=e^\alpha\nabla_\alpha \psi(y(\wt{s}))+\frac{1}{2}\left(\wt{\eta}_--\wt{\eta}\right)e^{\wt{s}}\tau^\alpha \nabla_\alpha \psi(y(\wt{s}))\\\nonumber
\{H_+,\psi_{\wt{s},\wt{\eta}}\}&=0\\
\{H_-,\psi_{\wt{s},\wt{\eta}}\}&=\frac{1}{2\phi_b}\left[\left(-e^{-\wt{s}}+\phi_b e^{\wt{s}}H_--\frac{1}{4}e^{\wt{s}}\wt{\eta}^2\right)\tau^\alpha \nabla_\alpha \psi(y(\wt{s}))+\wt{\eta}e^\alpha \nabla_\alpha\psi(y(\wt{s}))\right],\label{onesidealg}
\end{align}
where all two-sided observables are defined at times $t_-,t_+$ and $\psi_{\wt{s},\wt{\eta}}$ is defined at left-boundary time $t_-$.  In this section we study the physics of this algebra, understanding better its mathematical structure and using it to illustrate various phenomena in JT gravity.  In the course of our discussion we will need to use every single one of these brackets, and we will also show that the time-dependence of the observables $\wt{L}$, $\wt{\eta}_\pm$, $H_\pm$ can be completely solved using the algebra \eqref{gravalg}.

\subsection{Pure JT gravity}\label{trivsec}
We'll begin by comparing this algebra to results obtained in \cite{Harlow:2018tqv} for the special case where the matter theory is trivial (meaning that there is no matter theory).  The full set of solutions of the equations of motion are then given by \eqref{nomattsol}, and from equations \eqref{Hnomatt}-\eqref{etatdef} we have
\begin{align}\nonumber
H_\pm&=\frac{\Phi_h^2}{\phi_b}\\\nonumber
\wt{L}(t_-,t_+)&=2\log\left(\frac{1}{\Phi_h}\cosh \left(\frac{\Phi_h}{\phi_b}\frac{t_-+t_+}{2}\right)\right)\\
\wt{\eta}_\pm(t_-,t_+)&=2\Phi_h \tanh\left(\frac{\Phi_h}{\phi_b}\frac{t_-+t_+}{2}\right).
\end{align}
In \cite{Harlow:2018tqv} precisely these expressions for $\wt{L}$ and $\wt{\eta}\equiv \wt{\eta}_\pm$ were shown to obey the canonical commutation relation
\be\label{canonical}
\{\wt{L},\wt{\eta}\}=1,
\ee
which is thus compatible with the first line of \eqref{gravalg}.  Moreover we can use these equations to express $H_\pm$ as a function of $\wt{L}$ and $\wt{\eta}$, giving the mechanics of a particle in an exponential potential \cite{Harlow:2018tqv}:\footnote{In \cite{Harlow:2018tqv} what was studied was the full Hamiltonian, given by $H=H_++H_-=2H_\pm$, $\wt{\eta}$ was called $P$, and $\frac{t_-+t_+}{2}$ was called $\delta$.}
\be\label{expH}
H_\pm=\frac{1}{\phi_b}\left(\frac{\wt{\eta}^2}{4}+e^{-\wt{L}}\right).
\ee
Using \eqref{canonical} and \eqref{expH} we may then easily verify that the rest of the gravitational algebra \eqref{gravalg} holds, for example
\be
\{\wt{L},H_\pm\}=\frac{\wt{\eta}}{2\phi_b}\{\wt{L},\wt{\eta}\}=\frac{\wt{\eta}}{2\phi_b}.
\ee

We can also demonstrate the equivalences $\wt{\eta}_+=\wt{\eta}_-$ and $H_+=H_-$ in a more abstract way by integrating the constraint equations. Indeed consider the two-sided geodesic $y^\mu_{t_-t_+}(\lambda)$, with unit tangent vector $e^\mu(\lambda)$ and unit normal vector $\tau^\mu(\lambda)$.  Contracting the vacuum equations of motion
\be
\nabla_\mu \nabla_\nu \Phi=g_{\mu\nu}\Phi
\ee
with $e^\mu \tau^\nu$ and using \eqref{etdiv}, we have
\be
e^\mu \tau^\nu \nabla_\mu \nabla_\nu \Phi(y(\lambda))=e^\mu \nabla_\mu \left(\tau^\nu\nabla_\nu\Phi(y(\lambda))\right)=\frac{1}{L}\frac{d}{d\lambda}\big(\tau^\mu\nabla_\mu\Phi(y(\lambda))\big)=0.
\ee
Integrating this from $\lambda=\lambda_-=0$ to $\lambda=\lambda_+=1$ we therefore have
\be
\tau^\mu(\lambda_+)\nabla_\mu\Phi(y(\lambda_+))=\tau^\mu(\lambda_-)\nabla_\mu\Phi(y(\lambda_-)).
\ee
From the inverse of \eqref{undef1} we have
\be
\tau^\mu(\lambda_\pm)=\cosh \eta_\pm u_\pm^\mu-\sinh \eta_\pm n_\pm^\mu,
\ee
so since the boundary conditions require $u_\pm^\mu\nabla_\mu \Phi(y(\lambda_\pm))=0$ we must have
\be\label{sinheq}
\sinh \eta_+n_+^\mu\nabla_\mu\Phi(y(\lambda_+))=\sinh\eta_- n_-^\mu\nabla_\mu \Phi(y(\lambda_-)).
\ee
Similarly contracting with $e^\mu e^\nu$ gives
\be
\frac{1}{L^2}\frac{d^2\Phi}{d\lambda^2}=\Phi,
\ee
which we can solve together with the boundary conditions \eqref{BC} to find that
\be
\Phi(\lambda)=\frac{\phi_b}{\epsilon}\frac{\cosh(L(\lambda-1/2))}{\cosh (L/2)}.
\ee
This implies that
\be
e^\mu \nabla_\mu \Phi(y(\lambda_+))=-e^\mu \nabla_\mu \Phi(y(\lambda_-)),
\ee
or equivalently
\be\label{cosheq}
\cosh \eta_+n_+^\mu\nabla_\mu\Phi(y(\lambda_+))=\cosh\eta_- n_-^\mu\nabla_\mu \Phi(y(\lambda_-)).
\ee
\eqref{sinheq} and \eqref{cosheq}, together with \eqref{Hpm} and \eqref{etatdef}, then tell us that
\be
\wt{\eta}_+=\wt{\eta}_-\label{etaeq}
\ee
and
\be
H_+=H_-.\label{Heq}
\ee

\subsection{Traversable wormhole}
In \cite{Gao:2016bin} it was pointed out that the wormhole in the thermofield-double state of AdS/CFT can be made traversable by temporarily turning on a simple interaction between the two CFTs which are dual to the left and right boundaries.  This idea was explored in more detail in the context of JT gravity coupled to matter in \cite{Maldacena:2017axo}.  In both treatments the interaction was of the form
\be\label{twosideint}
H_{int}=\sum_i\int dt d\Omega h(t,\Omega)O_{-i}(t,\Omega)O_{+i}(t,\Omega),
\ee
where $t$ and $\Omega$ are boundary time and space coordinates, $O_{-i}$ is a local operator on the left boundary that is dual to some bulk field $\phi_i$, $O_{+i}$ is the same operator on the right boundary, and $h$ is some function that vanishes outside of a narrow time window.  To justify their calculations the authors of \cite{Gao:2016bin,Maldacena:2017axo} had to make various special assumptions about the matter fields $\phi_i$: in \cite{Gao:2016bin} they needed to assume that the dual operator $O$ (they didn't sum over $i$) was rather relevant, with $\Delta<\frac{d}{2}$, while in \cite{Maldacena:2017axo} it was necessary to take the number of fields $\phi_i$ to be large.  On the other hand the basic idea of traversability is quite general, and it would be a pity if it genuinely relied on such arbitrary features of the available matter.  In this section we will use the gravitational algebra \eqref{gravalg} to show that the essential mechanism behind the traversability phenomenon discovered in \cite{Gao:2016bin} is purely gravitational, with no special features of the matter fields required.  The only role for the matter fields $\phi_i$ is to translate this gravitational mechanism into the dual CFT in a way that avoids the factorization problem of \cite{Harlow:2015lma,Harlow:2018tqv}, and the special features needed in \cite{Gao:2016bin,Maldacena:2017axo} arise only if one wishes to make this boundary translation particularly simple.

\bfig
\includegraphics[height=7cm]{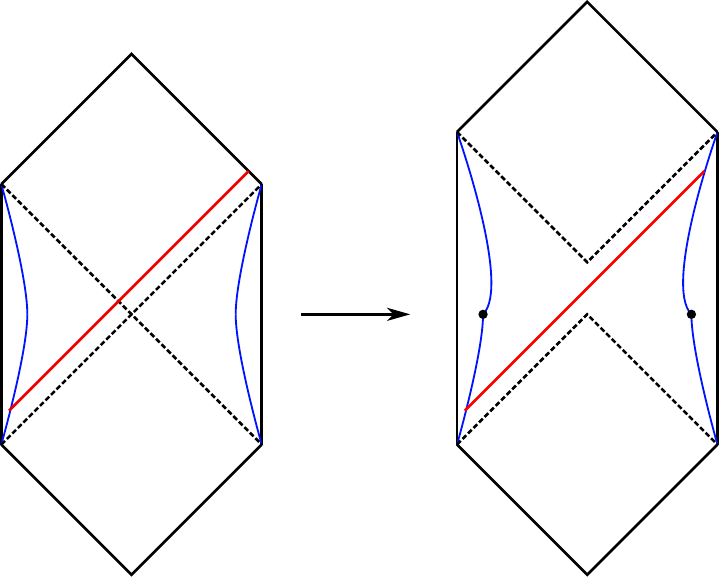}
\caption{The traversability picture of \cite{Maldacena:2017axo}: on the left an observer who attempts to cross from the left to the right boundary cannot do so, but once we introduce a perturbation that ``kicks'' the regulated boundaries inward, they take longer to get back to the $AdS_2$ boundaries and the red observer now can make it through.}\label{traversefig}
\efig
In \cite{Maldacena:2017axo} a simple picture of how wormhole traversability works in JT gravity was given: we just need to deform the evolution by an interaction which gives an ``inward kick'' to the regulated boundaries.  The basic idea is illustrated in figure \ref{traversefig}.  Much of the technical work done in \cite{Gao:2016bin,Maldacena:2017axo} went into showing that an appropriately constructed interaction of the form \eqref{twosideint} indeed has this effect.  Our gravitational algebra \eqref{gravalg} however suggests an immediate candidate for this interaction: the renormalized length observable $\wt{L}$.  Indeed the boundary kicks in figure \ref{traversefig} are nothing but jumps in the relative boosts $\eta_\pm(t_-,t_+)$, where $t_\pm$ are the boundary times of the kicks, and traversability requires
\be
\Delta \eta_{\pm}<0.
\ee
The first line of the gravitational algebra tells us that $\wt{L}$ and $\wt{\eta}_\pm$ generate translations of each other, and thus that a deformation of the action by $\wt{L}(t_-,t_+)$ of appropriate sign will produce the desired kicks.  In fact the perturbed solution in the right diagram of figure \ref{traversefig} is precisely the retarded solution we constructed in section \ref{Lsec}, and the jumps in $\eta_\pm$ are precisely those given by \eqref{etaLjump}!\footnote{We emphasize that this retarded solution is \textit{not} a solution of JT gravity coupled to matter, it is a solution of equations of motion for non-local deformed action \eqref{LdefS}.  This non-locality is necessary for traversability, and is similar to the non-locality of the interaction \eqref{twosideint} used in \cite{Gao:2016bin}.}  We thus see that the real driver of the traversable wormhole phenomenon is a deformation by $\wt{L}$: this is natural, as it is a geometric effect and the matter fields should not play any intrinsic role.

\bfig
\includegraphics[height=4cm]{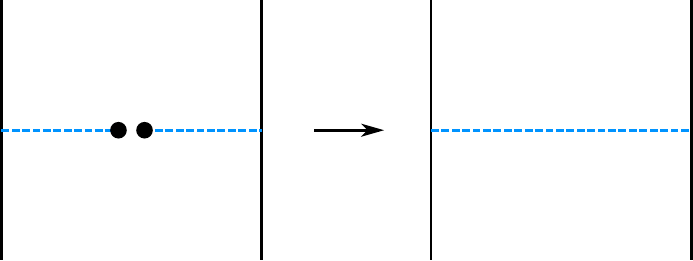}
\caption{Building a gauge or gravitational Wilson line by bringing together dressed matter fields.}\label{wilsonfig}
\efig
Why then did the authors of \cite{Gao:2016bin,Maldacena:2017axo} need to use matter fields and why did those matter fields have to obey certain special properties?  The reason is that they wanted a simple dual CFT description of the process, and at first it is not so clear what is the CFT description of $\wt{L}$.  In fact there is an analogous problem in electromagnetism: in the bulk there are Wilson line operators which extend from one boundary to the other, but there is no obvious way of ``reconstructing'' them from the dual CFT current since the Wilson line cannot be split into gauge-invariant parts using only the gauge field \cite{Harlow:2015lma}.  In  \cite{Harlow:2015lma} a way of fixing this was proposed: we can realize the Wilson line via an operator product of charged matter fields which are dressed to the left and right boundaries respectively (see figure \ref{wilsonfig}).  Moreover it was emphasized that a similar issue arises for gravitational Wilson lines, and in \cite{Harlow:2018tqv} this was discussed specifically for the renormalized length operator $\wt{L}$ and it was emphasized that its boundary representation in the SYK model involves bilinears of ``matter'' fermions on different boundaries.    This observation however does not quite get us to an interaction of the form \eqref{twosideint}, as to pick out $\wt{L}$ alone at low energies we need to bring the two matter fields in figure \ref{wilsonfig} quite close together in the bulk and therefore in the dual CFT we have to use rather nonlocal HKLL (or in general entanglement wedge) reconstructions of them \cite{Hamilton:2006az,Harlow:2018fse}.  If we keep the two matter operators near the boundary as in \cite{Gao:2016bin,Maldacena:2017axo}, then there will still be some projection onto $\wt{L}$ which implements the desired effect but the perturbation will also create other unwanted disturbances and work is needed to ensure these other effects do not overwhelm the traversability.  Ensuring this is the source of the complications and assumptions in  \cite{Gao:2016bin,Maldacena:2017axo}.

\subsection{SL(2,R) charges}
An $SL(2,\mathbb{R})$ algebra of charges in JT gravity coupled to matter was constructed in \cite{Lin:2019qwu}, with the basic idea being that they would implement the $AdS_2$ isometries on the matter fields while doing nothing to the metric and acting on the dilaton only to the extent necessary to continue solving the constraint equation.  More precisely there are various ways to satisfy the constraint equation, and what was constructed was a two-parameter family of $SL(2,\mathbb{R})$ algebras which are connected to each other by evolution by $H_+$ and $H_-$.  In this section we explain how these $SL(2,\mathbb{R})$ algebras fit into our gravitational algebra \eqref{gravalg}.

The first thing to notice is that if we define
\be\label{P0action}
P_0\equiv \wt{\eta}_+-\wt{\eta}_-,
\ee
from the brackets \eqref{twosidealg} we have
\be
\{\psi_{\wt\lambda},P_0\}=e^\mu\nabla_\mu\psi(y(\wt{\lambda})).
\ee
Thus $P_0(t_-,t_+)$ moves the dressed matter field $\psi_{\wt{\lambda}}(t_-,t_+)$ back and forth along its dressing geodesic $y^\mu_{t_-t_+}(\wt{\lambda})$.  Moreover $P_0$ vanishes by equation \eqref{etaeq} in the case where there are no matter fields, which makes it a natural candidate for an element of an $SL(2,\mathbb{R})$ algebra which is nontrivial only when there are nonvanishing matter fields.  Another two natural candidates, which vanish by equation \eqref{expH} when there are no matter fields, are
\be
\hat{P}_{\pm}\equiv H_{\pm}-\frac{\wt{\eta}_\pm^2}{4\phi_b}-\frac{1}{\phi_b}e^{-\wt{L}}.
\ee
Some calculation using the algebra \eqref{gravalg} shows that
\begin{align}\nonumber
\{P_0,\hat{P}_\pm\}&=\pm \hat{P}_{\pm}\\
\{\hat{P}_-,\hat{P}_+\}&=2 \frac{e^{-\wt{L}}}{4\phi_b^2}P_0.
\end{align}
This is almost an $SL(2,\mathbb{R})$ algebra, and noting that
\be
\left\{\wt{L},H_{\pm}-\frac{\wt{\eta}_\pm^2}{4\phi_b}\right\}=0,
\ee
we see that the rescaled observables
\be
P_{\pm}\equiv 2\phi_be^{\wt{L}/2}\hat{P}_{\pm}=2\phi_be^{\wt{L}/2}H_{\pm}-\frac{\wt{\eta}_\pm^2}{2}e^{\wt{L}/2}-2e^{-\wt{L}/2}
\ee
indeed obey the $SL(2,\mathbb{R})$ algebra
\begin{align}\nonumber
\{P_0,P_\pm\}&=\pm P_{\pm}\\
\{P_-,P_+\}&=2P_0.
\end{align}
$P_{\pm}$ also act simply on the two-sided dressed matter fields:
\be\label{Ppmaction}
\{\psi_{\wt{\lambda}},P_\pm\}=e^{\pm\wt{\lambda}}\tau^\mu\nabla_\mu \psi(y(\wt{\lambda})),
\ee
and indeed \eqref{P0action} and \eqref{Ppmaction} are precisely the actions on the matter fields required in \cite{Lin:2019qwu} (in their notation $P_0=\wt{P}$, $\frac{P_++P_-}{2}=\wt{E}$, $\frac{P_+-P_-}{2}=\wt{B}$, with $\wt{E}$ being a time translation at $\wt{\lambda}=0$ and $\wt{B}$ being a boost there).  Moreover since $P_0$ and $P_\pm$ depend on $t_+$ and $t_-$, what we really have is a two-parameter family of $SL(2,\mathbb{R})$ algebras which are carried into each other by conjugation by $H_{\pm}$, just as in \cite{Lin:2019qwu}.  The generators at different times can be explicitly related using the solution of the time-dependence presented in the following subsection, but we emphasize that the Hamiltonians $H_\pm$ are not in the $SL(2,\mathbb{R})$ algebra at any time.  In fact none of $\wt{L}$, $\wt{\eta}_\pm$, or $H_\pm$ is even in the algebraic union of all the $SL(2,\mathbb{R})$ algebras together, meaning the set of all observables generated by their mutual brackets, since any such observable would have to become trivial in the case of no matter fields and $\wt{L}$, $\wt{\eta}_\pm$, and $H_\pm$ are not trivial in that case as we saw in section \ref{trivsec}.  Thus our gravitational algebra \eqref{gravalg} is strictly larger than what can be obtained by combining all of the $SL(2,\mathbb{R})$ algebras constructed in \cite{Lin:2019qwu}.

\subsection{Solution of the time-dependence of the gravitational observables}
So far we have only computed brackets between observables which are evaluated at the same boundary times $t_\pm$.  In fact the time-dependence of the gravitational observables $\wt{L}$, $\wt{\eta}_\pm$, and $H_\pm$ can be solved in closed form, allowing us to drop this restriction and compute brackets involving observables at arbitrary times.  Indeed from the brackets of $\wt{L}$ and $\wt{\eta}_\pm$ with $H_\pm$ we have the differential equations
\begin{align}\nonumber
\partial_{\pm}\wt{L}&=\frac{\wt{\eta}_\pm}{2\phi_b}\\\nonumber
\partial_\pm \wt{\eta}_\pm&=H_{\pm}-\frac{\wt{\eta}_\pm^2}{4\phi_b}\\\nonumber
\partial_\pm \wt{\eta}_\mp&=\frac{1}{\phi_b}e^{-\wt{L}}\\\nonumber
\partial_{\pm}H_\pm&=0\\
\partial_\pm H_\mp&=0,
\end{align}
which can be combined to show that the quantity
\be
f\equiv e^{\wt{L}/2}
\ee
obeys the equations
\begin{align}\nonumber
\partial_\pm^2f&=\alpha_{\pm}^2 f\\
f\partial_+\partial_-f&=\partial_+f\partial_-f+\frac{1}{4\phi_b^2}\label{feq},
\end{align}
with
\be
\alpha_{\pm}\equiv \sqrt{\frac{H_{\pm}}{4\phi_b}}.
\ee
Given a solution for $f$ we can then determine $\wt{\eta}_{\pm}$ using
\be\label{etasol}
\wt{\eta}_\pm=4\phi_b\frac{\partial_{\pm}f}{f}.
\ee
The general solution of the first line of \eqref{feq} has the form
\begin{align}\nonumber
f(t_-,t_+)=&A\cosh(\alpha_+t_+)\cosh(\alpha_-t_-)+B\cosh(\alpha_+t_+)\sinh(\alpha_-t_-)\\
&+C\sinh(\alpha_+t_+)\cosh(\alpha_-t_-)+D\sinh(\alpha_+t_+)\sinh(\alpha_-t_-),
\end{align}
and the second line implies that
\be
AD-BC=\frac{1}{\phi_b\sqrt{H_+H_-}}.
\ee
The other three parameters are determined in terms of the initial values $\wt{L}_0$, $\wt{\eta}_{\pm0}$, giving
\begin{align}\nonumber
A&=e^{\frac{\wt{L}_0}{2}}\\\nonumber
B&=\frac{\wt{\eta}_{-0}}{\sqrt{4\phi_b H_-}}e^{\frac{\wt{L}_0}{2}}\\\nonumber
C&=\frac{\wt{\eta}_{+0}}{\sqrt{4\phi_b H_+}}e^{\frac{\wt{L}_0}{2}}\\
D&=\left(\frac{1}{\phi_b\sqrt{H_+H_-}}e^{-\wt{L}_0}+\frac{\wt{\eta}_{+0}\wt{\eta}_{-0}}{4\phi_b\sqrt{H_+H_-}}\right)e^{\frac{\wt{L}_0}{2}}.
\end{align}
Thus the full time-dependence of the renormalized length observable in any state is given by
\begin{align}\nonumber
\wt{L}(t_-,t_+)=\wt{L}_0+2\log\bigg[&\cosh(\alpha_+t_+)\cosh(\alpha_-t_-)+\frac{\wt{\eta}_{+0}}{\sqrt{4\phi_b H_+}}\sinh(\alpha_+t_+)\cosh(\alpha_-t_-)\\\nonumber
&+\frac{\wt{\eta}_{-0}}{\sqrt{4\phi_b H_-}}\cosh(\alpha_+t_+)\sinh(\alpha_-t_-)\\
&+\left(\frac{1}{\phi_b\sqrt{H_+H_-}}e^{-\wt{L}_0}+\frac{\wt{\eta}_{+0}\wt{\eta}_{-0}}{4\phi_b\sqrt{H_+H_-}}\right)\sinh(\alpha_+t_+)\sinh(\alpha_-t_-)\bigg].\label{Lsol}
\end{align}
Using this expression together with \eqref{etasol},  we can compute any of the brackets in \eqref{gravalg}, \eqref{twosidealg}, \eqref{onesidealg} with arbitrary times for both observables (we just use \eqref{Lsol} and \eqref{etasol} to rewrite each bracket in terms of equal time brackets).

\subsection{Fast Scrambling}
In \cite{Hayden:2007cs,Sekino:2008he} arguments were provided that black holes should scramble quantum information quite fast, at a timescale of order
\be
t_{scr}\sim \beta \log S,
\ee
where $\beta$ is the inverse temperature of the black hole and $S$ is its entropy.  In \cite{Shenker:2013pqa} holographic methods were used to refine this formula to
\be\label{scrambling}
t_{scr}=\frac{\beta}{2\pi}\log S,
\ee
and in \cite{Shenker:2014cwa,Maldacena:2015waa} this was shown to be an upper bound on the scrambling time of any roughly local system.  In this section we will show that the JT gravity version of the calculation of \cite{Shenker:2013pqa} can be done using only our algebra \eqref{gravalg}-\eqref{onesidealg} of diffeomorphism-invariant observables.

The main idea of \cite{Shenker:2013pqa} was to show that if we perturb the thermofield double state by some simple operator $O_-(-t_*)$ on the left boundary at time $t=-t_*$, with $t_*>0$, then the expectation value of the geodesic distance between the two boundaries at $t_-=t_+=0$ has a term that grows exponentially with $t_*$:
\be\label{disturbance}
\frac{\lan TFD|O_-^\dagger(-t_*)\wt{L}(0,0)O_-(-t_*)|TFD\ran}{\lan TFD|O_-^\dagger(-t_*)O_-(-t_*)|TFD\ran}=\lan TFD|\wt{L}(0,0)|TFD\ran+\frac{\#}{S}e^{\frac{2\pi }{\beta}t_*},
\ee
where $\#$ is some $O(1)$ constant and this equation holds until the second term becomes $O(1)$.  This is a signature of scrambling in the dual field theory because the correlation of simple operators between the two sides is controlled by this geodesic length, for example in the WKB approximation the left-right correlator of a scalar field $\phi$ of mass $m$ behaves like
\be
\lan \psi|\phi^\dagger_-(0)\phi_+(0)|\psi\ran\propto e^{-m\wt{L}}.
\ee
Increasing this length thus disrupts the correlation between the two sides, and \eqref{disturbance} shows that an early perturbation $O(-t_*)$ creates a disruption which starts out small but becomes dominant when
\be
t_*\approx\frac{\beta}{2\pi}\log S.
\ee
For near-extremal black holes such as those described by JT gravity, whose entropy does not vanish as we take the temperature to zero, these results are modified by replacing $S\to\Delta S$, where $\Delta S=S-S_0$ with $S_0$ being the zero-temperature entropy \cite{Leichenauer:2014nxa}.

To recast \eqref{disturbance} as an algebraic statement we can use the fact that the thermofield-double state which is dual to the wormhole solution \eqref{nomattsol} is annihilated by $H_+-H_-$, and thus we can rewrite the quantity on the left-hand side of \eqref{disturbance} as
\begin{align}\nonumber
\frac{\lan TFD|O_-^\dagger(0)\wt{L}(t_*,-t_*)O_-(0)|TFD\ran}{\lan TFD|O_-^\dagger(0)O_-(0)|TFD\ran}=&\frac{\lan TFD|O_-^\dagger(0)O_-(0)\wt{L}(t_*,-t_*)|TFD\ran}{\lan TFD|O_-^\dagger(0)O_-(0)|TFD\ran}\\
&+\frac{\lan TFD|O_-^\dagger(0)[\wt{L}(t_*,-t_*),O_-(0)]|TFD\ran}{\lan TFD|O_-^\dagger(0)O_-(0)|TFD\ran}.
\end{align}
The first term on the right-hand side is dominated by the short-distance behavior of $O_-^\dagger(0)O_-(0)$, and can be approximated as
\begin{align}\nonumber
\frac{\lan TFD|O_-^\dagger(0)O_-(0)\wt{L}(t_*,-t_*)|TFD\ran}{\lan TFD|O_-^\dagger(0)O_-(0)|TFD\ran}&\approx \frac{\lan TFD|O_-^\dagger(0)O_-(0)|TFD\ran\lan TFD|\wt{L}(t_*,-t_*)|TFD\ran}{\lan TFD|O_-^\dagger(0)O_-(0)|TFD\ran}\\
&=\lan TFD|\wt{L}(0,0)|TFD\ran,
\end{align}
which is independent of $t_*$ and accounts for the first term on the right-hand side of \eqref{disturbance}.  The exponentially growing term thus arises from the commutator
\be
[\wt{L}(t_*,-t_*),O_-(0)],
\ee
which we can study classically using our one-sided algebra \eqref{onesidealg} together with our solution \eqref{Lsol} for the time dependence of $\wt{L}$.  Indeed in the wormhole solution \eqref{nomattsol} we have
\begin{align}\nonumber
e^{-\wt{L}_0}&=\Phi_h^2\\\nonumber
\wt{\eta}_{\pm0}&=0\\
H_{\pm}&=\frac{\Phi_h^2}{\phi_b},
\end{align}
where $\Phi_h$ is the value of $\Phi$ at the horizon.  Using \eqref{Lsol} we can then compute the dependence of $\wt{L}(t_-,t_+)$ on the initial data evaluated at this solution:\footnote{In using these expressions to evaluate \eqref{scresult}, we are working in a classical approximation which ignores the fluctuations of $\wt{L}$, $\wt{\eta}_{\pm0}$, and $H_\pm$ about their expectation values.  This is a good approximation when the second term in \eqref{disturbance} is small, but once it becomes $O(1)$ then these fluctuations become important so we should not trust the exponential growth beyond this point.  In fact \eqref{Lsol} tells us that eventually the growth of $\wt{L}$ should be linear.}
\begin{align}\nonumber
\frac{\partial \wt{L}}{\partial \wt{L}_0}&=\frac{\cosh\left(\frac{\Phi_h}{2\phi_b}(t_+-t_-)\right)}{\cosh\left(\frac{\Phi_h}{2\phi_b}(t_++t_-)\right)}\\\nonumber
\frac{\partial \wt{L}}{\partial\wt{\eta}_{\pm 0}}&=\frac{1}{\Phi_h}\frac{\sinh\left(\frac{\Phi_h}{2\phi_b}t_{\pm}\right)\cosh\left(\frac{\Phi_h}{2\phi_b}t_{\mp}\right)}{\cosh\left(\frac{\Phi_h}{2\phi_b}(t_++t_-)\right)}\\\nonumber
\frac{\partial \wt{L}}{\partial H_-}&=\frac{1}{\cosh\left(\frac{\Phi_h}{2\phi_b}(t_++t_-)\right)}\Bigg[\frac{t_-}{2\Phi_h}\sinh\left(\frac{\Phi_h}{2\phi_b}(t_++t_-)\right)\\
&-\frac{\phi_b}{\Phi_h^2}\sinh\left(\frac{\Phi_h}{2\phi_b}t_+\right)\sinh\left(\frac{\Phi_h}{2\phi_b}t_-\right)\Bigg].
\end{align}
In particular if we take $t_-=-t_+=t_*$ we have
\begin{align}\nonumber
\frac{\partial \wt{L}}{\partial \wt{L}_0}&=\cosh\left(\frac{\Phi_h}{\phi_b}t_*\right)\approx \frac{1}{2}e^{\frac{2\pi}{\beta}t_*}\\\nonumber
\frac{\partial \wt{L}}{\partial\wt{\eta}_{\pm0}}&=\mp \frac{1}{\Phi_h}\sinh\left(\frac{\Phi_h}{2\phi_b}t_*\right)\cosh\left(\frac{\Phi_h}{2\phi_b}t_*\right)\approx \mp \frac{1}{4\Phi_h}e^{\frac{2\pi}{\beta}t_*}\\
\frac{\partial \wt{L}}{\partial H_-}&=\frac{\phi_b}{\Phi_h^2}\sinh\left(\frac{\Phi_h}{2\phi_b}t_*\right)\sinh\left(\frac{\Phi_h}{2\phi_b}t_*\right)\approx \frac{\phi_b}{4\Phi_h^2}e^{\frac{2\pi}{\beta}t_*},\label{growth}
\end{align}
where we have shown also the large $t_*$ approximations and used the relationship
\be\label{Phiheq}
\Phi_h=\frac{2\pi}{\beta}\phi_b
\ee
between the dilaton value at the horizon $\Phi_h$ and the inverse temperature $\beta$.  Note in particular the same exponential time-dependence as in \eqref{disturbance}.  To make use of our algebra we can take
\be
O_-(t_-)=\psi_{\wt{s},\wt{\eta}}(t_-),
\ee
so, for simplicity taking $\wt{\eta}=0$, by \eqref{growth} and \eqref{onesidealg} we have
\begin{align}\nonumber
\{\wt{L}(t_*,-t_*),\psi_{\wt{s},0}(0)\}&=\frac{\partial \wt{L}}{\partial \wt{L}_0}\{\wt{L}_0,\psi_{\wt{s},0}(0)\}+\frac{\partial \wt{L}}{\partial\wt{\eta}_{-0}}\{\wt{\eta}_{-0},\psi_{\wt{s},0}(0)\}+\frac{\partial \wt{L}}{\partial H_-}\{H_-,\psi_{\wt{s},0}(0)\}\\
&\approx \left(\frac{1}{4\Phi_h}e^\alpha\nabla_\alpha\psi-\frac{1}{8}e^{\wt{s}}\tau^\alpha\nabla_\alpha\psi+\frac{1}{8\Phi_h^2}e^{-\wt{s}}\tau^\alpha\nabla_\alpha\psi\right)e^{\frac{2\pi}{\beta}t_*}.\label{scresult}
\end{align}
If we locate the operator a few Schwarzschild radii outside of the horizon (so that the time and space derivatives just contribute factors of the temperature or AdS scale, which we think of as comparable), by \eqref{horizonst} we have $\wt{s}\sim -\log \Phi_h$ and thus
\be
\{\wt{L}(t_*,-t_*),\psi_{\wt{s},0}(0)\}\propto \frac{1}{\Phi_h}e^{\frac{2\pi}{\beta}t_*}\psi_{\wt{s},0}(0),
\ee
which can be promoted to a commutator to give
\be\label{scrambleresult}
\frac{\lan TFD |\psi_{\wt{s},0}^\dagger [L(t_*,-t_*),\psi_{\wt{s},0}]|TFD\ran}{\lan TFD|\psi_{\wt{s},0}^\dagger \psi_{\wt{s},0}]|TFD\ran}\propto \frac{1}{\Delta S}e^{\frac{2\pi}{\beta}t_*}.
\ee
Here
\be
\Delta S\equiv S-S_0= 4\pi \Phi_h
\ee
is the entropy beyond extremality in JT gravity (see e.g. \cite{Maldacena:2016upp} or \cite{Harlow:2018tqv}), so we indeed match \eqref{disturbance} with the near-extremal replacement $S\to\Delta S$.

\section{Energy to create an excitation}\label{energysec}
In quantum mechanics there are many situations where the amount that the action of an operator $O$ on a state $|\phi\ran$ changes the energy is controlled by commutator of $O$ with the Hamiltonian.  To see this, we first observe that
\be
\frac{\lan\phi|O^\dagger H O |\phi\ran}{\lan\phi|O^\dagger O|\phi\ran}=\frac{\lan\phi|O^\dagger [H,O] |\phi\ran}{\lan\phi|O^\dagger O|\phi\ran}+\frac{\lan\phi|O^\dagger OH|\phi\ran}{\lan\phi|O^\dagger O|\phi\ran}.\label{newenergy}
\ee
In any situation where we can approximate the second term on the right-hand side as
\be\label{appxH}
\frac{\lan\phi|O^\dagger OH|\phi\ran}{\lan\phi|O^\dagger O|\phi\ran}\approx \frac{\lan\phi|O^\dagger O|\phi\ran\lan\phi|H|\phi\ran}{\lan\phi|O^\dagger O|\phi\ran}=\lan\phi|H|\phi\ran,
\ee
the change in energy is given by the first term in \eqref{newenergy}.  One situation where \eqref{appxH} holds is when $|\phi\ran$ is well-localized in energy, e.g. if it is the ground state.  Another is if $O^\dagger O$ is close to a multiple of identity, as it will be either if $O$ is unitary or if $O$ is a smeared local operator (in the latter case the operator product is dominated by the leading term in the operator product expansion, which is usually given by the identity operator times the two-point function).  In general the sign and magnitude of the first term in \eqref{newenergy} is not clear, but if we can find an operator $O$ such that
\be
[H,O]\approx \Delta E O
\ee
for some $\Delta E\in \mathbb{R}$ then we have
\be
\frac{\lan\phi|O^\dagger [H,O] |\phi\ran}{\lan\phi|O^\dagger O|\phi\ran}\approx \Delta E.
\ee

For example let $\psi$ be a free massless scalar field in flat space.  For any test function $f$ we can define a smeared field
\be
\psi_f\equiv \int d^dxf(x)\psi(x).
\ee
In particular we can take
\be
f(\vec{x},t)=\frac{1}{(2\pi \delta^2)^{d/2}}e^{i\vec{k}\cdot \vec{x}-i|k|t}e^{-\frac{1}{2\delta^2}(|x|^2+t^2)},
\ee
with $|k|\delta\gg 1$, in which case
\be
\psi_f\sim \frac{1}{\delta^{d-1}|k|^{1/2}}a_{\vec{k}}^\dagger,
\ee
where $a_{\vec{k}}$ is the annihilation operator for the mode which is proportional to $e^{i\vec{k}\cdot \vec{x}-i|k|t}$.  We thus have
\be
[H,\psi_f]\approx |k|\psi_f,
\ee
so as expected smearing a field against a wave-packet of fairly definite momentum and frequency gives an operator which creates a quanta of that momentum and frequency in the vicinity of the spacetime support of the wave-packet.

For dressed matter operators in quantum gravity the situation is not so simple as for quantum fields in flat space.  In quantum gravity the total energy is given by a boundary term at spatial infinity, and the relationship between this ``asymptotic'' energy and the energy measured in some local inertial frame can be quite complicated due to gravitational red/blue shift effects.   Life is easy when expanding around a solution with an isometry $\xi^\mu$ that matches onto boundary time translations; in this case for any dressing locating a scalar field at $x$ we simply have
\be\label{KE}
[H,\psi(x)]=-i\xi^\mu\nabla_\mu \psi(x).
\ee
In the absence of such an isometry however, the commutator can depend sensitively on the choice of dressing and the details of the state.  We now study this phenomenon in more detail in JT gravity for an operator which creates a left or right moving matter excitation at some location determined using the one-sided geodesic $y^\mu_{t_-\wt{\eta}}(\wt{s})$, and also give a preliminary analysis of what happens in higher dimensions.

\subsection{General results and application to pure JT wormhole}\label{genEsec}
The basic idea is to smear our one-sided matter operator $\psi_{\wt{s},\wt{\eta}}(t_-)$ in Gaussian normal coordinates relative to $y^\mu_{t_-\wt{\eta}}(\wt{s})$ against a wave-packet which ensures that it creates an excitation of narrow width in momentum and frequency, and then compute its commutator with $H_-$ using \eqref{onesidealg}.  More concretely we can define
\be
\psi_{\wt{\eta},\wt{s}\tau}(t_-)\equiv \psi(\wt{s},\tau),
\ee
where $\wt{s},\tau$ are Gaussian normal coordinates defined relative to $y^\mu_{t_-\wt{\eta}}(\wt{s})$, and then we can define a smeared operator
\be
\psi_{k\wt{\eta}\wt{s}}(t_-)\equiv \int d\wt{s}' d\tau' f_{k\wt{s}}(\tau',\wt{s}')\psi_{\wt{\eta},\wt{s}'\tau'}(t_-),
\ee
with
\be
f_{k\wt{s}}(\tau',\wt{s}')\equiv\frac{1}{(2\pi \delta^2)}e^{ik(\wt{s}'-\wt{s})-i|k|\tau'}e^{-\frac{1}{2\delta^2}\left((\wt{s}'-\wt{s})^2+\tau^{\prime 2}\right)}
\ee
with $\frac{1}{|k|}\ll\delta\ll 1$.  In this regime we can compute the commutator of $\psi_{k\wt{\eta}\wt{s}}(t_-)$ with $H_-$ by simply replacing $e^\mu\nabla_\mu\psi(y(\wt{s}))\to -ik \psi_{k\wt{\eta}\wt{s}}(t_-)$ and $\tau^\mu \nabla_\mu\psi(\wt{s})\to i|k|\psi_{k\wt{\eta}\wt{s}}(t_-)$ in the fifth line of \eqref{onesidealg}:
\be
[H_-,\psi_{k\wt{\eta}\wt{s}}(t_-)]=i\{H_-,\psi_{k\wt{\eta}\wt{s}}(t_-)\}=\frac{|k|}{2\phi_b}\left[e^{-\wt{s}}+\left(\frac{1}{4}\wt{\eta}^2-\phi_bH_-\right)e^{\wt{s}}\pm \wt{\eta}\right]\psi_{k\wt{\eta}\wt{s}}(t_-),
\ee
where $\pm$ indicates the sign of $k$.  Thus $\psi_{k\wt{\eta}\wt{s}}(t_-)$ creates excitations which change $H_-$ by
\be\label{energyresult}
\Delta E_\pm\equiv\frac{|k|}{2\phi_b}\left[e^{-\wt{s}}+\left(\frac{1}{4}\wt{\eta}^2-\phi_bH_-\right)e^{\wt{s}}\pm \wt{\eta}\right].
\ee
We emphasize that $\pm$ indicates whether the particle is right or left moving, the commutator of $\psi_{k\wt{\eta}\wt{s}}(t_-)$ with $H_+$ vanishes. Thus we see that the relationship between the local energy $|k|$ seen by an observer in Gaussian normal coordinates and the change $\Delta E_\pm$ in the asymptotic energy measured by $H_-$ (or $H_-+H_+$) is indeed nontrivial.  In what follows we will refer to $\Delta E_\pm$ as the \textit{excitation energy}.

Perhaps the most interesting feature of \eqref{energyresult} is that $\Delta E_\pm$ has indefinite sign: sometimes creating a particle decreases the asymptotic energy!  Before analyzing this phenomenon in detail it is convenient to observe that \eqref{energyresult} is invariant under the operation of changing the sign of $\wt{\eta}$ and exchanging left/right movers, so without loss of generality we can assume that $\wt{\eta}\geq 0$.  This is also the situation which is relevant for addressing the firewall problem, as only one-sided geodesics with $\wt{\eta}\geq 0$ will cross the future event horizon.

We begin our study of the sign of $\Delta E_\pm$ by noting that it is always positive for sufficiently negative $\wt{s}$: this is reassuring, as particles which are created near the left boundary should have positive energy.  As we increase $\wt{s}$, what happens depends on whether we consider right or left movers.  The right-moving excitation energy $\Delta E_+$ changes sign only if
\be\label{etaless1}
\wt{\eta}<2\sqrt{\phi_b H_-},
\ee
in which case the change happens at
\be\label{Rflip}
\wt{s}=-\log\left(\sqrt{\phi_b H_-}-\frac{\wt{\eta}}{2}\right).
\ee
The left-moving excitation energy $\Delta E_-$ always changes sign, but if
\be\label{etaless2}
\wt{\eta}\leq 2\sqrt{\phi_b H_-}
\ee
this only happens once, at
\be\label{Lfirstflip}
\wt{s}=-\log\left(\sqrt{\phi_b H_-}+\frac{\wt{\eta}}{2}\right),
\ee
while if
\be\label{etagreater}
\wt{\eta}>2\sqrt{\phi_b H_-}
\ee
then the sign of $\Delta E_-$ again changes at \eqref{Lfirstflip} but then seems to flip back to positive when
\be\label{Lsecondflip}
\wt{s}=-\log \left(\frac{\wt{\eta}}{2}-\sqrt{\phi_b H_-}\right).
\ee
We expect however that this second flip never happens within the region that is dynamically determined from initial data on a Cauchy slice given the boundary conditions \eqref{BC}.

\bfig
\includegraphics[height=7cm]{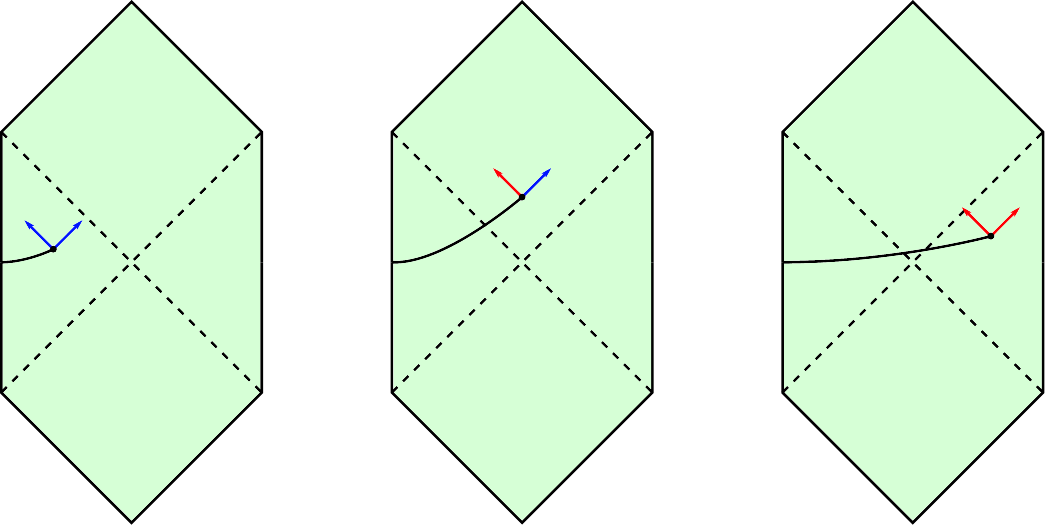}
\caption{The $H_-$ excitation energy of a particle with left-boundary gravitational dressing, meaning the change in $H_-$ if we create the particle, in the pure JT wormhole solution \eqref{nomattsol}.  Particles with positive $H_-$ are shaded blue, while particles with negative $H_-$ are shaded red.  All particles have $H_+=0$.  The change of the left-moving excitation energy from positive to negative as we cross the left future horizon is the essence of the firewall typicality arguments of \cite{Almheiri:2013hfa,Marolf:2013dba}.}\label{energiesfig}
\efig
We can get some intuition for these results by applying them to the special case of the pure JT wormhole solution \eqref{nomattsol} (similar results in this case were already obtained using the boundary-particle formalism in \cite{Almheiri:2018xdw}).  The relevant geodesics are shown in figure \ref{geodesicfig}.  Moreover for these solutions by \eqref{Hnomatt} we have
\be
2\sqrt{\phi_b H_-}=2\Phi_h,
\ee
so the conditions \eqref{etaless1}, \eqref{etaless2}, \eqref{etagreater} are precisely those which tell us whether or not the one-sided geodesic $y^\mu_{t_-\wt{\eta}}(s)$ makes it through the wormhole to the right exterior or stays behind the horizons.  Therefore by equations \eqref{rsPhi}, \eqref{alphaeta}, \eqref{horizonst}, and \eqref{horizonst2}, we see that \eqref{Lfirstflip} is precisely where the geodesic crosses the left future horizon and \eqref{Rflip} is precisely where the geodesic crosses the right future horizon if it does.  One can also easily check that \eqref{Lsecondflip} is indeed where the geodesic leaves the dynamically-determined region that is shaded green in figure \ref{geodesicfig}.  We summarize the situation in figure \ref{energiesfig}.

In fact to understand figure \ref{energiesfig}, the analysis we just went through is overkill: the pure JT wormhole \eqref{nomattsol} has a Killing symmetry which matches directly onto the evolution generated by $H_--H_+$, so we can determine the $H_--H_+$ energies of excitations using \eqref{KE}.  Moreover since our left-dressed operator $\psi_{k\wt{\eta}\wt{s}}$ commutes with $H_+$, we have
\be
[H_-,\psi_{k\wt{\eta}\wt{s}}]=[H_--H_+,\psi_{k\wt{\eta}\wt{s}}],
\ee
so the excitation energy defined using $H_-$ coincides with the excitation energy defined using $H_--H_+$.  The latter however is essentially obvious from figure \ref{energiesfig}: all blue-shaded particles propagate to or from the left boundary, while all red-shaded particles propagate to or from the right boundary, so via \eqref{KE} $H_--H_+$ (and thus $H_-$) should assign positive energy to the former and negative energy to the latter.  This argument however crucially relies on the symmetry of this state, so we now turn to studying \eqref{energyresult} in a class of geometries where there is no such symmetry.

\subsection{Multi-shockwave states of JT with conformal matter}
In higher dimensions the study of null shockwaves in gravity has a long history \cite{Aichelburg:1970dh,Dray:1984ha,Dray:1985yt,Hotta:1992qy,Sfetsos:1994xa,Cai:1999dz,Cornalba:2006xk}, and has recently been related to chaos \cite{Shenker:2013pqa,Shenker:2013yza,Shenker:2014cwa}, complexity \cite{Stanford:2014jda}, and traversable wormholes \cite{Gao:2016bin,Maldacena:2017axo,Hirano:2019ugo}.  In particular null shockwaves have been discussed in JT gravity in \cite{Maldacena:2016upp,Maldacena:2017axo,Kitaev:2017awl,Lin:2019qwu}.  In this subsection we give a more systematic treatment of the JT case, analogous to the treatment of $AdS_3$ in \cite{Shenker:2013pqa,Shenker:2013yza},\footnote{In fact this is more than an analogy, we will see in the following subsection that there is a simple mapping between our solutions and those constructed in \cite{Shenker:2013pqa,Shenker:2013yza}.} and then study the question of how much energy is created by the smeared observable $\psi_{k\wt{\eta}\wt{s}}$ we introduced in the previous subsection.  We have already seen from \eqref{energyresult} that the (possibly negative) energy it creates depends only on the left energy $H_-$ of the state, the dressing parameters $\wt{s}$, $\wt{\eta}$, and the smearing momentum $k$; it does \textit{not} depend on any details of how many shockwaves are present, how strong they are, or where they are located.  In this section we will show, perhaps surprisingly, that the value \eqref{Lfirstflip} of $\wt{s}$ at which the excitation energy of a left-moving particle vanishes always coincides with a point on the left future horizon, even when shockwaves are present.  For larger $\wt{s}$ the particle is in the wormhole interior and the excitation energy is negative, just as for the pure JT wormhole.  $\psi_{k\wt{\eta}\wt{s}}$ is therefore a diffeomorphism-invariant creation operator with just the energetic properties claimed in \cite{Almheiri:2013hfa,Marolf:2013dba}, including in situations where there is no boost isometry.

In JT gravity with matter, by the equations of motion \eqref{JTEOM} the metric is always just a piece of $AdS_2$.  In discussing null shockwaves it is convenient to adopt the Kruskal coordinates $X^\pm$,\footnote{$X^+$ is sometimes called $V$ and $X^-$ is sometimes called $U$.} in terms of which the metric is
\be\label{kruskg}
ds^2=-\frac{4dX^+dX^-}{(1+X^+X^-)^2}.
\ee
In these coordinates the $AdS_2$ boundaries are at $X^+ X^-=-1$, with $X^+>0$ and $X_-<0$ near the right boundary and $X^+<0$ and $X^->0$ near the left boundary.  In the left exterior they are related to the Schwarzschild coordinates \eqref{schwarzmetric} via
\begin{align}\nonumber
-X^+X^-&=\frac{r-r_s}{r+r_s}\\
\frac{-X^+}{X^-}&=e^{-2r_s\hat{t}},\label{kruskalsch}
\end{align}
with $X^+<0$ and $X^->0$, but since we will now consider situations with shockwaves the dilaton will \textit{not} just be $\Phi=\phi_b r$.  Instead we need to specify the locations and strengths of the shockwaves and then solve the equations of motion to determine the dilaton (and thus also the locations of the two regulated boundaries where $\Phi=\phi_b/\epsilon$).

\bfig
\includegraphics[height=8cm]{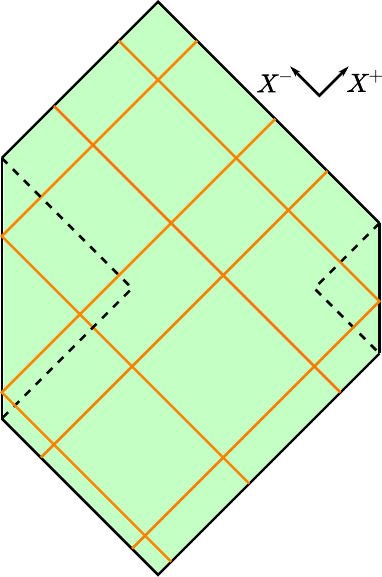}
\caption{A multi-shockwave configuration in JT gravity coupled to conformal matter.  The shockwaves are solid orange lines, while the future/past horizons of the left and right boundaries are dashed black lines.  Some shockwaves reflect off of the boundaries, while others live entirely in the wormhole interior.}\label{shocksfig}
\efig
We will take the energy-momentum tensor to be
\begin{align}\nonumber
T_{++}&=\sum_{i=1}^{N_L} 2k_{L,i}\delta(X^+-X^+_i)\\\nonumber
T_{--}&=\sum_{i=1}^{N_R} 2k_{R,i}\delta(X^--X^-_i)\\
T_{+-}&=0,\label{conformalT}
\end{align}
where $N_L$ is the number of left-moving shells, $N_R$ is the number of right-moving shells, $k_{L,i}>0$ is the strength of the $i$th left-moving shell, and $k_{R,i}>0$ is the strength of the $i$th right-moving shell.\footnote{For shells with $X^\pm_i=0$, $k_{L,i}$ and $k_{R,i}$ are the shell energies as seen by an inertial observer at rest in Kruskal coordinates at $X^+-X^-=0$ (such an observer would use coordinates $y^\pm=2 X^\pm$).}  In one of the unfortunate particularities of null coordinates, left-moving shells contributed to $T_{++}$ and right-moving shells contribute to $T_{--}$.  Quantum mechanically $T_{+-}$ is nonzero due to the Weyl anomaly, but this results only in a constant state-independent shift of $\Phi$ so we will ignore it.  One can easily check that  the energy-momentum tensor \eqref{conformalT} is conserved in the background \eqref{kruskg}.  Any classical state of conformal matter coupled to JT gravity has a stress tensor that can be well-approximated by something of the form \eqref{conformalT}.  We then need to solve the equations of motion
\begin{align}\nonumber
\partial_{\pm}^2\left((1+X^+X^-)\Phi\right)+\frac{1}{2}(1+X^+X^-)T_{\pm\pm}&=0\\
-\partial_+\partial_-\Phi-\frac{2}{(1+X^+X^-)^2}\Phi+\frac{1}{2}T_{+-}&=0\label{nulleom}
\end{align}
to determine the dilaton and the boundary locations.  Away from the shockwaves, the general solution of these equations is
\be\label{dilsol}
\Phi=\frac{a_+X^++a_-X^-+b(1-X^+ X^-)}{1+X^+X^-},
\ee
where $a_\pm$ and $b$ are arbitrary real parameters.  In particular for the solution \eqref{nomattsol} with no shockwaves we have $a_\pm=0$ and $b=\Phi_h$.  More generally the spacetime will be tiled by regions in which $\Phi$ takes this form, bounded by null shockwaves across which $a_\pm$ and $b$ will jump discontinuously.  An example of such a configuration is shown in figure \ref{shocksfig}.

Within each region where \eqref{dilsol} is obeyed, we have the convenient relation
\be\label{dilconstant}
\Phi^2-g^{\mu\nu}\nabla_\mu \Phi\nabla_\nu\Phi=a_+a_-+b^2.
\ee
In particular for any region which intersects either $AdS_2$ boundary we have
\be
(\Phi+n^\mu\nabla_\mu\Phi)(\Phi-n^\nu \nabla_\nu\Phi)|_\Gamma=a_+a_-+b^2,
\ee
where $\Gamma$ is the AdS boundary and we have used \eqref{uphi}.  The boundary conditions \eqref{BC} require that $\Phi|_\Gamma>0$, and finiteness of the energy \eqref{Hpm} requires that
\be\label{nfinite}
n^\mu\nabla_\mu \Phi=\frac{\phi_b}{\epsilon}+O(\epsilon),
\ee
which in particular is positive.  Finiteness and positivity of the Hamiltonians $H_\pm$ thus imply that
\be
a_+a_-+b^2\geq 0
\ee
in any region which intersects either AdS boundary.\footnote{Solutions with negative energy do exist, but they correspond to a ``linear dilaton vacuum'' which isn't accessible by starting with a wormhole solution and throwing in shock waves.}  In fact from \eqref{dilconstant} and \eqref{Hpm} the Hamiltonian at whichever boundary the region intersects is just given by
\be
H=\frac{a_+a_-+b^2}{\phi_b}.
\ee

The discontinuities in $a_\pm$ and $b$ across each shockwave are determined by integrating \eqref{nulleom} across it.  In the presence of a right-moving shockwave with
\be
T_{--}(X^-)=2k_R \delta(X^--X^-_0),
\ee
integrating the first line of \eqref{nulleom} leads to the discontinuities
\begin{align}\nonumber
\Delta (1+X^+X^-)\Phi&=0\\
\Delta\partial_-((1+X^+X^-)\Phi)&=-k_R(1+X^+X_0^-).\label{shockjumps}
\end{align}
Away from any shockwaves \eqref{dilsol} implies that
\be
\partial_\pm\left((1+X^+X^-)\Phi\right)=a_\pm-bX^{\mp},
\ee
so differentiating the first line of \eqref{shockjumps} with respect to $X^+$ and using the second line we have
\begin{align}\nonumber
\Delta a_+-X_0^-\Delta b &=0\\
\Delta a_--X^+\Delta b &=-k_R(1+X^+X_0^-)
\end{align}
for any $X^+$ along the shockwave (here $\Delta a_\pm$ and $\Delta b$ are the discontinuities across the shell from past to future).  We can solve these to find
\begin{align}\nonumber
\Delta a_+&=k_R(X_0^-)^2\\\nonumber
\Delta a_-&=-k_R\\
\Delta b&=k_R X_0^-.\label{rightjumps}
\end{align}
Similarly in the vicinity of a left-moving shockwave with
\be
T_{++}(X^+)=2k_L\delta(X^+-X_0^+),
\ee
integrating \eqref{nulleom} across this shockwave gives
\begin{align}\nonumber
\Delta a_--X^+_0 \Delta b&=0\\
\Delta a_+-X^-\Delta b&=-k_L(1+X_0^+ X^-)
\end{align}
and therefore
\begin{align}\nonumber
\Delta a_+&=-k_L\\\nonumber
\Delta a_-&=k_L(X_0^+)^2\\
\Delta b&=k_L X_0^+.\label{leftjumps}
\end{align}
These discontinuities are again the jumps from past to future across the shock wave.

\bfig
\includegraphics[height=5cm]{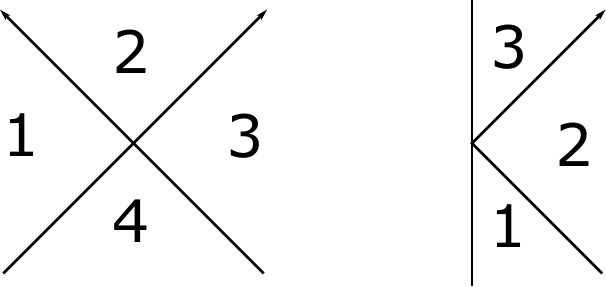}
\caption{Labeling the regions in the vicinity of a collision of two shock waves and the reflection of a shock wave off of the left $AdS$ boundary.}\label{collfig}
\efig
The expressions \eqref{rightjumps} and \eqref{leftjumps} are analogous to the recursion relations studied in \cite{Shenker:2013yza} for $AdS_3$, and they give us everything we need to determine the full dilaton solution for any energy-momentum tensor of the form \eqref{conformalT}.  In particular they imply that the dilaton itself is continuous across all of the shockwaves, it is only its gradient which jumps.  In the vicinity of a collision of two shock waves they also imply a JT version of the ``DTR relations'' of \cite{Dray:1985yt,redmount1985blue,Poisson:1990eh}: if we label the regions around the collision as in the left diagram of figure \ref{collfig}, then we have
\be
\left(\Phi_c^2-(a_+a_-+b^2)|_1\right)\left(\Phi_c^2-(a_+a_-+b^2)|_3\right)=\left(\Phi_c^2-(a_+a_-+b^2)|_2\right)\left(\Phi_c^2-(a_+a_-+b^2)|_4\right),
\ee
where $\Phi_c$ is the value of the dilaton at the collision.  We can also study what happens when a shock wave reflects off of the boundary in more detail.  Taking the incoming shock wave to have strength $k_L$ and location $X^+_0$ and the outgoing shock wave to have strength $k_R$ and location $X^-_0=-\frac{1}{X^+_0}$, and labeling the various regions as in the right diagram of figure \ref{collfig}, from equations \eqref{rightjumps} and \eqref{leftjumps} we have
\begin{align}\nonumber
a_+|_3&=a_+|_1-\frac{(X_0^+)^2k_L-k_R}{(X_0^+)^2}\\\nonumber
a_-|_3&=a_-|_1+(X_0^+)^2k_L-k_R\\
b|_3&=b|_1+\frac{(X_0^+)^2k_L-k_R}{X_0^+}.
\end{align}
The reflection must conserve energy, so we need to have
\be
\left(a_+a_-+b^2\right)|_3=\left(a_+a_-+b^2\right)|_1.
\ee
For a shockwave which reaches the boundary at finite boundary time this is only possible if
\be
k_R=(X_0^+)^2k_L,
\ee
so we see that the dilaton solutions in regions $1$ and $3$ have exactly the same values of $a_\pm$ and $b$.

\bfig
\includegraphics[height=6cm]{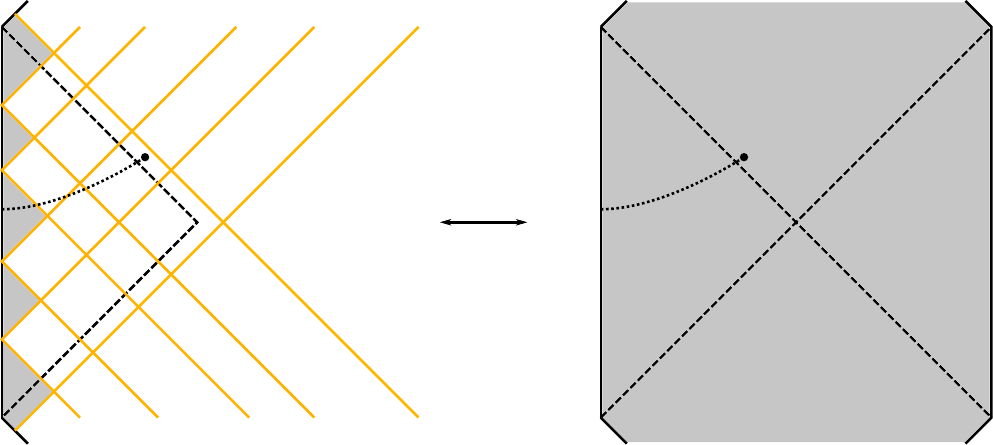}
\caption{For any shock wave solution, the dilaton in the shaded grey regions in the left diagram is identical to the dilaton in the vacuum solution shown on the right.  This is sufficient to determine the location of the left future/past horizons in Kruskal coordinates, shown as dashed lines, so these are the same in both spacetimes.  Moreover the dressing geodesic for the observable $\psi_{k\wt{\eta}\wt{s}}$ depends only on the regularized boundary location in the grey regions and the metric, which are the same in both solutions, so it will end at the same point in Kruskal coordinates in both.  We show such a dressing geodesic with the dotted line.}\label{replacefig}
\efig
We are now in a position to study the energy created by the dressed operator $\psi_{k\wt{\eta}\wt{s}}$ in a generic multi-shockwave state of left energy $H_-$.  The key point is that for any such configuration, the dilaton solution in all regions which are adjacent to the left boundary is the same as it would be if we were to remove all of the shock waves.  This is because, as we have just seen, the parameters $a_\pm$ and $b$ of the dilaton solution near the boundary do not change across any point where a shock wave reflects off of the boundary.  This implies that the trajectory of the left boundary is the same as it would be in the vacuum solution obtained by taking the values of $a_\pm$ and $b$ near any point on the left boundary and extending them throughout the spacetime (see figure \ref{replacefig}).  Moreover since the metric is also the same throughout (it is always \eqref{kruskg}), the location of the left future horizon in Kruskal coordinates is the same as for this vacuum solution and, for fixed $\wt{s}$ and $\wt{\eta}$, a left-boundary dressing geodesic will end at the same point in Kruskal coordinates in both solutions.  Therefore the energy \eqref{energyresult} created by $\psi_{k\wt{\eta}\wt{s}}$ with $\wt{\eta}>0$ and $k<0$ will change sign when the dressing geodesic crosses the left future horizon for \textit{any} multi-shockwave solution, just as it did in the pure JT wormhole.

\subsection{Higher-dimensional black holes}\label{higherdsec}
The results so far in this section have made use of many special features of JT gravity.  It is natural to wonder to what extent they hold in higher-dimensional gravity.  A complete study of this question is beyond the scope of this paper, but in this section we will present some preliminary results.  We will restrict to spherically-symmetric solutions, both for technical convenience and to facilitate the comparison to JT; to the extent that spherical symmetry is typical this is not an unreasonable restriction.\footnote{Spherical symmetry is certainly \textit{not} typical in the classical gravity sense: in the space of classical solutions there is no open set whose elements are all spherically-symmetric.  What matters for the firewall paradox however is typicality among pure quantum states in an energy band, and in AdS/CFT typical states in this sense are certainly expected to have spherical symmetry in the dual CFT.}  The conclusion so far is that in higher dimensions the excitation energy created by $\psi_{k\wt{\eta}\wt{s}}$ indeed depends on more details of the state than just its energy, but at least in three dimensions it still has the energetic properties claimed in \cite{Almheiri:2013hfa,Marolf:2013dba} (up to small corrections) in a wide variety of states.  In particular we will show that there is a precise mathematical map between the JT muti-shockwave states we constructed in the previous subsection and the $AdS_3$ muti-shockwave states studied in \cite{Shenker:2013pqa,Shenker:2013yza}, and that the excitation energies of radial quanta in these states are exactly the same as in JT gravity.  We will also consider an example of an asymptotically-$AdS_3$ state with more a general matter distribution that includes a non-vanishing transverse energy-momentum tensor, in which case there are now corrections to the JT results, but we will see that these corrections are small.

We begin by considering a rather general family of warped metrics in $D$ spacetime dimensions:
\be\label{warpedmet}
ds^2=\wt{g}_{\mu\nu}(x)dx^\mu dx^\nu+f(x)^2 \hat{g}_{ab}(y)dy^ady^b.
\ee
Here $\mu,\nu,\ldots$ run over $d$ spacetime coordinates and $a,b,\ldots$ run over $D-d$ spacetime coordinates.  In units where the AdS radius is set to one, the Einstein-Hilbert Lagrangian coupled to matter is
\be
\mathcal{L}=\frac{1}{16\pi G}\big(R+(D-1)(D-2)\big)+\mathcal{L}_{matter},
\ee
and we will assume the matter energy-momentum tensor has the form
\be\label{Tform}
T=\wt{T}_{\mu\nu}(x)dx^\mu dx^\nu+\hat{T}(x)\hat{g}_{ab}(y)dy^ady^b.
\ee
The nonzero Christoffel symbols for \eqref{warpedmet} are
\begin{align}\nonumber
\Gamma^\mu_{\alpha\beta}&=\wt{\Gamma}^\mu_{\alpha\beta}\\\nonumber
\Gamma^\mu_{ab}&=-f\wt{\nabla}^\mu f \hat{g}_{ab}\\\nonumber
\Gamma^a_{\mu b}=\Gamma^a_{b\mu}&=\frac{\wt{\nabla}_\mu f}{f}\delta^a_b\\
\Gamma^a_{bc}&=\hat{\Gamma}^a_{bc},
\end{align}
and the non-vanishing components of the Riemann tensor are
\begin{align}\nonumber
R_{\alpha\beta \mu\nu}&=\wt{R}_{\alpha\beta\mu\nu}\\\nonumber
R_{\alpha a \beta b}&=-f\hat{g}_{ab}\wt{\nabla}_\alpha\wt{\nabla}_\beta f\\
R_{abcd}&=f^2\left[\hat{R}_{abcd}+\wt{\nabla}_\mu f\wt{\nabla}^\mu f\left(\hat{g}_{ad}\hat{g}_{bc}-\hat{g}_{ac}\hat{g}_{bd}\right)\right]
\end{align}
and those related to these by the symmetries of the Riemann tensor.  The Ricci tensor is
\begin{align}\nonumber
R_{\mu\nu}&=\wt{R}_{\mu\nu}-(D-d)\frac{1}{f}\wt{\nabla}_\mu\wt{\nabla}_\nu f\\\nonumber
R_{\mu a}&=0\\
R_{ab}&=\hat{R}_{ab}-\hat{g}_{ab}\left(f\wt{\nabla}^2f+(D-d-1)\wt{\nabla}_\mu f\wt{\nabla}^\mu f\right),
\end{align}
and the Ricci scalar is
\be
R=\wt{R}+\frac{\hat{R}}{f^2}-\frac{2(D-d)}{f}\wt{\nabla}^2 f-\frac{(D-d)(D-d-1)}{f^2}\wt{\nabla}_\mu f \wt{\nabla}^\mu f.
\ee
The metric equations of motion then tell us that
\begin{align}\nonumber
\wt{R}_{\mu\nu}-(D-d)\frac{1}{f}\wt{\nabla}_\mu\wt{\nabla}_\nu f-&\frac{1}{2}\wt{g}_{\mu\nu}\Bigg(\wt{R}+\frac{\hat{R}}{f^2}-\frac{2(D-d)}{f}\wt{\nabla}^2 f\\
&-\frac{(D-d)(D-d-1)}{f^2}\wt{\nabla}_\alpha f \wt{\nabla}^\alpha f+(D-2)(D-1)\Bigg)=8\pi G \wt{T}_{\mu\nu}
\end{align}
and
\begin{align}\nonumber
\hat{R}_{ab}-\frac{1}{2}\hat{g}_{ab}\Bigg(&f^2 \wt{R}+\hat{R}-2(D-d-1)f\wt{\nabla}^2 f\\
&-(D-d-2)(D-d-1)\wt{\nabla}_\alpha f \wt{\nabla}^\alpha f+f^2(D-2)(D-1)\Bigg)=8\pi G \hat{T}\hat{g}_{ab}.
\end{align}
Specializing now to $d=2$ and taking $\hat{g}_{ab}$ to be the round metric on $\mathbb{S}^{D-2}$, we have
\begin{align}\nonumber
\wt{\nabla}_\mu \wt{\nabla}_\nu f+\left(\frac{D-1}{2}f-\wt{\nabla}^2 f+\frac{D-3}{2f}\left(1-\wt{\nabla}_\alpha f\wt{\nabla}^\alpha f\right)\right)\wt{g}_{\mu\nu}+\frac{8\pi G }{D-2}f\wt{T}_{\mu\nu}&=0\\
\wt{R}+(D-1)(D-2)+\frac{(D-3)}{f^2}\left((D-4)(1-\wt{\nabla}_\alpha f\wt{\nabla}^\alpha f)-2f\wt{\nabla}^2 f\right)+\frac{16\pi G}{f^2}\hat{T}&=0,\label{dimredeom}
\end{align}
which we can view as the equations of motion for a $1+1$ dimensional dilaton gravity theory with metric $\wt{g}_{\mu\nu}$ and dilaton $f$.  In particular when $D=3$ and $\hat{T}=0$ these equations become
\begin{align}\nonumber
\wt{\nabla}_\mu \wt{\nabla}_\nu f+\left(f-\wt{\nabla}^2 f\right)\wt{g}_{\mu\nu}+8\pi G f \wt{T}_{\mu\nu}&=0\\
\wt{R}+2&=0,
\end{align}
so comparing to \eqref{JTEOM} we see that after the replacements
\begin{align}\nonumber
f \to 16\pi G \Phi\\
f\wt{T}_{\mu\nu}\to T_{\mu\nu}
\end{align}
the spherical reduction of $2+1$ dimensional Einstein gravity with no transverse matter energy-momentum is precisely JT gravity coupled to matter with equations of motion \eqref{JTEOM}!  This means that any solution of JT gravity coupled to matter can be lifted to a spherically-symmetric solution of $2+1$ dimensional gravity with $\hat{T}=0$, and in fact the lifts of the null shockwave solutions we constructed in the previous subsection are precisely the solutions constructed in \cite{Shenker:2013pqa,Shenker:2013yza}.  The presentation obtained this way is perhaps more convenient than that in \cite{Shenker:2013pqa,Shenker:2013yza}, as the two-dimensional metric $\wt{g}_{\mu\nu}$ is given everywhere by \eqref{kruskg} for all shell configurations, while it is only the radial metric coefficient $f$ which is not smooth at the shockwave locations.\footnote{This did not happen in \cite{Shenker:2013pqa,Shenker:2013yza} because they constrained the form of $f$ to require $a_\pm=0$ away from the shocks (recall $a_\pm$ are defined in \eqref{dilsol}), which requires the introduction of singularities in $\wt{g}_{\mu\nu}$.}

We can now consider the question of the excitation energy created by a dressed matter operator in the multi-shockwave geometries of \cite{Shenker:2013pqa,Shenker:2013yza}.  Radial geodesics in these geometries are precisely the same as those in the two-dimensional metric $\wt{g}_{\mu\nu}$, and moreover the boundary locations are determined by surfaces of constant $f$ just as in JT gravity.  Since these are all the ingredients we need to determine the excitation energy created by a matter operator dressed by a radial geodesic using the method of subsections \ref{Hsec} and \ref{genEsec}, we thus see that the excitation energy will be exactly the same as what we found in the previous subsection for multi-shockwave solutions in JT gravity.  In particular we can still construct a diffeomorphism-invariant operator $\psi_{k\wt{\eta}\wt{s}}$ which creates negative energy just behind the horizon in any of these shockwave states, just as needed for the arguments of \cite{Almheiri:2013hfa,Marolf:2013dba}.

More generally, for $D>3$ and/or $\hat{T}\neq 0$ the equations \eqref{dimredeom} imply nontrivial backreaction by the dilaton and matter fields on the two-dimensional metric $\wt{g}_{\mu\nu}$.  The excitation energy created by a diffeomorphism-invariant operator such as $\psi_{k\wt{\eta}\wt{s}}$ then indeed depends on details of the state and changes sign at some location other than the horizon.  We will not attempt a general study of this phenomenon, but we will do one calculation which we expect is representative of the general situation.  The situation we will consider is $D=3$ but with $\hat{T}\neq 0$, in which case the equations of motion are
\begin{align}\nonumber
\wt{\nabla}_\mu \wt{\nabla}_\nu f+\left(f-\wt{\nabla}^2 f\right)\wt{g}_{\mu\nu}+8\pi G f \wt{T}_{\mu\nu}&=0\\
\wt{R}+2+\frac{16\pi G}{f^2}\hat{T}&=0.\label{hatTeom}
\end{align}
Spherically-symmetric matter with $\hat{T}\neq 0$ arises naturally from coherent waves of massless fields. For example the energy-momentum tensor of a conformally-coupled massless scalar field is
\begin{align}\nonumber
T_{mn}=&\frac{D}{2(D-1)}\nabla_m \psi\nabla_n \psi-\frac{1}{2(D-1)}g_{mn}\nabla_l\psi\nabla^l\psi+\frac{D-2}{2(D-1)}\left(g_{mn} \psi\nabla^2\psi-\psi\nabla_m\nabla_n\psi\right)\\
&+\frac{D-2}{4(D-1)}\left(R_{mn}-\frac{1}{2}g_{mn}R\right)\psi^2,\label{confT}
\end{align}
where $m,n,\ldots$ run over all $D$ spacetime coordinates, and this has $T_{ab}\neq 0$ even if $\psi$ depends only on the $x^\mu$ coordinates.  The spherical shockwaves of \cite{Shenker:2013pqa,Shenker:2013yza} represent incoherent distributions of massless ``dust'' particles, each of which is localized on the sphere, and thus have $\hat{T}=0$.  In black hole physics however the configurations relevant for Hawking radiation are coherent waves of low angular momentum, and these have $\hat{T}\neq 0$.

\bfig
\includegraphics[height=5cm]{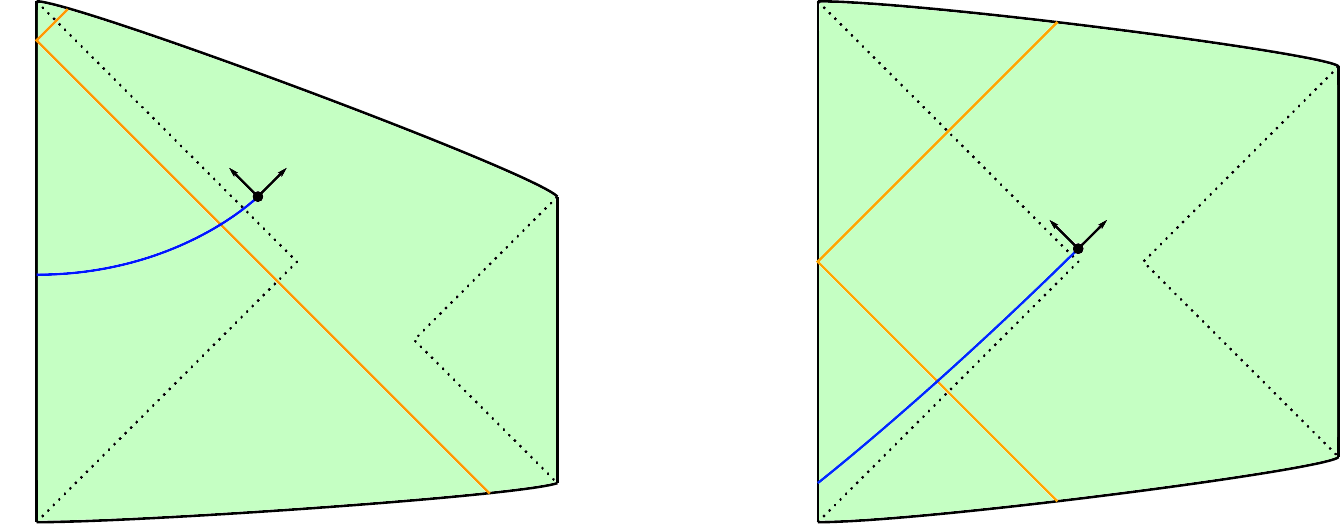}
\caption{A spherical null shell in the BTZ geometry.  In the left diagram we adopt a frame where the shell (shown in orange) reflects off of the left boundary at some late time, while on the right diagram we boost to a frame where it reflects near $t_-=0$.  We are interested in the excitation energy of a massless particle produced at the end of the blue dressing geodesic, in particular when the shell has nonzero $\hat{T}$ as in equation \eqref{shockansatz}.}\label{singleshellbackfig}
\efig
The situation we will consider is a single outgoing shockwave on the left side of the BTZ geometry
\be\label{btz}
ds^2=-\frac{4dX^+dX^-}{(1+X^+X^-)^2}+r_s^2\left(\frac{1-X^+X^-}{1+X^+X^-}\right)^2 d\phi^2,
\ee
with the shockwave energy-momentum tensor having the form \eqref{Tform} with
\begin{align}\nonumber
\wt{T}_{++}&=A(X^-)\delta(X^+-X_0^+)+B(X^-)\delta'(X^+-X_0^+)\\\nonumber
\wt{T}_{--}&=0\\\nonumber
\wt{T}_{-+}&=C(X^-)\delta(X^+-X_0^+)\\
\hat{T}&=D(X^-)\delta(X^+-X_0^+)\label{shockansatz}
\end{align}
and $X^+_0<0$.  This shell will eventually reflect off of the left boundary and fall back into the wormhole, but we will not need to model that explicitly.  We illustrate the situation in figure \ref{singleshellbackfig}.  We will work only to first order in the backreaction by this stress tensor, so we can require it to be conserved in the unperturbed geometry \eqref{btz}.    Conservation and tracelessness then tell us that we must have
\begin{align}\nonumber
A(X^-)&=A_0\frac{1+X_0^+X^-}{1-X_0^+ X^-}-2B_0\frac{X^-}{(1-X^+_0X^-)^2}-C_0\frac{1-X^+_0X^-(3+2X^+_0X^-)}{(X_0^+)^2(1-X^+_0X^-)^3}\\\nonumber
B(X^-)&=B_0\frac{1+X_0^+X^-}{1-X_0^+X^-}-C_0\frac{1+X_0^+X^-}{X_0^+(1-X_0^+X^-)^2}\\\nonumber
C(X^-)&=C_0\frac{1+X_0^+X^-}{(1-X_0^+X^-)^3}\\
D(X^-)&=C_0r_s^2\frac{1+X_0^+X^-}{1-X_0^+X^-},
\end{align}
where $A_0$, $B_0$, and $C_0$ are parameters.  $A_0$ parametrizes a spherical massless dust contribution, while $B_0$ represents something like two shells in rapid succession with opposite energy (this contribution vanishes when we feed a narrow symmetric wave packet into \eqref{confT}).  We are most interested in $C_0$, which gives a nonzero contribution to $\hat{T}$ and thus leads to nontrivial backreaction on $\wt{g}_{\mu\nu}$, and we will choose $C_0$ such that the energy of this shell near the boundary is of order the thermal scale $1/r_s$.  Naively this backreaction should only lead to corrections to the excitation energy $\Delta E_\pm$ which are of order $G/r_s$, but if our dressing geodesic is fired in at an early boundary time $-t_- r_s>>1$ then there is a large relative boost between the rest frame of the geodesic and the local frame in which the energy of the shell is thermal-scale so there is the possibility of an exponential enhancement.  Such an enhancement could potentially give a correction to $\Delta E_{\pm}$ which competes with the leading order result once $|t_-|$ is of order the scrambling time \eqref{scrambling}.  In the remainder of this section we will show that there is in fact no such enhancement, although at various intermediate steps it will appear as if there might be.

We can determine the backreaction by using the second line of \eqref{hatTeom}. Adopting conformal gauge we have
\be
\wt{g}_{\mu\nu}dx^\mu dx^\nu=-e^{2\omega}dX^+dX^-
\ee
and
\be
\wt{R}=8e^{-2\omega}\partial_+\partial_-\omega,
\ee
so we can rewrite the second line of \eqref{hatTeom} as
\be\label{omegaeq}
\partial_+\partial_-\omega=-e^{2\omega}\left(\frac{1}{4}+\frac{2\pi G}{f^2}\hat{T}\right)
\ee
with
\be
f=r_s\frac{1-X^+X^-}{1+X^+X^-}.
\ee
Away from $X^+=X^+_0$ the conformal factor $\omega$ obeys the Liouville equation
\be
\partial_+\partial_-\omega+\frac{1}{4}e^{2\omega}=0,
\ee
while the $\delta$-function in $\hat{T}$ leads to a singularity in $\omega$ at $X^+_0$.  Liouville solutions can always be written as lightlike coordinate transformations $h^\pm(X^\pm)$ of the Kruskal metric (see e.g. \cite{Seiberg:1990eb} or section 3.1 of \cite{Harlow:2011ny}),
\be\label{liouvillesol}
e^{2\omega}=\frac{4\partial_+h^+(X^+)\partial_-h^-(X^-)}{(1+h^+(X^+)h^-(X^-))^2},
\ee
so our goal is to determine $h^+$ and $h^-$ on both sides of the shockwave.  We will study the case where there is no perturbation to the left of the shell, in which case $h^\pm=X^\pm$ for $X^+<X^+_0$, so the question is what happens to the right of the shell ($X^+>X^+_0$).  To keep the shell at the same coordinate location on both sides we will take $h^+(X^+_0)=X^+_0$, and for $X^+>X^+_0$ we can simply choose $h^+(X^+)=X^+$ since this just implements an arbitrary diffeomorphism away from the shell.  We will not need to discuss the backreaction on the ``dilaton'' $f$, but we also take this to be unchanged for $X^+<X^+_0$.  The idea is that we want to construct a solution which has the same $H_-$ as the unperturbed solution, so that it can be viewed as another, possibly more typical, state in the same microcanonical ensemble.  With this prescription the boundary location prior to the boundary reflection of the shockwave will be the same as in the unperturbed solution, so dressing geodesics fired from below this reflection, such as in figure \ref{singleshellbackfig}, will be unmodified until they reach the shell.  For $h^-$ we adopt the following ansatz:
\be
h^-(X^-)=\begin{cases} X^- & X^+<X^+_0\\ X^-+\lambda(X^-) & X^+>X^+_0\end{cases},
\ee
where $\lambda(X^-)$ is a smooth function which we will treat as a small quantity of $O(G)$.  We may then integrate \eqref{omegaeq} across the $\delta$ function to obtain a differential equation for $\lambda(X^-)$:
\be
\frac{1}{2}\lambda''-\frac{X_0^+}{1+X_0^+X^-}\lambda'+\frac{(X_0^+)^2}{(1+X^-X^+_0)^2}\lambda=-8\pi G C_0\frac{1+X^-X^+_0}{(1-X^-X^+_0)^3}.
\ee
This has solution
\be\label{lambdasol}
\lambda(X^-)=M(1+X^-X^+_0)+N(1+X^-X^+_0)^2-\frac{8\pi G C_0}{(X_0^+)^2}\frac{1+X^-X^+_0}{1-X^-X^+_0},
\ee
where $M$ and $N$ are constants.
We can conveniently package this backreaction as a metric variation
\be
\delta \wt{g}_{\mu\nu}=\Theta(X^+-X^+_0)\Lambda(X^+,X^-)\wt{g}_{\mu\nu}, \label{deltagback}
\ee
with
\be\label{Lambdaeq}
\Lambda(X^+,X^-)\equiv \lambda'(X^-)-\frac{2X^+\lambda(X^-)}{1+X^-X^+}.
\ee

The constants $M$ and $N$ arise from a gauge ambiguity: they can be removed by the linearized diffeomorphism
\begin{align}\nonumber
X^+&=\wt{X}^++\Theta(\wt{X}^+-X^+_0)\left(N(X_0^+)^2-X_0^+(M+2N)\wt{X}^++(N+M)(\wt{X}^+)^2\right)\\
X^-&=\wt{X}^-,\label{Xpdiff}
\end{align}
which has no other effect on the metric (in particular it preserves conformal gauge and is continuous at $X^+=X^+_0$).  The physical interpretation of this diffeomorphism is as follows:  the unperturbed metric is invariant under the $SL(2,\mathbb{R})$ isometry group of $AdS_2$, which in Kruskal coordinates has the infinitesimal form
\begin{align}\nonumber
X^+&=\wt{X}^++a+b \wt{X}^++c(\wt{X}^+)^2\\
X^-&=\wt{X}^-+c-b\wt{X}^-+a(\wt{X}^-)^2.\label{isometry}
\end{align}
Restricting to transformations which preserve $X^+_0$, we have
\be
b=-\left(\frac{a}{X^+_0}+c X^+_0\right)
\ee
and thus
\be
X^-=\wt{X}^-+\left(c-\frac{a}{(X_0^+)^2}\right)\left(1+\wt{X}^-X_0^+\right)+\frac{a}{(X^+_0)^2}\left(1+\wt{X}^-X_0^+\right)^2.
\ee
Comparing to \eqref{lambdasol}, we see that if we choose $a=N(X^+_0)^2$ and $c=N+M$ then this diffeomorphism merely shifts the $N$ and $M$ terms in $\lambda$.  We do not actually want to act with this isometry however, as that would affect the coordinates to the left of the shell and would not actually change the metric.  Instead we want a diffeomorphism which does nothing to the left of the shell but removes $M$ and $N$ to the right of it.  Since the combined transformation \eqref{isometry} leaves the metric invariant, and the $X^-$ transformation with these choices of $a$ and $c$ simply adds the $M$ and $N$ terms in \eqref{lambdasol}, if we do only the $X^+$ part of this isometry then this must remove them, hence \eqref{Xpdiff}.  This can of course just be checked by inserting \eqref{Xpdiff} into the backreacted metric.

We now consider the excitation energy $\Delta E_\pm$ created by the dressed operator $\psi_{k\wt{\eta}\wt{s}}(t_-)$, as shown in figure \ref{singleshellbackfig}.  This is most easily computed using the algorithm from section \ref{Hsec}:
\be
\{\psi_{k\wt{\eta}\wt{s}},H_-\}=\dot{\psi}_{k\wt{\eta}\wt{s}}=i \Delta E_\pm \psi_{k\wt{\eta}\wt{s}},
\ee
where the dot indicates the derivative with respect to $t_-$ and as before $\pm$ indicates whether the particle we create is right or left moving.  In order to compute this time derivative to first order in the backreaction, we need to know how the dressing geodesic $y^\mu_{t_-\wt{\eta}}(\wt{s})$ changes under both a small change in $t_-$ \textit{and} the backreaction due to the shell.  The unperturbed geodesic in the BTZ geometry is given by
\be
X^\pm(\wt{s})=\mp \frac{1-\frac{1}{16\pi G}e^{\wt{s}}\left(r_s\pm 8\pi G\wt{\eta}\right)}{1+\frac{1}{16\pi G}e^{\wt{s}}\left(r_s\mp 8\pi G\wt{\eta}\right)}e^{\mp r_s t_-},\label{kruskalgeod}
\ee
which is obtained from \eqref{schgeod} using \eqref{kruskalsch} and replacing $\phi_b\to \frac{1}{16\pi G}$.  We have written this in a somewhat unusual way because in our conventions $\frac{e^{\wt{s}}}{16\pi G}$ and $8\pi G \wt{\eta}$ should be thought of as being $O(1)$ in the semiclassical expansion.  We can simplify our expressions by defining
\be
\gamma\equiv 8\pi G \wt{\eta}
\ee
and
\be\label{sigmaldef}
e^{\wt{s}}\equiv \frac{16\pi G}{r_s+\gamma}e^\sigma,
\ee
in terms of which we have
\begin{align}\nonumber
X^+&=-\frac{(r_s+\gamma)(1-e^{\sigma})}{r_s+\gamma+e^{\sigma}(r_s-\gamma)}e^{-r_s t_-}\\
X^-&=\frac{r_s+\gamma-e^{\sigma}(r_s-\gamma)}{(r_s+\gamma)(1+e^\sigma)}e^{r_s t_-}.\label{kruskalgeod2}
\end{align}
In particular note that the geodesic crosses the horizon $X^+=0$ at $\sigma=0$.  One can easily check that these obey the geodesic equation and normalization condition:
\begin{align}\nonumber
X^{\pm\prime\prime}-\frac{2X^\mp}{1+X^+X^-}(X^{\pm \prime})^2&=0\\
\frac{4X^{+\prime}X^{-\prime}}{(1+X^+X^-)^2}&=-1.\label{geodeqnorm}
\end{align}
The tangent and normal vectors are
\begin{align}\nonumber
e&=\frac{2 r_s (r_s+\gamma)e^{\sigma-r_s t_-}}{(r_s+\gamma+e^\sigma(r_s-\gamma))^2}\partial_+-\frac{2 r_s e^{\sigma+r_s t_-}}{(r_s+\gamma)(1+e^\sigma)^2}\partial_-\\
\tau&=\frac{2 r_s (r_s+\gamma)e^{\sigma-r_s t_-}}{(r_s+\gamma+e^\sigma(r_s-\gamma))^2}\partial_++\frac{2 r_s e^{\sigma+r_s t_-}}{(r_s+\gamma)(1+e^\sigma)^2}\partial_-,\label{kruskalvecs}
\end{align}
and from these expressions we can extract the deviation (here we suppress the explicit $\wt{\eta}$ dependence of $y^\mu_{t_-\wt{\eta}}$ to simplify notation)
\be
y_{t_-+\Delta t}^\mu(\sigma)\equiv y^\mu_{t_-}(\sigma)+\Delta y^\mu_{t_-}(\sigma)\equiv y^\mu_{t_-}(\sigma)+\hat{\alpha}(\sigma)e^\mu(\sigma)+\hat{\beta}(\sigma)\tau^\mu(\sigma),
\ee
with
\begin{align}\nonumber
\hat{\alpha}(\sigma)&=-\gamma \Delta t\\
\hat{\beta}(\sigma)&=\left(\gamma \cosh \sigma-r_s \sinh \sigma\right)\Delta t.\label{hatalphabeta}
\end{align}
This geodesic also undergoes a deviation due to the metric backreaction \eqref{deltagback}:
\be
\delta y_{t_-}^\mu(\sigma)=\alpha(\sigma)e^\mu(\sigma)+\beta(\sigma) \tau^\mu(\sigma),
\ee
with $\alpha$ and $\beta$ obtained from \eqref{oneabsol} by shifting from $s$ to $\sigma$ using \eqref{sshift} and \eqref{sigmaldef}.  More concretely we have
\be\label{alphaback}
\alpha(\sigma)=-\frac{1}{2}\int_{\sigma_0}^{\sigma}d\sigma' \Lambda(X^+(\sigma'),X^-(\sigma'))
\ee
and
\be
\beta(\sigma)=-\frac{1}{2}\sinh\left(\sigma-\sigma_0\right)\Lambda(X_0^+,X^-(\sigma_0))-\frac{1}{2}\int_{\sigma_0}^{\sigma}d\sigma'\sinh(\sigma-\sigma')\tau^\mu \partial_\mu \Lambda(X^+(\sigma'),X^-(\sigma')),\label{betaback}
\ee
where $\sigma_0$ is the value of $\sigma$ at which the unperturbed geodesic reaches $X_0^+$ and we always assume $\sigma>\sigma_0$.  More explicitly we have
\be\label{sigmasmall}
\sigma_0=\log \left(\frac{(r_s+\gamma)(1+X_0^+e^{r_s t_-})}{r_s+\gamma-(r_s-\gamma)X_0^+e^{r_s t_-}}\right),
\ee
which vanishes exponentially for $-r_s t_-\gg 1$.  These expressions for $\alpha$ and $\beta$ can be integrated: by \eqref{geodeqnorm} and \eqref{Lambdaeq} we have
\begin{align}\nonumber
\frac{d}{d\sigma'}\left(\frac{\lambda(X^-(\sigma'))}{X^{-\prime}(\sigma')}\right)=&\Lambda(X^+(\sigma'),X^-(\sigma'))\\\nonumber
\frac{d}{d\sigma'}\left(\cosh(\sigma'-\sigma)\frac{\lambda(X^-(\sigma'))}{X^{-\prime}(\sigma')}\right)=&\cosh(\sigma'-\sigma)\Lambda(X^+(\sigma'),X^-(\sigma'))\\
&-\frac{4\sinh(\sigma'-\sigma)\lambda(X^-(\sigma'))X^{+\prime}(\sigma')}{(1+X^+(\sigma')X^-(\sigma'))^2},
\end{align}
which can be used in \eqref{alphaback}, \eqref{betaback} to see that
\begin{align}\nonumber
\alpha(\sigma)&=-\frac{1}{2}\frac{\lambda(X^-(\sigma))}{X^{-\prime}(\sigma)}+\frac{1}{2}\frac{\lambda(X^-(\sigma_0))}{X^{-\prime}(\sigma_0)}\\
\beta(\sigma)&=\frac{1}{2}\frac{\lambda(X^-(\sigma))}{X^{-\prime}(\sigma)}-\frac{1}{2}\cosh(\sigma-\sigma_0)\frac{\lambda(X^-(\sigma_0))}{X^{-\prime}(\sigma_0)}-\sinh(\sigma-\sigma_0)\Lambda(X^+_0,X^-(\sigma_0)).\label{integratedab}
\end{align}

To compute $\Delta E_\pm$ however, the deviation we are really interested in is
\be\label{tildedev}
y^\mu_{t_-+\Delta t}(\wt{s})+\delta y^\mu_{t_-+\Delta t}(\wt{s})=y^\mu_{t_-}(\wt{s})+\delta y^\mu_{t_-}(\wt{s})+\wt{\alpha}(\wt{s})\wt{e}^\mu(\wt{s})+\wt{\beta}(\wt{s})\wt{\tau}^\mu(\wt{s}),
\ee
where $\wt{e}^\mu$ and $\wt{\tau}^\mu$ are the tangent and normal vectors to $y^\mu_{t_-}+\delta y^\mu_{t_-}$, in terms of which we have
\be
\Delta E_\pm=\frac{|k|}{\Delta t}\left(\mp \wt{\alpha}+\wt{\beta}\right)\label{DeltEab}
\ee
by the same argument as led to \eqref{energyresult}.  Therefore to find $\Delta E_\pm$ we simply need to work out $\wt{\alpha}$ and $\wt{\beta}$ in terms of $\alpha, \hat{\alpha}, \beta,\hat{\beta}$, each of which we know, focusing on terms which are of order $\Delta t \delta g$ since these are what will modify $\Delta E_\pm$ at leading order in $\delta g$.  To proceed, we need to know how $e^\mu$ and $\tau^\mu$ change under both the variation $\Delta$ which changes $t_-$ to $t_-+\Delta t$ and the variation $\delta$ which turns on the backreaction \eqref{deltagback}.  We have already worked out $\Delta e^\mu$ and $\delta e^\mu$ in \eqref{ecovjump} and \eqref{ecovjumpvar},
\begin{align}
\nonumber
\Delta^{(c)} e^\mu&\equiv\Delta e^\mu+\Gamma^\mu_{\alpha\beta}e^\alpha \Delta y^\beta_{t_-}\\\nonumber
&=\hat{\beta}'\tau^\mu\\\nonumber
\delta^{(c)} e^\mu&\equiv \delta e^\mu+\Gamma^\mu_{\alpha\beta}e^\alpha \delta y^\beta_{t_-}\\
&=-\frac{1}{2}e^\alpha e^\beta \delta g_{\alpha\beta}e^\mu+\beta'\tau^\mu,\label{evars}
\end{align}
and we can determine the variations of $\tau^\mu$ by requiring that acting with $\Delta^{(c)}$ and $\delta^{(c)}$ on $(e^\mu \tau_\mu)$ and $(\tau^\mu \tau_\mu)$ give zero, leading to
\begin{align}\nonumber
\Delta^{(c)} \tau^\mu&\equiv \Delta \tau^\mu+\Gamma^\mu_{\alpha\beta}\tau^\alpha \Delta y^\beta_{t_-}\\\nonumber
&=\hat{\beta}'e^\mu\\\nonumber
\delta^{(c)} \tau^\mu&\equiv \delta \tau^\mu+\Gamma^\mu_{\alpha\beta}\tau^\alpha \delta y^\beta_{t_-}\\
&=\left(\beta'-e^\alpha\tau^\beta\delta g_{\alpha\beta}\right)e^\mu+\frac{1}{2}\tau^\alpha\tau^\beta \delta g_{\alpha\beta}\tau^\mu.\label{tauvars}
\end{align}
The tangent and normal vectors appearing in the definitions of $\wt{\alpha}$ and $\wt{\beta}$ are given by
\begin{align}\nonumber
\wt{e}^\mu&=e^\mu+\delta e^\mu\\
\wt{\tau}^\mu&=\tau^\mu+\delta \tau^\mu,
\end{align}
so we can rewrite \eqref{tildedev} as
\be
\hat{\alpha}e^\mu+\hat{\beta}\tau^\mu+\alpha\Delta e^\mu+\beta \Delta \tau^\mu+\Delta t\left(\dot{\alpha}e^\mu+\dot{\beta}\tau^\mu\right)=\wt{\alpha}(e^\mu+\delta e^\mu)+\wt{\beta}(\tau^\mu+\delta \tau^\mu),\label{tildevar2}
\ee
where $\dot{\alpha}$, $\dot{\beta}$ are the derivatives of $\alpha$, $\beta$ with respect to $t_-$.  Taking the inner products of \eqref{tildevar2} with $e^\mu$ and $\tau^\mu$, using \eqref{evars} and \eqref{tauvars}, and neglecting terms of order $\delta g^2$, we arrive at
\begin{align}\nonumber
\wt{\alpha}&=\hat{\alpha}+\frac{1}{2}\hat{\alpha}e^\alpha e^\beta \delta g_{\alpha\beta}+\hat{\beta}e^\alpha \tau^\beta \delta g_{\alpha\beta}+\beta \hat{\beta}'-\hat{\beta}\beta'+\dot{\alpha}\Delta t\\
\wt{\beta}&=\hat{\beta}-\frac{1}{2}\hat{\beta}\tau^\alpha\tau^\beta \delta g_{\alpha\beta}+\alpha \hat{\beta}'-\hat{\alpha}\beta'+\dot{\beta}\Delta t.\label{tilderesults}
\end{align}

Now at last applying these results to the situation shown in figure \ref{singleshellbackfig}, our first goal is to see if $\Delta E_{\pm}$ has a term which is proportional to $e^{-r_s t_-}$, as if it does then this exponential growth can overcome the suppression by $G$ at time separations which are of order the scrambling time.  Keeping only terms which contain such an exponential, from \eqref{Lambdaeq}, \eqref{kruskalgeod2}, \eqref{kruskalvecs}, and \eqref{sigmasmall} we have
\begin{align}\nonumber
\Lambda(X^+(\sigma),X^-(\sigma))&\approx -\lambda(0)e^{-r_s t_-}\frac{r_s+\gamma}{r_s}\sinh \sigma\\
\tau^\mu \partial_\mu \Lambda(X^+(\sigma),X^-(\sigma))&\approx-\lambda(0) e^{-r_s t_-}\frac{r_s+\gamma}{r_s},
\end{align}
which we can use in \eqref{alphaback}, \eqref{betaback} to find that
\be\label{alphabetabackres}
\alpha(\sigma)\approx \beta(\sigma)\approx\lambda(0)e^{-r_s t_-}\frac{r_s+\gamma}{2r_s}\big(\cosh \sigma-1\big).
\ee
Using these expressions together with \eqref{deltagback} and \eqref{hatalphabeta} in \eqref{tilderesults}, we find that all new terms cancel in $\wt{\alpha}$ and $\wt{\beta}$: there is no exponentially-growing term in $\Delta E_\pm$ at this order in $\delta g_{\mu\nu}$.  This cancellation can be predicted directly from \eqref{alphabetabackres}: the value of $\lambda$ at any particular value of $\sigma$ can be changed by adjusting the constants $M$ and $N$ in \eqref{lambdasol}, but we saw that these constants could be removed by the diffeomorphism \eqref{Xpdiff} so there cannot be any physical effect which is proportional to $\lambda(0)$.  By using our integrated expressions \eqref{integratedab} for $\alpha$ and $\beta$ in \eqref{tilderesults} we can also give explicit forms for $\wt{\alpha}$ and $\wt{\beta}$: using also \eqref{lambdasol}, \eqref{kruskalgeod2}, \eqref{hatalphabeta}, and \eqref{sigmasmall}, we find
\begin{align}\nonumber
\wt{\alpha}(\sigma)=&-\gamma \Delta t\\\nonumber
\wt{\beta}(\sigma)=&(\gamma\cosh\sigma-r_s\sinh \sigma)\Delta t\\\nonumber
&+16\pi G C_0 r_s e^{r_s t_-}\frac{\gamma(r_s+\gamma)-2(r_s^2-\gamma^2)X^+_0 e^{r_s t_-}-\gamma(r_s-\gamma)(X_0^+)^2e^{2r_s t_-}}{(r_s+\gamma)\left(r_s+\gamma+(r_s-\gamma)(X_0^+)^2e^{2 r_s t_-}\right)^3}\\\nonumber
&\times\Big[\left((r_s+\gamma)^2+2\gamma(r_s+\gamma)X^+_0e^{r_s t_-}+(r_s^2+\gamma^2)(X_0^+)^2e^{2r_s t_-}\right)\sinh \sigma\\
&\phantom{++}-2r_sX_0^+e^{r_s t_-}\left(r_s+\gamma+\gamma X_0^+e^{r_s t_-}\right)\cosh \sigma\Big]\Delta t,
\end{align}
which may then be used in \eqref{DeltEab} to give
\begin{align}\nonumber
\Delta E_{\pm}=&|k|\left(\pm \gamma+\gamma\cosh\sigma-r_s\sinh \sigma\right)\\\nonumber
&+16\pi G C_0 r_s e^{r_s t_-}|k|\frac{\gamma(r_s+\gamma)-2(r_s^2-\gamma^2)X^+_0 e^{r_s t_-}-\gamma(r_s-\gamma)(X_0^+)^2e^{2r_s t_-}}{(r_s+\gamma)\left(r_s+\gamma+(r_s-\gamma)(X_0^+)^2e^{2 r_s t_-}\right)^3}\\\nonumber
&\times\Big[\left((r_s+\gamma)^2+2\gamma(r_s+\gamma)X^+_0e^{r_s t_-}+(r_s^2+\gamma^2)(X_0^+)^2e^{2r_s t_-}\right)\sinh \sigma\\
&\phantom{++}-2r_sX_0^+e^{r_s t_-}\left(r_s+\gamma+\gamma X_0^+e^{r_s t_-}\right)\cosh \sigma\Big].\label{finalresult}
\end{align}
Up to notation the first line of \eqref{finalresult} is the same as \eqref{energyresult}, so we see that the matter shell in figure \ref{singleshellbackfig} has only a small effect on the excitation energy, even when there is a large time separation. In fact when $-t r_s>>1$, the effect of the shell is exponentially small!

The exponentially small result \eqref{finalresult} for the correction to the excitation energy is in sharp contrast to the exponential growth found in \eqref{scrambleresult} for the scrambling phenomenon.  The reason for this distinction is the following: in the calculation leading to \eqref{scrambleresult}, we were interested in the effect of a matter observable dressed to the left boundary on the renormalized length $\wt{L}$.  As the definition of $\wt{L}$ involves both boundaries, we cannot use a diffeomorphism such as \eqref{Xpdiff} to remove the $SL(2,\mathbb{R})$ transformation across the shockwave created by the matter operator: such a diffeomorphism would act nontrivially at the right boundary, and thus affect $\wt{L}$.  In the calculation we just did, exponentially growing terms are related to $SL(2,\mathbb{R})$ transformations but we saw these could be removed by a diffeomorphism which did not affect the rest of the calculation.  We have some expectation this is a general phenomenon: in order to get an exponentially-growing effect in some correlation function in the JT or BTZ wormhole background, we need to ask a question which is non-local enough to prevent the removal of the relevant $SL(2,\mathbb{R})$ transformation by a diffeomorphism (see section 6.3 of \cite{Maldacena:2016upp} for some similar remarks).

\section{Discussion}\label{dsec}
We've now learned that the algebra of a wide variety of diffeomorphism-invariant observables in JT gravity coupled to matter is exactly computable, leading to the brackets \eqref{gravalg}, \eqref{twosidealg}, and \eqref{onesidealg}.  Moreover we have seen that much of the interesting physics in this theory can be explained using this algebra.  We expect many more calculations can be done using our techniques, both in JT gravity and also in higher dimensions.  In this final section we discuss our results from section \ref{energysec} from a more general point of view.

\bfig
\includegraphics[height=6cm]{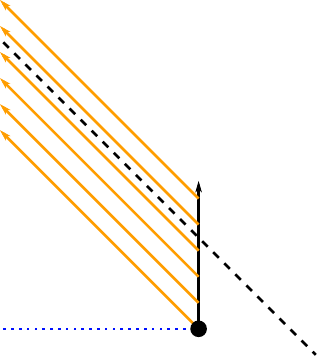}
\caption{The gravitational redshift of a laser signal near a horizon.  The horizon is the dashed line, the worldline of the laser is black, and the dotted blue line is the dressing geodesic for the laser location.}\label{laserfig}
\efig
One thing we have learned is that in gravitational theories there is a close connection between the asymptotic energy and the presence of an event horizon.  In a wide variety of states, including those with no boost isometry, the location of a horizon as measured by renormalized proper distance from the boundary can be diagnosed using the commutator of a dressed matter operator and the Hamiltonian: we simply look for the place where the asymptotic energy of an outgoing excitation vanishes.  This statement may seem to be in tension with the fact that finding the location of an event horizon requires knowledge of the full spacetime, while the algebra can be computed using only data on a fixed Cauchy slice, but of course these are not really contradictory since knowing the full set of data on a Cauchy slice by definition determines the full spacetime.  What is perhaps more surprising is that this simple bracket already seems to have enough information to locate the horizon in many cases.  One way to think about this is as a manifestation of gravitational redshift: by definition, an event horizon is a boundary between points which can send signals to the asymptotic boundary and points which cannot.  If we imagine placing a laser at rest at the end of our dressing geodesic and then letting it go while it is firing a periodic sequence of pulses out towards the asymptotic boundary, only a finite number of pulses will be fired before the laser crosses the horizon (see figure \ref{laserfig}).  Since signals fired from very near the horizon come out much later, this will cause the observed frequency at infinity to be less than it was in the rest frame of the laser.  Moreover this effect will be more and more severe as we move the dropping-off point closer to the horizon along the dressing geodesic, as fewer and fewer pulses will make it out.  Since energy is conserved, in situations where the outgoing pulse does not substantially interact with other matter the asymptotic energy of the pulses when they arrive at infinity must be the same as their asymptotic energy right when they were produced, and thus the excitation energy must become zero right as our dressing geodesic crosses the horizon. In this way the Hamiltonian can access just the right non-local information which is necessary to identify the location of the horizon.  In JT gravity coupled to conformal matter we saw that this test works in all states.  In particular direct interactions of the outgoing pulses with other shells do not happen, since in 2D CFT left- and right-movers do not interact and for the gravity fields only the dilaton, which does not interact with the matter fields, is singular across the shells.  In higher dimensions the situation is less clear, as we need to worry both about the possibility of interactions with exterior matter and also the fact that when spherical symmetry is not present we need to be more careful about the words ``ingoing'' and ``outgoing''.\footnote{We emphasize that the rigorous geodesic-deviation method we used for computing the excitation energy in the main text automatically takes such effects into account, so it is more powerful than the intuitive argument given here.}  On the other hand we expect that typical black hole microstates will be spherically symmetric, and also not have large amounts of exterior matter present,\footnote{Thermal-scale shells may indeed be present, but we saw in section \ref{higherdsec} that these have only a small effect \eqref{finalresult} on the excitation energy even at large time separation.} so this test may yet work in higher dimensions if we restrict to appropriate states.

To make contact with the firewall paradoxes of \cite{Almheiri:2013hfa,Marolf:2013dba} however, we need more than just a way of diagnosing the location of the horizon.  We also need to argue that this location is ``robust'', in the sense that we can find a single diffeomorphism-invariant operator whose commutator with the Hamiltonian vanishes for \textit{most} of the states in the microcanonical ensemble of a black hole.  For example if the renormalized proper distance $\wt{s}$ to the horizon varied substantially from microstate to microstate, then, although we could use the single-side operator $\psi_{k\wt{\eta}\wt{s}}$ to create particles with small excitation energy near the horizon of any particular microstate, we'd have to use operators with different values of $\wt{s}$ for different microstates and thus we wouldn't have an argument that this operator generically creates structure at the horizon.  In JT gravity coupled to conformal matter we learned that in fact the value of $\wt{s}$ at which the horizon is located depends \textit{only} on the total energy of the state (see equation \eqref{Lfirstflip}), and thus that we can use exactly the same operator to create structure at the horizon in any state with this energy.  In higher dimensions this will not in general be true, although we did see that it remains the case for states constructed from spherical massless dust shells in the BTZ geometry such as those studied in \cite{Shenker:2013pqa,Shenker:2013yza}.  More generally however we saw that for states with  massless matter shells with nonzero transverse pressure (what we called $\hat{T}$ in \eqref{Tform}), there \textit{is} a nonzero state-dependent correction \eqref{finalresult} to the excitation energy created by $\psi_{k\wt{\eta}\wt{s}}$. The question of whether or not this change is sufficient to destroy the ability of $\psi_{k\wt{\eta}\wt{s}}$ to generically create structure at the horizon without costing energy is then the question of what types of shell configurations are typical.

The nature of typical states at fixed energy in quantum gravity is a long-standing difficult conceptual problem, about which we do not have too much concrete to say.  Most likely typical microstates do not have a globally-valid geometric description, but on the other hand to study the excitation energy of $\psi_{k\wt{\eta}\wt{s}}$ near the horizon we only need to know what they are like outside the horizon and just a little bit inside.  One kind of state which \textit{does} change the value of $\wt{s}$ at which the excitation energy vanishes at leading order is a state with a single shell reflected from the boundary as in figure \ref{singleshellbackfig}, but with a shell energy that is of order the energy of the black hole.  This however is unlikely to be typical, as it should be much more entropically favorable to leave most of that energy in the black hole.  On the other hand if we take the shell to have thermal-scale energy near the boundary, as we did in our calculations in subsection \ref{higherdsec}, then there is no particularly good reason why such a shell should not be present at some time.  After all the Hawking process should roughly speaking produce a regular flux of such shells.  Indeed we can just begin with any microstate and then add such a shell, which only changes the energy by the temperature scale.

\bfig
\includegraphics[height=5cm]{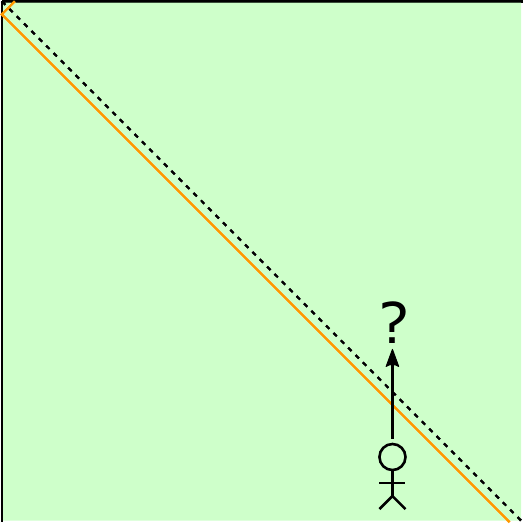}
\caption{A puzzling situation: an AdS black hole is in a state where at some time in the distant future, a thermal scale shell (or for that matter a bullet) will definitely come out.  What happens to an observer who jumps in now?  Classically they experience a high-energy collision, but do they really?  A ``stretched horizon'' picture would instead suggest that the shell does not exist until it crosses a surface roughly a Planck distance outside of the horizon.}\label{infallerfig}
\efig
There is a particularly puzzling aspect of black hole states with such shells present. Indeed say that we have a black hole which will definitely emit a thermal-scale shell at some time in the distant future (see figure \ref{infallerfig}).  What happens if we jump in now?  Classical gravity would say that we encounter a highly-boosted version of that shell very close to the horizon, which sounds unpleasant, but in the usual view of unitary black hole evaporation we would just say that as we evolve backwards in time from the point where the shell emerged it would eventually be thermalized into the microstate degrees of freedom, perhaps at a Planck distance from the horizon, and no longer exist.  Which statement is true?  We do not currently have the tools to answer this question, which seems to illustrate a major gap in our understanding of the quantum physics of black holes even outside of the horizon.  In this paper we have answered a more limited question: we have shown that the effect of such a shell on the excitation energy of the dressed operator $\psi_{k\wt{\eta}\wt{s}}$ is small, even when the time separation is large.  To the extent that we can think of a typical state as having an $O(1)$ number (or even an $O(\sqrt{S})$ number) of these shells scattered about at various times, this shows that the firewall typicality paradoxes of \cite{Almheiri:2013hfa,Marolf:2013dba} apply.  But on the other hand we do not really expect typical states to have so simple a description: at best we can say that these states are somewhat more typical than the thermofield double, so we have probed a potential failure mode of the paradoxes and found that it is not realized.

\paragraph{Acknowledgments} We thank Chris Akers, Ahmed Almheiri, Bin Chen, Liangyu Chen, Lin-Qing Chen, Netta Engelhardt, Steve Giddings, Yingfei Gu, Muxin Han, Ruizhen Huang, Gary Horowitz, Ling-Yan Hung, Don Marolf, Lampros Lamprou, Yang Lei, Henry Lin, Hong Liu, Juan Maldacena, Henry Maxfield, Cheng Peng, Eric Perlmutter, Wei Song, Antony Speranza, Aron Wall, Diandian Wang, Huajia Wang, Hongbao Zhang, Ying Zhao, and Yang Zhou for useful discussions.  We do not thank Covid-19 for greatly complicating the logistics of this project.  JQW would like to thank Peking University, Fudan University, and KITS for hospitality.  Especially JQW would like to acknowledge the friendly and creative atmosphere at KITS where much of this work was completed.  DH is supported by the Simons Foundation as a member of the ``It from Qubit'' collaboration, the Sloan Foundation as a Sloan Fellow, the Packard Foundation as a Packard Fellow, and the Air Force Office of Scientific Research under the award number FA9550-19-1-0360.  JQW is supported by the Packard foundation, a Len DeBenedictis Postoctoral Fellowship, and funds from the University of California.

\appendix
\section{The matter energy-momentum tensor from covariant phase space}\label{Tapp}
In discussions of energy in gravitational theories, it is often the case that the expression for the total energy does not depend on the details of the matter theory.  For example the ADM energy in asymptotically-flat $3+1$ dimensional gravity looks the same whether or not a minimally-coupled dynamical scalar field is present.  This is analogous to the fact that in electromagnetism the total charge is given by the electric flux at infinity, regardless of whether the charged matter consists of fermions, bosons, etc.  On the other hand there are some situations where the expression for the energy \textit{is} modified, and so in this appendix we explore the question of when it is and when it isn't using the version of the covariant phase space formalism that we developed in \cite{Harlow:2019yfa}.  We will also use this formalism to explain the general relationship between the matter energy-momentum tensor and the canonical Hamiltonian, systematizing previous discussions of ``improvement terms'' and a difference between ``canonical'' and ``gravitational'' energy-momentum tensors \cite{Callan:1970ze,Weinberg:1995mt}. Our motivation for discussing this here is that in some places in the main text we restricted to matter theories for which the standard expression \eqref{Hpm} for the energy of JT gravity is unmodified, so we want to explore how strong this assumption is.  Readers who are happy to accept it without question need not read this appendix.  We will not review the formalism of \cite{Harlow:2019yfa} in detail, but instead just give a rough sketch and quote results as needed.  Our conventions for differential forms, orientations, etc are as in \cite{Harlow:2019yfa}.

Indeed consider any Lagrangian field theory with dynamical fields $\phi^a$ and fixed background fields $\chi_m$.  For now the spacetime metric $g_{\mu\nu}$ can be either dynamical or background.  We can write the action as
\be
S=\int_M L(\phi,\chi)+\int_\Gamma \ell(\phi,\chi),
\ee
where $L$ is a $d$-form, $\ell$ is a $(d-1)$-form, and $\Gamma$ is the spatial boundary.  Note that we do not include boundary terms at future/past boundaries: this is because we do not want to specify boundary conditions there, as that would be a choice of state but we only wish to discuss the theory as a whole.  The variation of $L$ can always be written as
\be
\delta L=E_a\delta \phi^a+d\Theta,
\ee
where
\be\label{mattereom}
E_a=0
\ee
are the equations of motion and $\Theta(\phi,\chi,\delta\phi)$ is a $(d-1)$-form on spacetime and a one-form on the space of dynamical field configurations obeying the boundary conditions (we call this latter space $\mathcal{C}$).  In order for the action to be stationary about solutions of the equations of motion (up to boundary terms at the future/past boundaries), we need the boundary action $\ell$ to be chosen such that
\be\label{ThetaC}
(\Theta+\delta \ell)|_{\Gamma}=dC
\ee
for some $(d-2)$-form $C(\phi,\chi,\delta\phi)$ which is also a one-form on $\mathcal{C}$.  The central idea of covariant phase space is to define ``pre-phase space'', denoted $\wt{\mathcal{P}}$, to be the set of solutions of the equation of motion obeying the boundary conditions (or equivalently the set of stationary points of $S$ up to future/past terms).  On this pre-phase space we can define a ``pre-symplectic form'', given by
\be\label{presymp}
\wt{\Omega}=\int_\Sigma \delta \Theta-\int_{\partial\Sigma}\delta C,
\ee
where $\Sigma$ is any Cauchy slice, and then define the physical phase space $\mathcal{P}$ to be the quotient of $\wt{\mathcal{P}}$ by the zero modes of $\wt{\Omega}$ and the symplectic form $\Omega$ to be the quotient of $\wt{\Omega}$ by the same.  The boundary term in \eqref{presymp} was first introduced in the context of Einstein gravity with Neumann boundary conditions in \cite{Compere:2008us}, and then studied systematically in \cite{Harlow:2019yfa}.  In particular in \cite{Harlow:2019yfa} it was shown that for any diffeomorphism generator $\xi^\mu$ which respects the boundary conditions and preserves the background fields, the evolution by $\xi^\mu$ on this phase space is generated by the conserved Hamiltonian
\be\label{Hmatt}
H_\xi=\int_\Sigma J_\xi+\int_{\partial \Sigma}\left(\xi\cdot \ell-X_\xi\cdot C\right),
\ee
where the ``Noether current'' $J_\xi$ is given by
\be
J_\xi\equiv X_\xi\cdot \Theta-\xi\cdot L
\ee
and $X_\xi$ is defined to be the vector field on $\mathcal{C}$ which replaces each field variation $\delta\psi^i$ by the Lie derivative $\mathcal{L}_\xi \psi^i$.  The symbol ``$\cdot$'' indicates the insertion of a vector field into the first argument of a differential form; we use this notation both for vector fields on $\mathcal{C}$ and vector fields on spacetime.  The test that this Hamiltonian indeed generates $\xi^\mu$ evolution is that it obeys
\be\label{hamilton}
\delta H_\xi=-X_\xi\cdot \wt{\Omega},
\ee
which is what was shown in \cite{Harlow:2019yfa} (earlier versions of covariant phase space did not systematically address boundary terms and thus couldn't unambiguously verify \eqref{hamilton}).  Many more details of this formalism are explained in \cite{Harlow:2019yfa}, where the phase space and Hamiltonian constructed in this way are shown to possess all their needed physical properties and to agree with standard constructions wherever they are applicable.

We now explain how the expression \eqref{Hmatt} for the Hamiltonian is related to the energy-momentum tensor in a field theory with dynamical ``matter'' fields $\psi^i$ coupled to a fixed background metric $g_{\mu\nu}$.  We define the energy-momentum tensor via
\be
\delta^{tot}L=E_i\delta\psi^i+d\Theta+\frac{1}{2}T^{\mu\nu}\delta g_{\mu\nu} \epsilon+d\Theta^g,
\ee
where $\delta^{tot}$ indicates a variation which acts also on the background metric $g_{\mu\nu}$, $\epsilon$ is the volume form on spacetime,  and $\Theta^g(\psi,g,\delta g)$ is a $(d-1)$-form on spacetime and a one-form on a larger configuration space $\mathcal{C}^g$ consisting of both the matter field configurations obeying their boundary conditions and metric configurations obeying what will be the metric  boundary conditions once we make it dynamical.  Note that $\Theta$ does not depend on $\delta g_{\mu\nu}$ and $\Theta^{g}$ does not depend on $\delta \psi$.  $\Theta^g$ vanishes automatically in theories where the action does not depend on derivatives of the metric. To get a natural relationship between the energy-momentum tensor $T^{\mu\nu}$ and the energy $H_\xi$, we need to assume that $L$ is a covariant function of $\psi^i$ and $g_{\mu\nu}$.  This means that for any diffeomorphism generator $\xi^\mu$ we must have
\be
d(\xi\cdot L)=\mathcal{L}_\xi L=X_\xi\cdot \delta L+T^{\mu\nu}\nabla_\mu\xi_\nu\epsilon+d(X_\xi^g\cdot \Theta^g),
\ee
where $X_\xi^g$ is the vector field on $\mathcal{C}^g$ which replaces $\delta g_{\mu\nu}$ by $\mathcal{L}_\xi g_{\mu\nu}=\nabla_\mu\xi_\nu+\nabla_\nu\xi_\mu$.  We can rewrite this as
\be\label{consarg}
\xi_\nu\nabla_\mu T^{\mu\nu}\epsilon=E_i\mathcal{L}_\xi \psi^i+d\left(J_\xi+\xi\cdot T\cdot \epsilon+X_\xi^g\cdot \Theta^g\right),
\ee
where we have defined
\be
(\xi\cdot T \cdot \epsilon)_{\mu_1\ldots \mu_{d-1}}\equiv\xi_\mu T^{\mu\nu}\epsilon_{\nu \mu_1\ldots \mu_{d-1}}.
\ee
Integrating \eqref{consarg} on-shell over any spacetime region $R$ we have
\be
\int_R\xi_\nu\nabla_\mu T^{\mu\nu}\epsilon=\int_{\partial R}(J_\xi+\xi\cdot T\cdot \epsilon+X_\xi^g\cdot \Theta^g).
\ee
Since $\xi^\mu$ is an arbitrary vector field, we can choose it to pick out any particular component of $\nabla_\mu T^{\mu\nu}$ in the center of $R$, and also to vanish in the vicinity of $\partial R$, so on-shell the conservation equation
\be
\nabla_\mu T^{\mu\nu}=0
\ee
must hold throughout spacetime.  We therefore learn that on-shell we have
\be
d\left(J_\xi+\xi\cdot T\cdot \epsilon+X_\xi^g\cdot \Theta^g\right)=0.
\ee
Since $\xi$ is arbitrary, by lemma one of \cite{wald1990identically} there must be a $d-2$ form $Z_\xi$ which is a local functional of $\psi$, $\xi$, and $g$ such that
\be
J_\xi+\xi\cdot T\cdot \epsilon+X_\xi^g\cdot \Theta^g=dZ_\xi.
\ee
If we now specialize to diffeomorphisms which preserve the background metric $g$, then $X_\xi^g\cdot \Theta^g=0$ and we are left with
\be\label{JTrel}
J_\xi+\xi\cdot T\cdot \epsilon=dZ_\xi.
\ee
In simple cases where $J_\xi$ does not involve derivatives of $\xi^\mu$ it can be interpreted as
\be
J_\xi=-\xi\cdot T^{\{c\}}\cdot \epsilon,
\ee
where $T^{\{c\}}_{\mu\nu}$ is a ``canonical energy-momentum tensor'' that is not necessarily symmetric (see sections 7.3-7.4 of \cite{Weinberg:1995mt}). In this case \eqref{JTrel} tells us how this energy-momentum tensor is related to the true energy-momentum tensor $T_{\mu\nu}$ (sometimes called the ``gravitational'' or ``Belinfante'' energy-momentum tensor) by what are sometimes called ``improvement terms'' \cite{Callan:1970ze,Weinberg:1995mt}.  The canonical Hamiltonian \eqref{Hmatt} for these diffeomorphisms is thus
\be\label{Hmatt2}
H_\xi=-\int_\Sigma \xi\cdot T \cdot \epsilon+\int_{\partial \Sigma}\left(\xi\cdot \ell-X_\xi \cdot C+Z_\xi\right),
\ee
which is the usual expression in terms of an integral of $T_{\mu\nu} \xi^\mu \tau^\mu$ on $\Sigma$ (where $\tau^\mu$ is the future-pointing normal vector to $\Sigma$) together with some boundary terms that will in general be nonzero.  In theories where we do have
\be\label{boundaryvanish}
\left(\xi\cdot \ell-X_\xi \cdot C+Z_\xi\right)|_\Gamma=0,
\ee
conservation of $H_\xi$ apparently requires that
\be
\xi\cdot T \cdot \epsilon |_\Gamma=0\label{boundaryT}.
\ee
This can indeed be shown to follow from \eqref{ThetaC}, \eqref{JTrel}, and \eqref{boundaryvanish}, together with the fact that preserving the boundary conditions requires $\xi^\mu$ to be parallel to $\Gamma$.

A useful example of a theory where the boundary terms do not all vanish is the Maxwell Lagrangian
\be
L=-\frac{1}{4}F_{\alpha\beta}F^{\alpha\beta}\epsilon
\ee
with Dirichlet boundary conditions fixing the pullback of $A$ to $\Gamma$, in which case we can take $\ell=0$ and $C=0$, and for which the energy momentum tensor is
\be
T^{\mu\nu}=F^{\mu\lambda}F^\nu_{\phantom{\nu}\lambda}-\frac{1}{4}g^{\mu\nu}F_{\alpha\beta}F^{\alpha\beta}.
\ee
A short calculation then shows that
\be
J_\xi=-\xi\cdot T\cdot \epsilon-d((\xi\cdot A) \star F),
\ee
so in this theory we have
\be
Z_\xi=-(\xi\cdot A) \star F
\ee
and the Hamiltonian will include a nontrivial boundary term $\int_{\partial \Sigma}Z_\xi$ if the (fixed) pullback of $A$ to $\Gamma$ has nonzero overlap with $\xi^\mu$ (as it will for example if we set $A_0=\mu$ at the boundary of $AdS$).

We now turn to the question of what happens when the metric is dynamical.  We will take the full action to be
\be
S=\int_M(L+\wt{L})+\int_\Gamma(\ell+\wt{\ell}),
\ee
where $\wt{L}$ and $\wt{\ell}$ are bulk and boundary ``gravitational'' Lagrangians which depend on the metric $g_{\mu\nu}$, and possibly also some dynamical ``dilatonic'' fields $\Phi^m$ which do not appear in the matter Lagrangians, but not on the matter fields $\psi^i$.  For example $\wt{L}$ could be the Einstein-Hilbert Lagrangian or the pure JT Lagrangian.  We can write the variation of $\wt{L}$ as
\be
\delta \wt{L}=\wt{E}^{\mu\nu}\delta g_{\mu\nu}\epsilon+\wt{E}_m\delta\Phi^m+d\wt{\Theta},
\ee
where for future convenience we have extracted a factor of the volume form $\epsilon$ from $\wt{E}^{\mu\nu}$, and we assume the boundary conditions are such that
\be
(\wt{\Theta}+\delta\wt{\ell})|_\Gamma=d\wt{C}
\ee
for some boundary $(d-2)$-form $\wt{C}$ constructed out of the metric and the dilatonic fields (this ensures the gravity theory by itself has a stationary action about solutions of its equations of motion).  Stationarity of the full action then requires
\be
(\Theta^g+\delta_g \ell)|_\Gamma=dC^g
\ee
for some $C^g(g,\psi,\delta g)$.  We can introduce a ``gravitational'' Noether current
\be
\wt{J}_\xi=X_\xi^g\cdot \wt{\Theta}-\xi\cdot \wt{L},
\ee
where now we extend $X_\xi^g$ to act on both $\delta g_{\mu\nu}$ and $\delta \Phi^m$ by replacing them $\mathcal{L}_\xi g_{\mu\nu}$ and $\mathcal{L}_\xi \Phi^m$.  On shell we have $d \wt{J}_\xi=0$, so using again lemma one of \cite{wald1990identically} we have that \cite{Iyer:1994ys}
\be\label{noetherQ}
\wt{J}_\xi=d\wt{Q}_\xi+\wt{f},
\ee
where $\wt{Q}_\xi$ and $\wt{f}$ are local functionals of the metric, the dilatonic fields, and $\xi^\mu$, and $\wt{f}$ vanishes if the gravitational equations of motion $\wt{E}^{\mu\nu}=0$ and $\wt{E}_m=0$ hold.  For example in Einstein gravity with a cosmological constant, for which
\be
\wt{L}=\frac{1}{16\pi G}(R-2\Lambda)\epsilon,
\ee
\eqref{noetherQ} holds with
\be
\wt{Q}_\xi=-\frac{1}{16\pi G}\star d\xi
\ee
and
\be
\wt{f}_{\nu_1\ldots \nu_{d-1}}=\frac{1}{8\pi G}\xi_\mu(R^{\mu\nu}-\frac{R}{2}g^{\mu\nu}+\Lambda g^{\mu\nu})\epsilon_{\nu \nu_1\ldots \nu_{d-1}},
\ee
while for pure JT gravity with
\be
\wt{L}=\Phi(R+2)\epsilon
\ee
\eqref{noetherQ} holds with
\be\label{JTQ}
\wt{Q}_\xi=-\Phi\star d\xi+2\star(d\Phi\wedge \xi)
\ee
and
\be
\wt{f}_\alpha=-2\xi_\mu\left(\nabla^\mu\nabla^\nu\Phi+g^{\mu\nu}\left(\Phi-\nabla^2\Phi\right)\right)\epsilon_{\nu\alpha}.
\ee
The total Noether current for the theory in which $\psi^i$, $g_{\mu\nu}$, and $\Phi^m$ are all dynamical is given by
\begin{align}\nonumber
J^{tot}_\xi&=J_\xi+\wt{J}_\xi+X_\xi^g\cdot \Theta^g\\\nonumber
&=d(\wt{Q}_\xi+Z_\xi)+\wt{f}-\xi\cdot T \cdot \epsilon\\\nonumber
&=d(\wt{Q}_\xi+Z_\xi)+(\wt{f}+2\xi\cdot\wt{E}\cdot \epsilon)-\xi\cdot(2 \wt{E}+T)\cdot \epsilon\\
&=dQ^{tot}_\xi+f^{tot},\label{Jtot}
\end{align}
where in the last line we have again used lemma one of \cite{wald1990identically} to argue that $J^{tot}_\xi$ must be equal to an exact form $dQ^{tot}$ plus something which vanishes once we impose the full set of equations of motion
\begin{align}\nonumber
E_i&=0\\\nonumber
\wt{E}^{\mu\nu}+\frac{1}{2}T^{\mu\nu}&=0\\
\wt{E}_m&=0.
\end{align}
In fact the equivalence of the last two lines  of \eqref{Jtot} implies that we can take
\be\label{Qtot}
Q_\xi^{tot}=\wt{Q}_\xi+Z_\xi,
\ee
since the last term in the third line of \eqref{Jtot} vanishes by the metric equation of motion $\wt{E}^{\mu\nu}+\frac{1}{2}T^{\mu\nu}=0$ and the only way for $(\wt{f}+2\xi\cdot\wt{E}\cdot \epsilon)$ to vanish on-shell is for it to vanish identically once the dilatonic equations $\wt{E}_m=0$ are imposed (it doesn't depend on the matter fields and therefore can't make use of the metric equation of motion).  Therefore applying \eqref{Hmatt} now to the full system we arrive at the total on-shell energy
\begin{align}\nonumber
H_\xi^{tot}&=\int_{\partial \Sigma}\left(\wt{Q}_\xi+\xi\cdot(\ell+\wt{\ell})-X_\xi \cdot C-X_\xi^g\cdot (\wt{C}+C^g)+Z_\xi\right)\\
&=\wt{H}_\xi+\int_{\partial\Sigma}\left(\xi\cdot \ell-X_\xi \cdot C-X_\xi^g\cdot C^g+Z_\xi\right),\label{Htot}
\end{align}
where
\be
\wt{H}_\xi=\int_{\partial \Sigma}\left(\wt{Q}_\xi+\xi\cdot \wt{\ell}-X_\xi^g\cdot \wt{C}\right)
\ee
is the Hamiltonian for the purely gravitational system with bulk and boundary Lagrangians $\wt{L}$ and $\wt{\ell}$.  Except for the term involving $C^g$,  the additional boundary terms in the second line of \eqref{Htot} are precisely those appearing in our expression \eqref{Hmatt2} for the matter Hamiltonian in a fixed background. Thus we see that the full Hamiltonian will agree with the purely gravitational Hamiltonian provided that the matter Hamiltonian \eqref{Hmatt2} has no boundary terms when written in terms of an integral of the energy-momentum tensor over space and also $C^g=0$.  $C^g$ is automatically zero if the matter Lagrangian does not depend on derivatives of the metric, and we have checked that it also vanishes for a conformally-coupled scalar field with Dirichlet boundary conditions $\Phi|_\Gamma=0$.

\section{The case of non-vanishing matter energy-momentum tensor at the boundary}\label{mattapp}
In subsections \ref{onesidebracketsec} and \ref{etabracketsec} we for simplicity assumed that the matter energy-momentum tensory $T_{\mu\nu}$ vanished at the $AdS_2$ boundaries.  We argued in footnote \ref{stressfootnote} that this is typically true automatically in the $\epsilon\to 0$ limit, but for completeness we here briefly sketch how this assumption can be relaxed for the specific case of computing $\{H_-,\psi_{s_0,\eta}\}$ for a free massless scalar field $\psi$ coupled to JT gravity.  In subsection \ref{onesidebracketsec} we found that, after deforming the action by $-k\psi_{s_0,\eta}$, for $s>0$ we have the dilaton discontinuities
\begin{align}\nonumber
\Delta \Phi(s)&=\frac{k}{2}\theta(s_0-s)\sinh(s_0-s)\tau^\alpha\nabla_\alpha \psi(y(s_0))\\
\Delta \dot{\Phi}&=-\frac{k}{2}\theta(s_0-s)e^\alpha\nabla_\alpha \psi(y(s_0)).
\end{align}
The dilaton boundary conditions then led to a naive jump
\be
\Delta \hat{y}^\mu(t_-)=-\frac{k}{2}\sinh s_0 \frac{\tau^\alpha\nabla_\alpha \psi(y(s_0))}{n^\beta_-\nabla_\beta\Phi}n_-^\mu(t_-)
\ee
in the boundary trajectory.  As shown in figure \ref{jumpfig} we then used a discontinuous diffeomorphism to move this ``new'' boundary location back to the ``old'' boundary location, giving a retarded solution which obeys the correct boundary conditions.  In the unperturbed solution our scalar $\psi$ obeys Dirichlet boundary conditions, and thus will be constant along the boundary trajectory:
\be\label{upsi}
u^\mu_-\nabla_\mu \psi=0
\ee
This condition will be violated by the boundary jump, so to restore it we need to modify $\psi$ in some way which also obeys the wave equation to linear order in $k$ in the future of the discontinuity, which leads to a matter shockwave as in figure \ref{boundaryfig}.  To simplify notation we will assume that the boundary jumps outwards as in figure \ref{jumpfig}, the inward case is treated similarly.  Our ansatz for $\Delta \psi$ near the jump is then
\be
\Delta \psi=C \theta(\tau)\theta(\tau-s),
\ee
which is nonzero only in between the ``old'' and ``new'' boundaries, and to ensure the boundary conditions are restored we need
\be
C=-\Delta \hat{y}^\mu \nabla_\mu \psi=\frac{k}{2}\sinh s_0 \frac{\tau^\alpha\nabla_\alpha \psi(y(s_0))}{n^\beta_-\nabla_\beta\Phi}n_-^\mu\nabla_\mu \psi.
\ee
To proceed, we need to compute the backreaction of this matter discontinuity on the dilaton.  The energy-momentum tensor is
\be
T_{\mu\nu}=\nabla_\mu \psi \nabla_\mu \psi-\frac{1}{2}g_{\mu\nu}\nabla_\alpha \psi \nabla^\alpha \psi,
\ee
so we have
\begin{align}\nonumber
e^\mu e^\nu T_{\mu\nu}&=\frac{1}{2}\left((\tau^\mu\nabla_\mu \psi)^2+(e^\mu\nabla_\mu \psi)^2\right)\\\nonumber
&=e^\mu e^\nu T_{\mu\nu}^{\{0\}}+C\Big\{\big(\delta(\tau)\theta(-s)+\theta(\tau)\delta(\tau-s)\big)\tau^\mu\nabla_\mu \psi-\theta(\tau) \delta(\tau-s)e^\mu\nabla_\mu \psi\Big\}+O(k^2)\\\nonumber
e^\mu \tau^\nu T_{\mu\nu}&=e^\mu\nabla_\mu \psi\tau^\nu\nabla_\nu \psi\\
&=e^\mu \tau^\nu T_{\mu\nu}^{\{0\}}+C\Big\{ \big(\delta(\tau)\theta(-s)+\theta(\tau)\delta(\tau-s)\big)e^\mu\nabla_\mu \psi-\theta(\tau) \delta(\tau-s)\tau^\mu\nabla_\mu \psi\Big\}+O(k^2),\label{Tformulas}
\end{align}
where $T_{\mu\nu}^{\{0\}}$ is the unperturbed stress tensor.  Away from $s=0$ we only need to consider the terms involving $\delta(\tau)\theta(-s)$, which via the equations of motion \eqref{JTEOM} give additional contributions to the dilaton discontinuities. Using these equations in the form
\begin{align}\nonumber
\ddot{\Phi}+\Phi=-\frac{1}{2}e^\mu e^\nu T_{\mu\nu}\\
\frac{d}{d\tau}(e^\mu \nabla_\mu \Phi)=-\frac{1}{2}e^\mu \tau^\nu T_{\mu\nu},
\end{align}
we find the additional dilaton discontinuities
\begin{align}\nonumber
\Delta \dot{\Phi}&\supset -\frac{C}{2}\tau^\alpha\nabla_\alpha \psi\theta(-s)\\\nonumber
&=\frac{k}{4}\sinh s_0 \sinh\eta \frac{\tau^\alpha\nabla_\alpha \psi(y(s_0))}{n^\beta_-\nabla_\beta\Phi}(n_-^\mu\nabla_\mu \psi)^2\theta(-s)\\
&=\frac{k}{2}\sinh s_0 \sinh\eta \frac{\tau^\alpha\nabla_\alpha \psi(y(s_0))}{n^\beta_-\nabla_\beta\Phi} n_-^\mu n_-^\nu T_{\mu\nu}^{\{0\}}\theta(-s)\label{newdil1}
\end{align}
and
\begin{align}\nonumber
e^\mu\nabla_\mu \Delta \Phi&\supset -\frac{C}{2}e^\alpha \nabla_\alpha \psi\theta(-s)\\\nonumber
&=\frac{k}{4}\sinh s_0 \cosh\eta \frac{\tau^\alpha\nabla_\alpha \psi(y(s_0))}{n^\beta_-\nabla_\beta\Phi}(n_-^\mu\nabla_\mu \psi)^2\theta(-s)\\
&=\frac{k}{2}\sinh s_0 \cosh\eta \frac{\tau^\alpha\nabla_\alpha \psi(y(s_0))}{n^\beta_-\nabla_\beta\Phi} n_-^\mu n_-^\nu T_{\mu\nu}^{\{0\}}\theta(-s),\label{newdil2}
\end{align}
where we have used
\begin{align}\nonumber
\tau^\mu \nabla_\mu \psi&=-\sinh \eta\, n_-^\mu \nabla_\mu \psi\\
e^\mu \nabla_\mu \psi&=-\cosh \eta \,n_-^\mu \nabla_\mu \psi,
\end{align}
which follow from \eqref{undef2} and \eqref{upsi}.  Interestingly there is no extra discontinuity in $\Phi$ even though there is an extra discontinuity in $e^\mu\nabla_\mu \Phi$: this is because the spatial region where the extra discontinuity in $e^\mu \nabla_\mu \Phi$ exists is small, so integrating this would give a discontinuity in $\Phi$ which is second order in $k$.  Right at $s=0$ there will be additional $\delta$-function type singularities in $\Phi$ from the other terms in \ref{Tformulas}, but we will not need to determine these.  Indeed we can now see that the only modification in the calculation of $\{H_-,\psi_{\eta,s_0}\}$ is that in going from the third to the fourth line in equation \eqref{Hcalc}, the equations of motion need to be modified from \eqref{vacEOM} to include the unperturbed stress tensor and the dilaton discontinuities need to be modified to include \eqref{newdil1}, \eqref{newdil2}.  These together generate a new term
\be
\Delta^{(c)} H_-\supset \frac{k}{2\epsilon}\sinh s_0\frac{\tau^\alpha\nabla_\alpha \psi(y(s_0))}{n^\beta_-\nabla_\beta\Phi} n_-^\mu n_-^\nu T_{\mu\nu},
\ee
and therefore a new term
\be
\{H_-,\psi_{\eta,s_0}\}\supset \frac{1}{2\epsilon}\sinh s_0\frac{\tau^\alpha\nabla_\alpha \psi(y(s_0))}{n^\beta_-\nabla_\beta\Phi} n_-^\mu n_-^\nu T_{\mu\nu}
\ee
in the Peierls bracket. This however is precisely the term we found in \eqref{Honepsi2} by computing the bracket in the other order using $H_-$ as the generator.

\section{Boundary terms and the symplectic form on covariant phase space}\label{boundaryapp}
In this appendix we give an explicit demonstration that the discontinuities \eqref{psisjump} and the discontinuous diffemorphism $\xi^\mu$ which undoes the boundary jump \eqref{dely} as in figure \ref{jumpfig} together define a flow on covariant phase space which is indeed the one generated by the one-sided matter observable $\psi_{s_0,\eta}(t_-)$.  We will be careful about all boundary terms, and in particular we will see the importance of the somewhat mysterious boundary term in our expression \eqref{oneabsol} for $\beta(s)$.  Readers who are unfamiliar with covariant phase space may wish to first consult the partial review at the beginning of appendix \ref{Tapp}, although for a deeper understanding they may wish to consult \cite{Harlow:2019yfa} as well.

In computing the Peierls bracket, we introduced the linearized solution $\delta_g \phi \equiv \delta_R\phi-\delta_A \phi$ of the undeformed equations of motion.  In the covariant phase space formalism ``pre-phase space'' is defined as the set of solutions of the equations of motion which obey the boundary conditions, so any linearized solution about some particular solution can be interpreted as tangent vector at a point on pre-phase space and an assignment of a linearized solution at each point on pre-phase space can be thought of as a vector field.  Moreover variations $\delta\phi^a(x)$ are interpreted as differential forms. For our present purposes however it is convenient to consider a larger solution space where the boundary conditions only need to be obeyed at zeroth order in $k$, which we will call extended pre-phase space.  In particular we can then interpret the discontinuities \eqref{psisjump} as defining a vector field $X$ on extended pre-phase space, which obeys
\begin{align}\nonumber
X\cdot \delta \Phi(y(s))&=\frac{1}{2}\theta(s_0-s)\sinh(s_0-s)\tau^\alpha \nabla_\alpha \psi(y(s_0))\equiv \Delta \Phi\\\nonumber
X\cdot \delta \dot{\Phi}(y(s))&=-\frac{1}{2}\theta(s_0-s)e^\alpha \nabla_\alpha \psi(y(s_0))\equiv \Delta \dot{\Phi}\\\nonumber
X\cdot \delta g_{\mu\nu}(y(s))&=0\\
X\cdot \delta \psi(y(s))&=\delta_R\psi(y(s))-\delta_A\psi(y(s)),\label{Xdef}
\end{align}
where we have not been explicit about the theory-dependent action of $X$ on $\psi$ as we are primarily interested in boundary effects.  Similarly the diffeomorphism $\xi^\mu$, which moves the ``new'' boundary back to the ``old'' boundary in figure \ref{jumpfig}, is also associated to a vector field $X_\xi$ on extended pre-phase space whose action on the dynamical fields $\phi^a(x)$ is
\be
X_\xi\cdot \delta \phi^a(x)=\mathcal{L}_\xi \phi^a(x),
\ee
where $\mathcal{L}_\xi$ is the Lie derivative.  Separately $X$ and $X_\xi$ are not tangent to the true pre-phase space in which all solutions obey the boundary conditions, but their sum $X+X_\xi$ is and thus defines a natural vector field there.  To show that this vector field is indeed the one which implements the Hamiltonian evolution generated by $\psi_{s_0,\eta}(t_-)$, we need to show they obey\footnote{If the matter stress tensor does not vanish at the boundary there may be additional terms elated to the matter field jump we discussed in the previous appendix, but we will not discuss these again.}
\be\label{Cgoal}
(X+X_\xi)\cdot \wt{\Omega}=-\delta \psi_{s_0,\eta}(t_-),
\ee
where $\wt{\Omega}$ is the pre-symplectic form given by \eqref{presymp} (see for example the analogous equation \eqref{hamilton} for the Hamiltonian).  In JT gravity coupled to matter we have \cite{Harlow:2019yfa}
\be
\Theta=\theta\cdot \epsilon+\Theta_{matter},
\ee
where
\be
\theta^\mu=\Phi\left(g^{\mu\alpha}\nabla^\beta-g^{\alpha\beta}\nabla^\mu\right)\delta g_{\alpha \beta}+\left(\nabla^\mu\Phi g^{\alpha\beta}-\nabla^\alpha \Phi g^{\beta \mu}\right)\delta g_{\alpha \beta},
\ee
and
\be
C=c\cdot \epsilon_{\partial M} +C_{matter},
\ee
with
\be
c^\mu=\Phi u^\mu u^\nu n^\alpha \delta g_{\nu\alpha}.
\ee
Our variations obey the boundary conditions
\begin{align}\nonumber
\delta\Phi|_{\Gamma}&=0\\
u^\mu u^\nu \delta g_{\mu\nu}|_\Gamma&=0,\label{varbc}
\end{align}
where $\Gamma$ is the spatial boundary, and from the discussion around \eqref{dely} we also should require that
\begin{align}\nonumber
\xi^\mu u_\mu&=0\\
u^\mu u^\nu \nabla_\mu \xi_\nu&=0.
\end{align}
with the former following from \eqref{dely} and the latter ensuring that the induced metric on the boundary is preserved.  Using our one-sided geodesic $y^\mu_{t_-\eta}(s)$ to define a Cauchy slice $\Sigma$ (at least out to $s=s_0$, for larger $s$ we can bend the slice as we like), we then have
\begin{align}\nonumber
X\cdot \delta \Theta|_\Sigma=&X\cdot \delta \Theta_{matter}|_\Sigma-X\cdot\delta(\tau_\mu\theta^\mu\epsilon_\Sigma)\\\nonumber
=&X\cdot \delta \Theta_{matter}|_\Sigma\\\nonumber
&+\Big[\Delta\Phi\left(g^{\alpha\beta}\tau^\mu\nabla_{\mu}\delta g_{\alpha\beta}-\tau^\alpha \nabla^\beta\delta g_{\alpha\beta}\right)+\left(\tau^\beta \nabla^\alpha \Delta \Phi-\tau^\mu\nabla_\mu \Delta \Phi g^{\alpha\beta}\right)\delta g_{\alpha\beta}\Big]\epsilon_\Sigma\\\nonumber
=&X\cdot \delta \Theta_{matter}|_\Sigma+e^\alpha(e^\beta \tau^\mu-e^\mu \tau^\beta)\left(\Delta \Phi \nabla_\mu \delta g_{\alpha \beta}-\nabla_\mu \Delta\Phi \delta g_{\alpha\beta}\right)\epsilon_\Sigma
\end{align}
and
\begin{align}\nonumber
-X\cdot \delta C|_{\partial \Sigma}&=-X\cdot \delta C_{matter}|_{\partial \Sigma}-X\cdot \delta(u_\mu c^\mu\epsilon_{\partial \Sigma})\\
&=-X\cdot \delta C_{matter}|_{\partial \Sigma}+\Delta \Phi u^\alpha n^\beta \delta g_{\alpha \beta}\epsilon_{\partial \Sigma}.\label{CvarX}
\end{align}
These calculations greatly benefit from the fact that $X\cdot \delta g_{\mu\nu}=0$, as this means no metric variations need to be computed.  Combining these results we then have
\begin{align}\nonumber
X\cdot \wt{\Omega}=&\int_0^{s_0}ds\left(\Delta \Phi e^\alpha e^\beta \tau^\mu \nabla_\mu \delta g_{\alpha\beta}+2 e^\mu \nabla_\mu \Delta \Phi e^\alpha \tau^\beta \delta g_{\alpha \beta}-\Delta \dot{\Phi}e^\alpha e^\beta \delta g_{\alpha \beta}\right)\\\nonumber
&+\Delta \Phi(y(0))(e^\alpha \tau^\beta+u^\alpha n^\beta) \delta g_{\alpha\beta}(y(0))-\delta \psi(y(s_0))\\\nonumber
=&-\delta \psi(y(s_0))-\alpha(s_0)e^\mu\nabla_\mu \psi(y(s_0))-\beta(s_0)\tau^\mu \nabla_\mu \psi(y(s_0))\\\nonumber
&+\frac{1}{2}\tau^\mu \nabla_\mu \psi(y(s_0))\sinh s_0 \tanh \eta u^\alpha u^\beta \delta g_{\alpha\beta}(y(0))\\
=&-\delta \psi_{s_0,\eta}(t_-),
\end{align}
where we have integrated by parts and used \eqref{Xdef}, \eqref{oneabsol}, \eqref{undef2}, \eqref{varbc}, and \eqref{psi1var}, and in particular we have included the boundary term in $\beta(s_0)$.  Thus we see that \eqref{Cgoal} will follow provided we can show that $X_\xi\cdot \wt{\Omega}=0$.

In the covariant phase space formalism there is a standard set of techniques for computing quantities involving the vector field $X_\xi$ which implements a diffeomorphism on the set of solutions.  In our case however the diffeomorphism $\xi^\mu$ is field-dependent, so some extra care is needed.  In particular a little bit of thought shows that the action of the Lie derivative with respect to $X_\xi$ on the dynamical field differentials is given by
\be
\mathcal{L}_{X_\xi}\delta \phi^a(x)=\mathcal{L}_\xi\delta \phi^a(x)+\mathcal{L}_{\delta \xi}\phi^a(x),
\ee
which after some more thought tells us that for covariant $\Theta$ we have
\be
\mathcal{L}_{X_\xi}\Theta=\mathcal{L}_\xi \Theta+X_{\delta \xi}\cdot\Theta.
\ee
Making repeated use of Cartan's magic formula (see \cite{Harlow:2019yfa} for more details on such manipulations), we have
\begin{align}\nonumber
X_\xi \cdot \delta \Theta&=\mathcal{L}_{X_\xi}\Theta-\delta(X_\xi\cdot \Theta)\\\nonumber
&=\mathcal{L}_\xi \Theta+X_{\delta \xi}\cdot\Theta-\delta J_\xi-\delta(\xi \cdot L)\\\nonumber
&=d(\xi\cdot \Theta)+J_{\delta \xi}+\xi \cdot (d\Theta-\delta L)-\delta J_\xi\\
&=-\xi\cdot E_a\delta\phi^a+d\big(Q_{\delta \xi}-\delta Q_\xi+\xi\cdot \Theta\big).
\end{align}
Thus on-shell we have\footnote{The integral over $\partial \Sigma$ here is zero-dimensional, but we still write it since there can be multiple boundary components and we need to be careful about orientation.  We wrote the factors of $\epsilon_{\partial \Sigma}$ in \eqref{CvarX} for the same reason.}
\be
\int_\Sigma X_\xi\cdot \delta \Theta=\int_{\partial \Sigma}\big(Q_{\delta \xi}-\delta Q_\xi+\xi\cdot \Theta\big),
\ee
where in JT coupled to conformal matter $Q_\xi$ is given by \eqref{JTQ} (here we are assuming the matter fields do not contribute an additional $Z_\xi$ term to $Q_\xi$ a la equation \eqref{Qtot}, as we are assuming throughout the paper by using the Hamiltonian \eqref{Hpm}).  Evaluating this in JT gravity is logically straightforward but somewhat laborious, the result is
\be
X_\xi\cdot\int_\Sigma \delta \Theta=\int_{\partial \Sigma}\left[-\frac{1}{2}\Phi n^\alpha n^\beta \delta g_{\alpha\beta}u^\mu n^\nu\left(\nabla_\mu\xi_\nu+\nabla_\nu \xi_\mu\right)+u^\alpha n^\beta \delta g_{\alpha \beta}n^\mu n^\nu\left(\Phi\nabla_\mu \xi_\nu-\nabla_\mu \Phi \xi_\nu\right)\right]\epsilon_{\partial \Sigma}.
\ee
We also need to compute
\begin{align}\nonumber
-X_\xi\cdot\delta C&=X_\xi\cdot \delta\left(\Phi n^\alpha u^\beta \delta g_{\alpha\beta}\epsilon_{\partial \Sigma}\right)\\
&=\left(X_\xi\cdot \delta \Phi n^\alpha u^\beta \delta g_{\alpha\beta}+\Phi u^\beta (X_\xi \cdot\delta n^\alpha)\delta g_{\alpha \beta}-\Phi u^\beta \delta n^\alpha X^\xi\cdot \delta g_{\alpha\beta}\right)\epsilon_{\partial \Sigma},
\end{align}
which can be done using (see \cite{Harlow:2019yfa})
\begin{align}\nonumber
X_\xi\cdot \delta g_{\alpha\beta}&=\nabla_\alpha\xi_\beta+\nabla_\beta\xi_\alpha\\
\delta n^\mu&=(u^\mu u^\alpha n^\beta-\frac{1}{2}n^\mu n^\alpha n^\beta)\delta g_{\alpha \beta}.
\end{align}
After some simplification we then have
\be
-X_\xi\cdot \delta C=\left[\xi^\mu \nabla_\mu \Phi n^\alpha u^\beta \delta g_{\alpha \beta}-\Phi u^\alpha n^\beta \delta g_{\alpha \beta}n^\mu n^\nu \nabla_\mu \xi_\nu+\frac{1}{2}\Phi n^\alpha n^\beta \delta g_{\alpha\beta} u^\mu n^\nu (\nabla_\mu\xi_\nu+\nabla_\nu \xi_\mu)\right]\epsilon_{\partial \Sigma},
\ee
and thus
\be
X_\xi\cdot \wt{\Omega}=\int_{\partial \Sigma}\left[n^\alpha u^\beta \delta g_{\alpha \beta}\left(\xi^\mu \nabla_\mu \Phi -n^\mu n^\nu\nabla_\mu \Phi \xi_\nu\right)\right]\epsilon_{\partial \Sigma}.
\ee
Finally recalling that $\xi^\mu \propto n^\mu$, we see that these two terms cancel and thus we indeed have
\be
X_\xi\cdot \wt{\Omega}=0.
\ee
We view this as a strong check of our method in section \ref{onesidebracketsec}, and also a useful illustration of the power of the covariant phase space formalism and the importance of the $C$ term in the symplectic form \cite{Compere:2008us,Harlow:2019yfa}.

\bibliographystyle{jhep}
\bibliography{bibliography}
\end{document}